\documentclass[aps, prd, twocolumn, amsmath, floats,floatfix, superscriptaddress, nofootinbib]{revtex4}

\usepackage{epsfig}
\usepackage{amsmath,amssymb,mathrsfs,mathtools}
\usepackage[caption=false]{subfig}
\usepackage{verbatim}
\usepackage{float}
\usepackage{morefloats}
\usepackage[dvipsnames]{xcolor}
\usepackage{booktabs}
\usepackage[normalem]{ulem}
\usepackage{multirow}
\usepackage{makecell}
\usepackage{tabularx}
\usepackage{cancel}
\usepackage{hyperref}
\usepackage[T1]{fontenc}

\usepackage[capitalise]{cleveref}

\usepackage{comment}

\crefname{section}{Sec.}{Secs.}
\crefname{table}{Tab.}{Tabs.}
\crefname{figure}{Fig.}{Figs.}
\crefname{equation}{Eq.}{Eqs.}
\crefname{appendix}{Appendix\ }{Appendix\ }

\providecommand{\openone}{\leavevmode\hbox{\small1\kern-3.8pt\normalsize1}}

\usepackage{xspace}
\usepackage{mathrsfs}
\DeclareSymbolFontAlphabet{\mathrsfs}{rsfs}
\newcommand{\scri}{\mathrsfs{I}}
\newcommand{\scripx}{$\scri^+$}
\newcommand{\scrip}{\scripx\xspace}

\usepackage{booktabs}
\usepackage[Q=yes,pverb-linebreak=no]{examplep}
\usepackage{scrextend}

\allowdisplaybreaks[4]

\newcommand{\orcid}[1]{\href{https://orcid.org/#1}{\includegraphics[width=8pt]{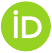}}}

\begin{document}

\title{\boldmath Free Hyperboloidal Evolution of the Einstein-Maxwell-Klein-Gordon System \unboldmath}

\author{Jo\~ao~D.~\'Alvares\,\orcid{0000-0001-5501-9014}}
\email{joaodinis01@tecnico.ulisboa.pt (corresponding author)}
\affiliation{CENTRA, Departamento de F\'{\i}sica do Instituto Superior T\'{e}cnico (IST), Universidade de Lisboa, 1049-001 Lisboa, Portugal}\
\affiliation{Instituto de Telecomunicaç\~oes (IT), Universidade de Aveiro Campus Universitário de Rua Santiago, 3810-193 Aveiro}

\author{Alex Va\~no-Vi\~nuales\,\orcid{0000-0002-8589-006X}}
\email{alex.vano@uib.es}
\affiliation{CENTRA, Departamento de F\'{\i}sica do Instituto Superior T\'{e}cnico (IST), Universidade de Lisboa, 1049-001 Lisboa, Portugal}
\affiliation{Departament de Física, Universitat de les Illes Balears, IAC3, Carretera Valldemossa km 7.5, E-07122 Palma, Spain}

\begin{abstract}
We present simulations of the Einstein-Maxwell-Klein-Gordon system on compactified hyperboloidal slices. To the best of our knowledge, these are the first hyperboloidal evolutions of this system that employ a common formulation like BSSN/Z4. 
Hyperboloidal slices smoothly reach future null infinity, the only location in spacetime where radiation (such as gravitational waves) is unambiguously defined. We are thus able to reach null infinity and extract signals there.  
We showcase the capabilities of our implementation in spherical symmetry with the evolution of a charged scalar field perturbing a regular spacetime and near an electrically charged black hole. We also present the collapse of a charged scalar field into a Reissner-Nördstrom black hole. Here we show sample evolutions and validation tests for these setups.
\end{abstract}

\maketitle

\section{Introduction}

The Einstein-Maxwell-Klein-Gordon (EMKG) system provides the simplest solutions to the Einstein equations beyond electrovacuum. Even in spherical symmetry, the EMKG system allows us to model interesting scenarios, such as a Reissner-Nördstrom (RN) (charged) black hole (BH) \cite{wald_general_1984}. Despite not having been detected in the universe, RN BHs include the extremal behaviour also present in the more astrophysically relevant Kerr (rotating) BHs, and thus serve as simpler models to study it. Another option is to evolve a charged scalar field determined by the Klein-Gordon (KG) equation. If massive, scalar fields are potential dark matter candidates, and can also be used to create boson stars \cite{lopez_charged_2023,jaramillo_full_2024,jetzer_charged_1989, zilhao_nonlinear_2015}. A previous numerical work evolving the Einstein-Maxwell equations in a formulation similar to ours include \cite{alcubierre_einstein-maxwell_2009}. The recent \cite{Patino:2025epy} evolves the Einstein-Maxwell system on a pseudospectral code. Instead of on spacelike slices, \cite{madler_characteristic_2025} evolves the EMKG system on a double null foliation. 
Such a null setup is also used in mathematical works that study (in)stability of the system motivated by the strong cosmic censorship conjecture. For instance, a charged massive scalar field perturbing RN is considered in~\cite{VandeMoortel:2017ztd}, which focuses on the behaviour of the Cauchy horizon. In our work, the initial slices considered for the RN spacetime are qualitatively like those in Fig.~3.13 in~\cite{vano-vinuales_free_2015}, so they do not cross the Cauchy horizon. We believe it is unlikely that they do even during the dynamical part of the evolution, but this has not been determined. 
A different setup (generalized wave coordinates adapted to the outgoing Schwarzschild light cones) is used in the global existence proofs for EMKG in~\cite{Kauffman:2021hgs}, although the spacetime considered here is close to Minkowski without matter. Our numerical simulations cannot proof existence or stability in the mathematical sense, but they show how the system behaves for the given initial setup.

The innovative aspect of our implementation is that it evolves the EMKG system on hyperboloidal slices \cite{zenginoglu_hyperboloidal_2008}. 
These slices are spacelike, but they become tangent to null rays at future null infinity, $\mathscr{I}^+$. The relevance for including \scrip in the numerical integration domain is that it is the only location in spacetime where gravitational radiation is unambiguously defined, and so it can be extracted there without systematic errors. Also, \scrip corresponds to the idealization of astrophysical observers measuring the emission of gravitational waves and electromagnetic radiation \cite{frauendiener_numerical_2000,leaver_spectral_1986,purrer_news_2005}.
Besides the hyperboloidal, other methods that also reach future null infinity are Cauchy-characteristic extraction/evolution \cite{Bishop:1996gt,Taylor:2013zia,moxon_spectre_2021}, whose evolution is limited in time, and Cauchy-characteristic matching \cite{bishop_cauchy-characteristic_1997,ma_fully_2023,Ma:2024hzq}, which currently employs a partially ill-posed formulation. 

The first approach to hyperboloidal evolution followed conformal compactification \cite{PhysRevLett.10.66} on which the Conformal Einstein Field Equations \cite{friedrich1983} are constructed. Since their first numerical implementations \cite{Hubner:1998hn,Frauendiener:1997zc}, hyperboloidal evolution has been revisited several times: with constrained evolutions \cite{Rinne:2009qx,Rinne:2013qc,Morales:2016rgt}, in a modern setup with the Conformal Einstein Field Equations \cite{Frauendiener:2021eyv,Frauendiener:2022bkj,Frauendiener:2023ltp,Frauendiener:2025xcj}, and with the dual-frame approach \cite{Hilditch:2015qea,Gasperin:2019rjg,Peterson:2023bha,Peterson:2024bxk}. This work builds upon an approach that expresses the conformally rescaled Einstein equations in BSSN \cite{NOK,PhysRevD.52.5428,Baumgarte:1998te} and Z4 \cite{bona_general-covariant_2003} form, and which has provided a complete understanding of the hyperboloidal problem in spherical symmetry \cite{Vano-Vinuales:2014koa,vano-vinuales_free_2015,Vano-Vinuales:2017qij,Vano-Vinuales:2023yzs,Vano-Vinuales:2023pum,Vano-Vinuales:2024tat} by evolving the Einstein's field equations (EFEs) coupled to a real massless KG equation and considering both regular and BH initial data. 
The purpose of this work is thus to extend this research line by simulating the EMKG system on hyperboloidal slices.

While the Maxwell equations are conformally invariant, coupling them to the EFEs and KG makes the relations between the quantitites more complicated. We focus on spherical symmetry, because the extensions of hyperboloidal slices to three dimensions are still under development. Spherical symmetry also implies that we do not need to consider the magnetic field \cite{alcubierre_einstein-maxwell_2009, torres_gravitational_2014}, making the electromagnetic coupling to the EFEs and KG equation solely through the electric field (due to the vanishing magnetic field). The asymptotic behaviour of hyperboloidal slices is not suitable for the inclusion of massive terms in the scalar field equation \cite{Gautam:2021ilg}, so our charged complex scalar field will also be massless. Here we will present results for several simulations achievable by our implementation, namely a charged scalar field perturbing a regular spacetime and a RN BH, and the gravitational collapse of a charges scalar field. 
In a companion paper \cite{resultspaper} we will exploit our access to future null infinity to study the behaviour of the scalar field there in the fully nonlinear regime. 

A setup similar to ours has recently been presented in \cite{masterlachlancampion}. The significant differences are that they use the Conformal Einstein Field Equations together with the conformal Gau{\ss} gauge and couple them to the conformally invariant wave equation, whereas we chose the commonly used BSSN/Z4 formulations with moving-puncture-type gauges and our wave equation is not conformally invariant. 

This paper is organised as follows: we will first show the equations that govern the EMKG system (Section~\ref{sec:EMKG}); then present the formulation of the Einstein equations to be used (Section~\ref{sec:Formulation}), followed by Maxwell's equations (Section~\ref{sec:Maxwell}) and the charged scalar field (Section~\ref{eq:chargedKG}). Then there is a discussion on the gauge conditions used during the simulations (Section~\ref{sec:gauge}). In this section we include the lapse and shift gauge conditions, and the version of the Lorenz gauge we chose. As a final ingredient for the simulations, Section~\ref{sec:initialdata} presents the initial data for the several cases to be considered in this paper, namely, charged scalar field perturbations of Minkowski and RN spacetimes and also the collapse of the charged scalar field into a BH. The numerical setup is described in Section~\ref{sec:numericalsetup}, and the results are shown in Section~\ref{sec:results}. The conclusions of the work are presented in Section~\ref{sec:conclusions}. Appendix \ref{appendix:gbssneinstein} presents the evolution equations of the system and the stress-energy tensor projections.

\section{Einstein-Maxwell-Klein-Gordon System}
\label{sec:EMKG}
The action describing the EMKG system is given by:
\begin{equation}
    S = \int \left(R - 8\pi\left[(\mathcal{D}_\mu \phi)^*(\mathcal{D}^\mu \phi) \right] - F^{\mu\nu}F_{\mu\nu}\right)\,\sqrt{-g}\,d^4x\;,
    \label{eq:actionfull}
\end{equation}
where $R$ is the Ricci scalar, $\mathcal{D}_\mu = \nabla_\mu + iq A_\mu$ is the gauge invariant covariant derivative, $F_{\mu\nu}$ is the Faraday/electromagnetic tensor and $\phi$ is the charged complex scalar field. $A_\mu$ is the four-potential and $q$ the scalar field's charge. We assume a four-dimensional spacetime, so the abstract indices sum over time and 3 spatial dimensions. The fundamental constants $G$ and $c$ have been set to unity. The Faraday tensor is defined as \cite{lopez_charged_2023,john_david_jackson_classical_1975},
\begin{equation}
    F_{\mu\nu} = \partial_\mu A_\nu - \partial_\nu A_\mu\, .
    \label{eq:electromagnetic_tensor}
\end{equation}
The energy-momentum tensor describing the contributions from Maxwell's and massless KG fields is given by
\begin{equation}
    \begin{aligned}
        T_{\mu\nu} &= F_{\mu\alpha}F_{\nu}^{\alpha} - \frac{1}{4}g_{\mu\nu}F_{\alpha\beta}F^{\alpha\beta}\\
        &+\mathcal{D}_{\mu}\phi \mathcal{D}^{\mu}\phi - \frac{1}{2}g_{\mu\nu}\mathcal{D}^{\alpha}\phi \mathcal{D}_{\alpha}\phi\,.
    \end{aligned}
\end{equation}
Due to the coupling between electromagnetism and the charged scalar field, the system does not change under the following transformations,
\begin{equation}
    \phi \rightarrow e^{iq \theta(x^\mu)} \phi \, , \quad A_\mu \rightarrow A_\mu - \partial_\mu \theta(x^\mu)\,. 
\end{equation}
This provides the following conserved current, according to Noether's theorem \cite{jaramillo_full_2024,lopez_charged_2023}:
\begin{equation}
    J_\mu = \frac{iq}{2}\left[\phi^* \mathcal{D}_\mu \phi - \phi (\mathcal{D}_\mu \phi)^* \right]\,.
    \label{eq:conservedcurrent}
\end{equation}
The asterisk * denotes the complex conjugate. The vanishing covariant derivative of the stress-energy tensor, $\nabla_\mu T^{\mu\nu} = 0$, yields the KG and Maxwell's equations
\begin{equation}
    \begin{aligned}
        \mathcal{D}_\mu \mathcal{D}^\mu \phi &= 0\,,\\
        \nabla^\mu F_{\nu\mu} &= 4 \pi J_\nu\,,\\
        \nabla_{[\mu}F_{\nu\sigma]} &= 0\,. 
    \end{aligned}
    \label{eq:kgandmax}
\end{equation}
The last equation involves the magnetic field. As our setup is in spherical symmetry, the magnetic part will not play a role and this equation can be ignored. 
Note that the current $J_\nu$ in \eqref{eq:kgandmax} is the same as the one in \eqref{eq:conservedcurrent}, denoting the coupling between electromagnetism and the charged scalar field.\par

\section{Formulation of the Einstein Field Equations}
\label{sec:Formulation}

Throughout this work, we will need to distinguish between different manifolds. We shall: start with the physical metric; rescale it via a conformal transformation (subsection~\ref{sec:conftransf}); decompose it in the time and spatial parts (ADM equations) and change them into a well-posed 3+1 decomposed formulation (BSSN/Z4) (subsections~\ref{sec:z4} and~\ref{sec:BSSNform}). Thus, we have three different stages, and moreover quantities that are time-independent.
We express each of them in the following notation:
\begin{itemize}\itemsep0em
\item physical quantities, written with a tilde: $\tilde{g}_{ab}, \tilde{K}_{ab}, \tilde{\nabla}_a, \tilde{n}_a$, ...; 
\item conformally rescaled quantities, with a bar: $\bar{g}_{ab}$, ...; 
\item 3+1 decomposed and spatially conformally rescaled quantities, without anything on top: $\gamma_{ab}$, $A_{ij}$, ...; 
\item time-independent quantities, with a hat: $\hat{g}_{ab}$, ...
\end{itemize}

According to this notation, every quantity appearing in Section~\ref{sec:EMKG} should have a tilde on top.

\subsection{Conformal Transformation}
\label{sec:conftransf}
Our method method requires being able to cover an infinite spacetime distance with finite values of our coordinates. This is possible by compactifying them (see~\eqref{eq:compacfactor}). However, the metric components now blow up in the asymptotic limit, and a regularization procedure is needed. The option we take is a conformal transformation following the usual procedure in~\cite{PhysRevLett.10.66},
\begin{equation}
    \bar{g}_{\mu\nu} = \Omega^2\tilde{g}_{\mu\nu},
    \label{eq:conformal_transformation}
 \end{equation}
 where $\Omega$ is the conformal factor, which goes to 0 at $\scri$ fast enough such that the rescaled metric is finite everywhere. The explicit expression for $\Omega$ will be given later (subsection~\ref{sec:regularinitialdata}). Writing the Einstein tensor $G_{\mu\nu}$ as a function of $\bar{g}_{\mu\nu}$ ($G_{\mu\nu}[\bar{g}]$), the EFEs become:
 \begin{equation}
    \begin{aligned}
        G_{\mu\nu}[\bar{g}] + \frac{2}{\Omega}(\bar{\nabla}_\mu \bar{\nabla}_\nu \Omega - \bar{g}_{\mu\nu} \bar{\nabla}^\mu \bar{\nabla}_\nu \Omega) &+ \\
        + \frac{3}{\Omega^2}\bar{g}_{\mu\nu} \bar{\nabla}^\alpha \Omega \bar{\nabla}_\alpha \Omega =& 8\pi \tilde{T}_{\mu\nu}\,,
    \end{aligned}
    \label{eq:einsteindecomp}
 \end{equation}
 where $\bar{\nabla}_\mu$ is the covariant derivative associated with $\bar{g}_{\mu\nu}$.
 
\subsection{Z4 Formalism}
\label{sec:z4}
We introduce here the Z4 formalism \cite{bona_symmetry-breaking_2004,bona_general-covariant_2003}, which adds a constraint-type variable $\tilde{Z}_\mu$ to the EFEs. This formalism includes terms that damp the constraints (those proportional to the $\kappa_i$'s) and others that propagate the constraint violations (those proportional to the derivatives of $\tilde{Z}_\mu$). This changes \eqref{eq:einsteindecomp} into:
\begin{equation}
    \begin{aligned}
        G_{\mu\nu}[\tilde{g}]& + 2\bar{\nabla}_{(\mu}\bar{Z}_{\nu)}-\bar{g}_{\mu\nu}\bar{\nabla}^\alpha \bar{Z}_\alpha + \frac{4}{\Omega}\bar{Z}_{(\mu}\bar{\nabla}_{\nu)}\Omega\\
        & - \frac{\kappa_1}{\Omega} (2 \bar{n}_{(\mu}\bar{Z}_{\nu)}+\kappa_2 \bar{g}_{\mu\nu}\bar{n}^\alpha \bar{Z}_\alpha) = 8 \pi \tilde{T}_{\mu\nu}\,,
    \label{eq:Z4EFE}
    \end{aligned}
 \end{equation}
where $\kappa_1$ and $\kappa_2$ are parameters to choose empirically \cite{vano-vinuales_free_2015} and we used $G_{\mu\nu}[\tilde{g}]$ instead of $G_{\mu\nu}[\bar{g}]$ for better readibility. $\tilde{n}^\alpha$ is the normal vector used in the 3+1 decomposition. It is a future-pointing unit vector, $\tilde{n}^\mu \tilde{n}_\mu = -1$ and under a conformal transformation, it changes as
\begin{equation}
    \tilde{n}^\mu = \Omega \bar{n}^\mu\,, \quad \tilde{n}_\mu = \Omega^{-1} \bar{n}_\mu\,.
    \label{eq:normaltransf}
\end{equation}
It is also important to define the 3+1 decomposition of the $\bar{Z}_\mu$ variable as 
\begin{equation}
    \bar{\Theta} = - \bar{n}^\mu \bar{Z}_\mu, \; Z^i = \bar{\gamma}^i_\nu \bar{Z}^\nu.
\end{equation}

\subsection{3+1 split into conformal Z4 / BSSN Formalism}
\label{sec:BSSNform}
For the 3+1 decomposition of the EFEs, we follow the generalized BSSN formalism \cite{brown_bssn_2008,brown_covariant_2009} compatible with the (conformal) Z4 described in the previous section. Let $\bar{\gamma}_{ij}$ be the purely spatial part of $\bar{g}_{\mu\nu}$, with associated covariant derivative $\bar D_i$. Then, the BSSN formalism introduces a new conformal factor, $\chi$, such that,
\begin{equation}
    \chi = \left( \frac{\gamma}{\bar{\gamma}}\right)\,,
\end{equation}
where $\gamma$ is the conformally rescaled metric from $\bar{\gamma}$. Our final line element reads \cite{zilhao_nonlinear_2015},
\begin{equation}
    \begin{aligned}
    ds^2 = -\alpha^2 dt^2 + \chi^{-1}\gamma_{ij}(dx^i + \beta^i dt)(dx^j + \beta^j dt)\,,
    \end{aligned}
    \label{eq:lineelement}
\end{equation} 
where $\alpha$ and $\beta^r$ are the (rescaled) lapse and the shift. The rescaled lapse relates to the physical one as $\tilde{\alpha} = \Omega^{-1} \alpha$. 
The conformal extrinsic curvature tensor $\bar{K}_{ij}$ is then divided into its trace and trace-free parts as,
\begin{equation}
    A_{ij} = \chi \bar{K}_{ij} - \frac{1}{3}\gamma_{ij}\bar{K}\,,
\end{equation}
with $\bar{K} = \bar{\gamma}^{ij}\bar{K}_{ij}$. We will however work with $K$, a mix of $\bar K$ and the Z4 quantity $\Theta$ given by
\begin{equation}
    K = \bar{K} - 2 \Theta\, .
    \label{eq:kdefinition}
\end{equation}
The relations between the conformal $\bar K$ and $\Theta$ and their physical counterparts $\tilde K$ and $\tilde \Theta$ are
\begin{equation}
    \tilde{K} = \Omega \bar{K} - \frac{3 \beta^i \partial_i \Omega}{\alpha}\,,\quad \tilde{\Theta} = \Omega \Theta\, .
    \label{eq:ktransformations}
\end{equation}
The variables evolved in the simulation (that come from our choice of formalism) are $A_{ij}$, $\tilde{K}$, $\tilde{\Theta}$, $\gamma_{rr}$, $\chi$, $\alpha$, $\beta^r$ and $\Lambda^r$. This last variable is defined as,
\begin{equation}
    \Lambda^i = \Delta \Gamma^i + 2 \gamma^{ij}Z_j\,,
\end{equation}
with $\Delta \Gamma^i = \Gamma^i - \hat{\Gamma}^i$ and $\Gamma^i = \gamma^{jk}\Gamma^i_{jk}$. $\hat{\Gamma}_{jk}^i$ is the connection coming from the time-independent spatial metric $\hat\gamma_{ij}$. For the details regarding the derivation of the conformally rescaled EFEs, see \cite{vano-vinuales_free_2015}. We leave the equations in their tensorial form in Appendix \ref{appendix:gbssneinstein}, together with the projections of the stress-energy tensor including the Maxwell and KG contributions.

\section{Maxwell's Equations}
\label{sec:Maxwell}
\subsection{Checking Conformal Invariance}
Maxwell's equations \cite{wald_general_1984} are expressed in terms of the electromagnetic tensor $\tilde{F}_{\mu\nu}$ and $\tilde{J}_{\nu}$ as defined in \eqref{eq:kgandmax}. To check that these equations are conformally invariant \cite{cote_revisiting_2019}, we propose that the electromagnetic tensor transforms as
\begin{equation}
   \bar{F}_{\mu\nu} = \Omega^s\tilde{F}_{\mu\nu},
\end{equation}
where $s$ is a constant to be determined. Looking at the covariant derivative of the electromagnetic tensor, as done in \cite{wald_general_1984}, in $n$ dimensions,
\begin{equation*}
   \bar{g}^{\beta \mu}\bar{\nabla}_{\beta}\bar{F}_{\mu\nu} = \Omega^{s+2}\tilde{g}^{\beta \mu}\tilde{\nabla}_\beta \tilde{F}_{\mu\nu} + (n-4+s)\Omega^{s+1}\tilde{g}^{\beta \mu}\tilde{F}_{\mu\nu}\tilde{\nabla}_\beta \Omega
\end{equation*}
In the physically relevant case, $n=4$, the right-most term disappears if we impose that $s=0$,
\begin{equation}
   \bar{g}^{\beta \mu}\bar{\nabla}_\beta \bar{F}_{\mu\nu} = \Omega^{2} \tilde{g}^{\beta \mu} \tilde{\nabla}_\beta\tilde{F}_{\mu\nu}.
   \label{eq:Ftransformation}
\end{equation}
This is enough to show the conformal invariance of Maxwell's equations in vacuum. If a non-zero four-current is present \cite{arbab_conformal_2021}, it must change according to
\begin{equation}
   \bar{J}_{\mu} = \Omega^{2} \tilde{J}_{\mu}
   \label{eq:4currentinv}
\end{equation}
to preserve the invariance of the equations, as was already observed in \cite{cote_revisiting_2019}. This will be confirmed later with the conserved current coming from the KG field. Note also that \eqref{eq:electromagnetic_tensor} implies that the electromagnetic potential $\tilde{A}_\mu$ does not change under conformal transformations.\par
As a side note, the value of $s-2$ is related to the mass of the photon: \cite{arbab_conformal_2021} proposes considering the conformal factor $\Omega$ as a background field related to electromagnetism. Our choice, $s = 0$, indicates we are assuming that the photon is massless.

\subsection{Electric Field}
To derive the electric field's evolution equation, we consider the alternative way of writing the Faraday tensor \cite{bona_elements_2009},
\begin{equation}
    \tilde{F}_{\mu\nu} = \tilde{n}_{\mu}\tilde{E}_\nu - \tilde{n}_\nu \tilde{E}_\mu + \tilde{\epsilon}_{\mu \nu \alpha \beta}\tilde{B}^{\alpha}\tilde{n}^\beta,
    \label{eq:maxwellelectric}
\end{equation}
where $\tilde{E}_\nu$ and $\tilde{B}_\alpha$ are the unrescaled electric and magnetic fields respectively. The quantity $\tilde{\epsilon}^{\mu \nu \alpha \beta}$ is the Levi-Civita tensor, defined with respect to the Levi-Civita symbol $\eta^{\mu\nu\alpha\beta}$ by:
\begin{equation}
    \tilde{\epsilon}_{\mu \nu \alpha \beta} = \frac{1}{\sqrt{\tilde{g}}}\eta_{\mu \nu \alpha \beta}.
    \label{eq:levicivitatensor}
\end{equation}
Using the conformal invariance of $\tilde{F}_{\mu\nu}$, we get that 
\begin{equation}
    \begin{aligned}
    \tilde{F}_{\mu\nu} &=\bar{F}_{\mu\nu} \qquad \textrm{can equivalently be written as}\\ %
    & \tilde{n}_{\mu}\tilde{E}_\nu - \tilde{n}_\nu \tilde{E}_\mu + \tilde{\epsilon}_{\mu \nu \alpha \beta}\tilde{B}^{\alpha}\tilde{n}^\beta\\
    & = \Omega^{-1} \bar{n}_{\mu}\tilde{E}_\nu - \Omega^{-1} \bar{n}_\nu \tilde{E}_\mu + \Omega^4 \bar{\epsilon}_{\mu \nu \alpha \beta} \tilde{B}^{\alpha}\Omega \bar{n}^\beta,
    \end{aligned}
    \label{eq:FaradayElectric}
\end{equation}
where we used \eqref{eq:levicivitatensor} to obtain the transformation law $\tilde{\epsilon}_{\mu \nu \alpha \beta} = \Omega^4 \bar{\epsilon}_{\mu \nu \alpha \beta}$ and $\tilde{\epsilon}^{\mu \nu \alpha \beta} = \Omega^{-4} \bar{\epsilon}^{\mu \nu \alpha \beta}$. We also used the transformation of the normal vector \eqref{eq:normaltransf}. Both the electric and magnetic fields are frame-dependent spatial vectors. Defining the rescaled electric and magnetic fields as
\begin{equation}
    \begin{aligned}
        \bar{E}_\mu = \frac{\tilde{E}_\mu}{\Omega},\; \bar{E}^\mu = \frac{\tilde{E}^\mu}{\Omega^3},\; \bar{B}^\mu = \Omega^5 \tilde{B}^\mu,\; \bar{B}_\mu =  \frac{\tilde{B}_\mu}{\Omega^9}\,,
    \end{aligned}
    \label{eq:elmagntransf1}
\end{equation}
allows for writing the rescaled electromagnetic tensor in the same functional form as the physical one \eqref{eq:maxwellelectric}. Since there is no magnetic field in spherical symmetry, from now on we will only consider the electric field, which we will use in its rescaled version, $\bar{E}^i$.

\subsection{Adapted Maxwell's Equations}
Following \cite{bona_elements_2009}, we shall use an adapted version of Maxwell's equations that reads:
\begin{equation}
    \tilde{\nabla}^\mu \left(\tilde{F}_{\nu\mu} + \tilde{g}_{\mu\nu}\Psi \right) = 4 \pi (\tilde{J}_\nu-  \tilde{n}_\nu k  \Psi) \;\;,
    \label{eq:faradaypsi}
\end{equation}
where $k$ is a constant to control the evolution of $\Psi$ and in the simulations shown in this paper it will be set to 1. $\Psi$ can be interpreted as the numerical deviation from the Poisson constraint associated with Maxwell's equations. Together with \eqref{eq:maxwellelectric}, this gives the evolution equations for the electric field and for $\Psi$:
\begin{subequations}
    \begin{align}
    \partial_t \bar{E}^i &= \beta^j D_j \bar{E}^i - \bar{E}^j D_j \beta^i + \bar{E}^i K \alpha + \nonumber\\
    & - 4 \pi \bar{j}^i \alpha + \frac{\gamma^{ik} \alpha \chi D_k \Psi}{\Omega^2}\\
    \partial_t \Psi &= \beta^j D_j \Psi + \alpha \Omega^2 D_i \bar{E}^i - 4\pi \bar{q}_{\text{dens.}} \alpha + \nonumber\\
    & + 4 \pi k \alpha \Omega^2 \Psi - \frac{3 \bar{E}^i \alpha \Omega^2 D_i \chi}{2 \chi} \label{eq:psieveq}\,,
    \end{align}
    \label{eq:ElPsieveq}
\end{subequations}
where $D_i$ is the covariant derivative defined from $\gamma_{ij}$, $K$ is the conformally rescaled trace of the extrinsic curvature and $\alpha = \Omega \tilde{\alpha}$. While these expressions are written tensorially, the magnetic field $\bar{B}^i$ has been set to 0. $\tilde{j}^i$ and $\tilde{q}_{\text{dens.}}$ come from the 3+1 decomposition of the four-current $\tilde{J}^\mu$:
\begin{align}
    \bar{j}_i &= \bar{\gamma}_i^\mu \bar{J}_\mu \;\;, \quad \bar{q}_{\text{dens.}} = -\bar{n}^\mu \bar{J}_\mu \;\;.
    \label{eq:currentdecomp}
\end{align}
If we set $\Psi$ to 0 in its evolution equation \eqref{eq:psieveq}, we  recover the Poisson/Gauss constraint \cite{alcubierre_einstein-maxwell_2009}, 
\begin{equation}
    D_i \bar{E}^i - \frac{3}{2}\frac{\bar{E}^i D_i \chi}{\chi} = \bar{D}_i \bar{E}^i = 4\pi \bar{q}_{\text{dens.}}\,.
    \label{eq:poissonconst}
\end{equation}

Regarding the four-potential $\tilde{A}_\mu = \bar{A}_\mu$, we 3+1 decompose it as well,
\begin{equation}
    \bar{\Phi} =  -\bar{n}_\mu \bar{A}^\mu \;\;, \quad \bar{A}_3^i = \bar{\gamma}^i_\mu \bar{A}^\mu \;\;.
    \label{eq:fourpotentialdecomp}
\end{equation}
$\bar{\Phi}$ is a scalar potential related to the electric potential, as we will see further, and $\bar{A}_3$ is the vector potential. The evolution equation for ${\bar{A}_3}^i$ is obtained from \eqref{eq:electromagnetic_tensor} and \eqref{eq:FaradayElectric},
\begin{equation}
    \partial_t \bar{A}_3^i = \mathcal{L}_\beta \bar{A}_3^i + \alpha \bar{E}^i + \bar{D}^i(\alpha \bar{\Phi})\,.
    \label{eq:A3reveq}
\end{equation}

\section{Charged-Scalar Field}
\label{eq:chargedKG}
We define the rescaled scalar field $\bar\phi$ as
\begin{equation}
    \bar{\phi} = \frac{\tilde{\phi}}{\Omega}\,.
    \label{eq:scalarfielconf}
\end{equation}
Expressed in terms of $\bar\phi$, the KG equation \eqref{eq:kgandmax} is changed to
\begin{equation}
    \bar{\mathcal{D}}_\mu \bar{\mathcal{D}}^\mu \tilde{\phi} - \frac{2}{\Omega}(\bar{\mathcal{D}}_\mu \tilde{\phi})(\bar{\mathcal{D}}^\mu \Omega)= 0\,,
\end{equation}
with $\bar{\mathcal{D}}_\mu = \bar{\nabla}_\mu + i q \bar{A}_\mu$. To simulate the complex scalar field, it is useful to separate it into its real and imaginary parts:
\begin{equation}
    \bar{\phi} = \bar{c}_\phi + i \bar{d}_\phi \;, \quad \bar{c}_\phi, \bar{d}_\phi \in \mathbb{R} \;\;,
\end{equation}
and to reduce the KG equation to four first-order in time equations, we define,
\begin{equation}
    \bar{c}_\Pi = \partial_t \bar{c}_\phi, \; \bar{d}_\Pi = \partial_t \bar{d}_\phi . 
\end{equation}
Note that by imposing \eqref{eq:scalarfielconf}, the four-current indeed changes according to \eqref{eq:4currentinv}. Explicitly, the four-current projections in terms of the scalar field quantities are given by
\begin{equation}
    \begin{aligned}
    \tilde{j}_i &= \Omega^2\left(-q^2 \tilde{A}_{3i} (\bar{c}_\phi^2 + \bar{d}_\phi^2) + q \bar{c}_\phi D_i \bar{d}_\phi + q \bar{d}_\phi D_i \bar{c}_\phi \right)\,,\\
    \tilde{q}_{\text{dens.}} &= \Omega^2 \bigg(-q^2 \bar{\Phi}(\bar{c}_\phi^2 + \bar{d}_\phi^2) + q\frac{\bar{c}_\phi \bar{d}_\Pi -\bar{d}_\phi \bar{c}_\Pi}{\alpha} + \\
    &+ q \frac{\bar{d}_\phi \beta^i D_i \bar{c}_\phi - \bar{c}_\phi \beta^i D_i \bar{d}_\phi}{\alpha}\bigg) . 
    \end{aligned}
    \label{eq:chargedens}
\end{equation}

\section{Gauge Conditions}
\label{sec:gauge}
\subsection{Lapse and Shift}
For the lapse and shift gauge conditions, we follow those presented in \cite{vano-vinuales_free_2015}, which are adapted to hyperboloidal slices:
\begin{subequations}
    \begin{align}
        \dot{\alpha} &= \beta^r \alpha^\prime - \hat{\beta}^r \hat{\alpha}^\prime - \frac{(n_{cK}(r_{\scri^+}^2-r^2)^4+\alpha^2) \Delta \tilde{K}}{\Omega} +\nonumber \\
        & + \frac{\Omega^\prime}{\Omega}(\hat{\beta}^r \hat{\alpha}-\beta^r \alpha) +\frac{\xi_{cK}(\hat{\alpha}-\alpha)}{\Omega}\,,\\
        \dot{\beta}^r &= \beta^r {\beta^r}^\prime - \hat{\beta}^r \hat{\beta}^{r\prime} + \frac{3}{4}\left(\lambda(r_{\mathscr{I}^+}^2-r^2)+\alpha^2 \chi \right)\Lambda^r + \nonumber \\
        &+ \eta(\hat{\beta}^r - \beta^r) + \xi_{\beta^r}\left(\frac{\hat{\beta}^r}{\Omega}-\frac{\beta^r}{\Omega} \right)\,.
    \end{align}
    \label{eq:gaugeconditions}
\end{subequations}
$n_{cK}, \lambda, \xi_{cK}, \eta, \xi_{\beta^r}$ are constants to be chosen, as the system to be evolved requires, and $\Delta \tilde{K} = \tilde{K} - K_{\text{CMC}}$, with $K_{\text{CMC}}$ defined later in Section~\ref{sec:initialdata}. The hatted variables are derived from a time-independent background metric, which we set to be that of Minkowski on a constant-mean-curvature hyperboloidal slice. Their explicit expressions will be given in Section~\ref{sec:initialdata}. $r_{\mathscr{I}^+}$ is the radial location of $\mathscr{I}^+$, set to 1 in our simulations. The gauge conditions considered allow the coordinate position of $\scri^+$ to remain the same throughout the evolutions, which is known as ``scri-fixing''. The following parameters were set in all simulations: $K_{\text{CMC}} = -1,\, \kappa_1 = 1.5,\, \kappa_2 = 0.1,\, \lambda = 1.0,\, \eta = 0.1,\, \xi_{\beta^r} = 0.0$. For regular and strong field initial data, $n_{\text{cK}} = 1.0,\, \xi_{\text{cK}} = 1.0$, while for collapse simulations $n_{\text{cK}} = 0.0,\, \xi_{\text{cK}} = 2.0$.
\subsection{Adapted Lorenz Gauge}
The last gauge condition to complete the system of evolution equations is the one for $\bar{\Phi}$. The most common choice is the Lorenz gauge \cite{zilhao_nonlinear_2015,lopez_charged_2023,jaramillo_full_2024}:
\begin{equation}
    \tilde{\nabla}_\mu \tilde{A}^\mu = 0\,.
    \label{eq:lorenzgauge}
\end{equation}
It is also possible to use the ``conformal'' Lorenz gauge, $\bar{\nabla}_\mu \bar{A}^\mu =0$, or the light-cone gauge \cite{cote_revisiting_2019, majumdar_residual_2023, mccartor_light-cone_1994}, which reads,
\begin{equation}
    l_\mu \tilde{A}^\mu = 0\,,
\end{equation}
where $l^\mu$ is an outgoing null vector. We have tested all of these possibilities and found that the electromagnetic potentials, $\bar{A}_3$ and $\bar{\Phi}$, steadily grow at $\mathscr{I}^+$. Inspired by \cite{hilditch_introduction_2013}, we propose an alternative Lorenz gauge, where the factors in the evolution equation that envolve $\tilde{A}_{3i}$ are made proportional to a factor $\mu$. Starting from \eqref{eq:lorenzgauge}, we obtain
\begin{equation}
    \begin{aligned}
        \dot{\bar{\Phi}} &= \beta^i \bar{D}_i \bar{\Phi} + \bar{K}\alpha\bar{\Phi} - \frac{2\bar{\Phi} \beta^i \bar{D}_i \Omega}{\Omega} + \\
        & - \mu(\alpha\bar{D}^i \bar{A}_{3i} + \bar{A}_3^i \bar{D}_i \alpha- \frac{2 \bar{A}_3^i \alpha\bar{D}_i \Omega}{\Omega})\,.
    \end{aligned}
    \label{eq:modifiedLG}
\end{equation}
For $\mu \leq 0.5$ the variables eventually reach a stationary state at $\mathscr{I}^+$. We found that making the change $\mu \rightarrow \mu (1-r^2_{\mathscr{I}^+})$ worked even better for a wider range of $\mu$, and we stick with this version of the adapted Lorenz gauge.\par
This change by hand of the principal part of the Lorenz gauge preserves the strong hyperbolicity of the system of partial differential equations (PDEs) describing the EMKG system. One would have to take care when extending beyond spherical symmetry, for the evolution equation of the magnetic field will make the PDE system become weakly hyperbolic. See \cite{alcubierre_einstein-maxwell_2009} for a brief discussion on that.
\section{Initial Data}
\label{sec:initialdata}

\subsection{Hyperboloidal Conformal Compactification}
Foliating spacetime along hyperboloidal slices can be done through a height function \cite{zenginoglu_gravitational_2008,gentle_constant_2001,malec_constant_2003}, $h(\tilde{r})$, as
\begin{equation}
    t = \tilde{t} - h(\tilde{r})\,,
    \label{eq:hypchange}
\end{equation}
with $\tilde{t}$ being the physical time and $t$ the new time coordinate whose level sets give hyperboloidal slices. This height function satisfies $\partial_{\tilde{r}}h(\tilde{r}) < 1$ everywhere \cite{vano-vinuales_wave_2023} and this derivative asymptotes to 1 at $\mathscr{I}^+$. The radial derivative of the height function is also called the ``boost'' function.\par
In a spherically symmetric spacetime, we can start with the generic line element:
\begin{equation}
    d\tilde{s}^2 = -A(\tilde{r})d\tilde{t}^2 + \frac{1}{A(\tilde{r})}d\tilde{r}^2 + \tilde{r}^2 d\tilde{\sigma}^2.
\end{equation}
When we introduce the height function \eqref{eq:hypchange},
the line element becomes,
\begin{equation}
    \begin{aligned}
    d\tilde{s}^2 &= -A(\tilde{r}) dt^2 - 2A(\tilde{r})h'(\tilde{r})dtd\tilde{r}+\\
    &+\frac{[1-(A(\tilde{r})h'(\tilde{r}))^2]}{A(\tilde{r})}d\tilde{r}^2 + \tilde{r}^2 d\sigma^2 . 
    \end{aligned}
    \label{eq:lineB}
\end{equation}
However, we are still lacking a functional form for $h(\tilde{r})$. From the definition of the extrinsic curvature \cite{vano-vinuales_free_2015},
\begin{equation}
    \begin{aligned}
        \tilde{\bar{K}} &= -\frac{1}{\sqrt{-\tilde{g}}}\partial_a (\sqrt{-\tilde{g}} \tilde{n}^a) \\
        &= -\frac{1}{\tilde{r}^2}\partial_{\tilde{r}} \left[\tilde{r}^2 \frac{A^{3/2}(\tilde{r})h'(\tilde{r})}{\sqrt{1-(A(\tilde{r})h'(\tilde{r}))^2}} \right]\,,
    \end{aligned}
    \label{eq:knormalmetric}
\end{equation}
where in the second line we have used the metric components from~\eqref{eq:lineB}. We will determine $h'(\tilde r)$ by imposing that the hyperboloidal slices it defines correspond to constant-mean-curvature (CMC) ones, which implies that the physical extrinsic curvature trace ($\tilde{K}$) is constant. Therefore, we can integrate \eqref{eq:knormalmetric}, choosing $\tilde{K} = K_{\text{CMC}} < 0$. In our convention following~\cite{Misner1973}, it is asymptotically negative for slices reaching \scrip, and positive for slices reaching $\scri^-$. Let $C_{\text{CMC}}$ be the constant coming from the integration and solving \eqref{eq:knormalmetric} for $h(\tilde{r})$, we get
\begin{equation}
    h'(\tilde{r}) = - \frac{\frac{K_{\text{CMC}}\tilde{r}}{3} + \frac{C_{\text{CMC}}}{\tilde{r}^2}}{A(\tilde{r})\sqrt{A(\tilde{r}) + \left(\frac{K_{\text{CMC}}\tilde{r}}{3} + \frac{C_{\text{CMC}}}{\tilde{r}^2}\right)^2}}.
    \label{eq:heightder}
\end{equation}
To put $\mathscr{I}^+$ at a finite radial coordinate, we introduce a compactification factor, $\bar{\Omega}$, such that
\begin{equation}
    \tilde{r} = \frac{r}{\bar{\Omega}(r)}\,,
    \label{eq:compacfactor}
\end{equation}
and note that $\bar{\Omega}$ and $\Omega$ are in principle different quantities. Inserting \eqref{eq:compacfactor} into \eqref{eq:lineB}, we get,
\begin{equation}
    \begin{aligned}
    d\tilde{s}^2 &= - A dt^2 - 2A h^\prime \frac{\bar{\Omega}-r\bar{\Omega}^\prime}{\bar{\Omega}^2}dtdr +\\
    & + \frac{1-(Ah^\prime)^2}{A}\frac{(\bar{\Omega}-r\bar{\Omega}^\prime)^2}{\bar{\Omega}^4}dr^2 + \frac{r^2}{\bar{\Omega}^2}d\sigma^2
    \end{aligned}
    \label{eq:dfinal}
\end{equation}
The final step is the conformal transformation coming from the metric, $ds^2 = \Omega^2 d\tilde{s}^2$.
Comparing \eqref{eq:lineelement}, in spherical symmetry, with \eqref{eq:dfinal}, we then have the following initial data,
\begin{gather}
    \alpha_0 = \Omega \tilde{\alpha}_0, \quad {\gamma_{rr}}_0 = \frac{(\bar{\Omega}-r\bar{\Omega}')^2}{\tilde{\alpha}_0^2 \bar{\Omega}^2}, \quad \chi_0 = \frac{\bar{\Omega}^2}{\Omega^2}\nonumber\\
     \beta^r_0 = \frac{\bar{\Omega}^2\left(\frac{K_{\text{CMC}}r}{3\bar{\Omega}} + \frac{C_{\text{CMC}}\bar{\Omega}^2}{r^2} \right)\tilde{\alpha}_0}{(\bar{\Omega}-r \bar{\Omega}')}\label{eq:indatagen2}\\
     \text{with } \tilde{\alpha}_0 = \sqrt{A\left(\frac{r}{\bar{\Omega}} \right) + \left(\frac{K_{\text{CMC}}r}{3\bar{\Omega}} + \frac{C_{\text{CMC}}\bar{\Omega}^2}{r^2} \right)^2} . \nonumber
     \label{eq:initialdataoverall}
\end{gather}
Together with this, we also take $\gamma_{\theta\theta} = \sqrt{\gamma_{rr}}$, to fix the freedom introduced with the evolution variable $\chi$. A convenient choice \cite{vano-vinuales_free_2015} for $\bar{\Omega}$ is taking the initial spatial metric to be conformally flat $\gamma_{rr} = 1$, which is the same as imposing initial isotropic coordinates. Initial data for the remaining evolution variables from the EFEs are: $\Lambda^r_0 = \tilde{\Theta}_0 = Z_{r0} = 0$ and $A_{rr} = -\frac{2C_{\text{CMC}\bar{\Omega}^3}}{r^3 \Omega}$.

\subsection{Regular Initial data}
\label{sec:regularinitialdata}
In flat spacetime, $A(\tilde{r})=1$. We additionally take $C_{\text{CMC}} = 0$. This allows \eqref{eq:heightder} to have an analytic solution,
\begin{equation}
    h(\tilde{r}) = \sqrt{\tilde{r}^2 + \left(\frac{3}{K_{\text{CMC}}} \right)^2}\,.
\end{equation}
The conformally flat condition then gives an analytical formula for $\bar{\Omega}$,
\begin{equation}
    \bar{\Omega} = -K_{\text{CMC}}\frac{r^2_{\mathscr{I}^+}-r^2}{6r_{\mathscr{I}^+}}\,.
    \label{eq:confexp}
\end{equation}
For simplicity, we further choose $\Omega = \bar{\Omega}$, for the expression in \eqref{eq:confexp} is also suited for the conformal factor. We will remain with this expression for $\Omega$ throughout the rest of the work. The background metric $\hat{\gamma}_{ij}$ will take the same values as the metric we describe here (corresponding to Minkowski spacetime). \eqref{eq:indatagen2} simplifies to,
\begin{gather}
    \alpha_0 = \sqrt{\Omega^2 + \left( \frac{K_{\text{CMC}}r}{3}\right)^2}, \quad \chi_0 = 1, \quad \gamma_{rr0}= 1,\nonumber\\
    \beta^r_0 =\frac{K_{\text{CMC}}r}{3}, \quad A_{rr0} = \Theta_ 0 = Z_{r0} = \Lambda_{r0} = 0\,.
    \label{eq:indatagen3}
\end{gather}
$\hat{\alpha}$ and $\hat{\beta}$ will take the values of $\alpha_0$ and $\beta^r_0$ as described above.

\subsection{Strong Field Initial Data}
\label{subsec:strongfield}
An important aspect in strong field initial data is the careful choice of $C_{\text{CMC}}$. CMC slices in general share the property of being horizon-penetrating slices, but the critical value of $C_{\text{CMC}}$ will give us trumpet initial slices \cite{baumgarte_analytical_2007, friedrich_cauchy_2000}, which are useful to avoid the singularity. The way to do this is by setting the discriminant of the denominator of $\tilde{\gamma}_{rr}$ to 0, already with the expression for $h^\prime(\tilde{r})$ substituted
\begin{equation}
    \tilde{\gamma}_{\tilde r\tilde r} = \frac{1}{A(\tilde{r}) + \left(\frac{K_{\text{CMC}}\tilde{r}}{3} + \frac{C_{\text{CMC}}}{\tilde{r}^2}\right)^2}\,,
\end{equation}
where we shall consider $A(\tilde{r}) = 1 - \frac{2M}{\tilde{r}}+\frac{Q^2}{\tilde{r}^2}$, for a RN BH. A vanishing discriminant is equivalent to having a double root $\tilde{r} = R_0$ in the denominator. $R_0$ corresponds to the innermost value of the Schwarzschild-like radial coordinate that the slices reach and thus can be interpreted as the throat of the trumpet. We redirect the interested reader to \cite{vano-vinuales_spherically_2023} to see how slices change when choosing a $C_{\text{CMC}}$ different from the critical value.\par
\begin{figure}
    \centering
    \includegraphics[width=0.4\textwidth]{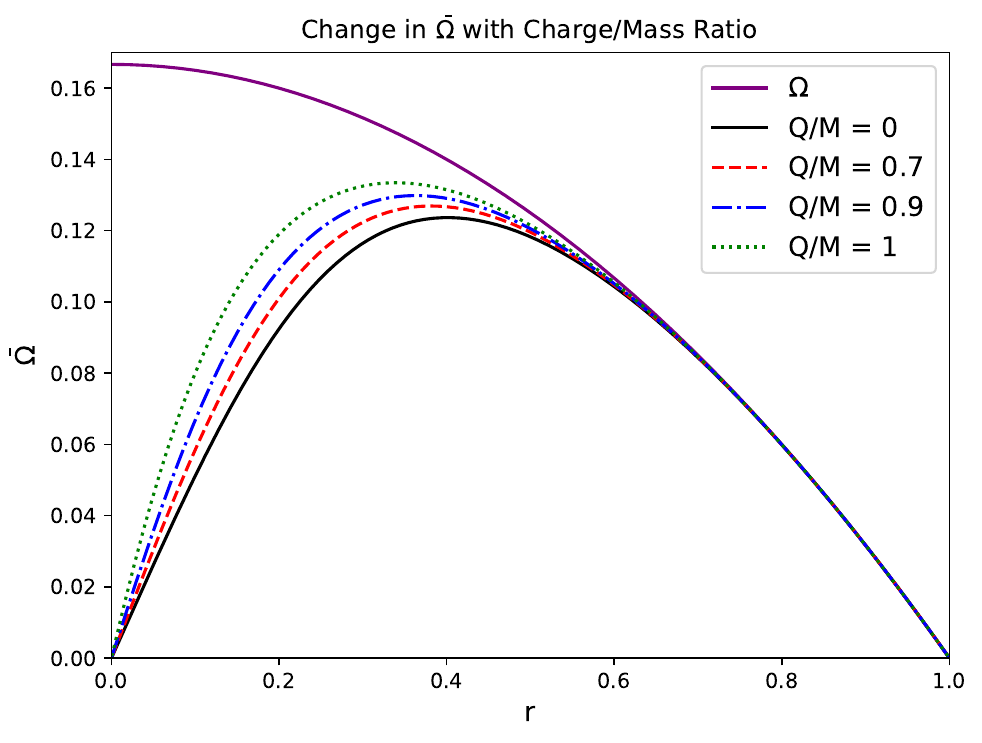}
    \caption{Variation of $\bar{\Omega}$ according to different values of the charge-to-mass ratio $Q/M$, as solution to~\eqref{eq:indatagen2} setting ${\gamma_{rr}}_0=1$. $M$ has been set to 1 and $Q$ is changing.}
    \label{fig:aconfQMratio}
\end{figure}

The conformally flat metric condition does not produce an analytical result now, therefore having to be numerically integrated by solving~\eqref{eq:indatagen2} with ${\gamma_{rr}}_0=1$. Fig.~\ref{fig:aconfQMratio} shows how $\bar{\Omega}$ changes with the charge to mass ratio, with the mass set to 1 and $K_{\text{CMC}} = -1$. Note how $\bar{\Omega}$ and $\Omega$ go to zero at the same rate as they approach \scrip.

\subsection{Charged Scalar Field Initial Data}
\label{subsec:chargedscalarinitialdata}
To make sure the constraints are initially satisfied, we have to make sure that the Hamiltonian, $\mathcal{H}$, momentum, $\mathcal{M}_r$, and Poisson, $\mathcal{P}$, constraints are solved (respectively, \eqref{eq:Ham}, \eqref{eq:Mom} and \eqref{eq:poissonconst}). Assuming we have a perturbation in the real part of the scalar field and another one in $\bar{A}_{3r}$, the condition that solves the momentum constraint reads
\begin{equation}
    \bar{c}_\Pi = \beta^r \bar{c}_\phi^\prime + \beta^r \bar{c}_\phi\frac{\Omega^\prime}{\Omega}\, .
    \label{eq:cpiinitial}
\end{equation}
Inserting this condition into the Hamiltonian constraint, for regular initial data, yields
\begin{equation}
    \begin{aligned}
        \mathcal{H} = & - \tfrac{8}{9} \pi r^2 \bigl(\bar{c}_\phi\bigr)^2 \chi + \frac{24 \bigl(1 -  r^2 \chi\bigr)}{(1 -  r^2)^2} + \tfrac{8}{9} \pi r (1 -  r^2) \bar{c}_\phi \chi \bar{c}_\phi^\prime \\
        &-  \tfrac{2}{9} \pi (1 -  r^2)^2 \chi \Bigl(q^2 \bigl(\bar{A}_{3r}\bigr)^2 \bigl(\bar{c}_\phi\bigr)^2 + \bigl(\bar{c}_\phi^\prime\bigr)^2\Bigr)\\
        & + \frac{4 \chi^\prime}{r} -  \frac{5 \bigl(\chi^\prime\bigr)^2}{2 \chi} + \frac{6 \bigl(-4 \chi + \tfrac{2}{3} r \chi^\prime\bigr)}{1 -  r^2} + 2 \chi^{\prime\prime}\, 
    \label{eq:hamchisimple}
    \end{aligned}
\end{equation}
which we solve for $\chi$. It is useful to do the following change,
\begin{equation}
    \begin{aligned}
    \chi \rightarrow \chi_0 \psi^{-4} = \frac{\bar{\Omega}^2}{\Omega^2}\psi^{-4}\,,
    \end{aligned}
    \label{eq:chichange}
\end{equation}
and then solve the Hamiltonian constraint for $\psi$. Note that, for regular initial data, we are choosing $\bar{\Omega}=\Omega$, simplifying the above expression. On the other hand, for strong field initial data, $\chi_0$ depends on the numerical integration of $\bar{\Omega}$, as shown in Fig.~\ref{fig:aconfQMratio}. We also have to take into account that ${A_{rr}}_0$ is non-vanishing, as described in subsection~\ref{subsec:strongfield}, and it is therefore also useful to do the following change
\begin{equation}
    A_{rr} \rightarrow {A_{rr0}} \psi^{-6} = -\frac{2 C_{\text{CMC}} \bar{\Omega}^3}{r^3 \Omega} \psi^{-6}\, ,
    \label{eq:Arrchange}
\end{equation}
Substituting \eqref{eq:chichange} and \eqref{eq:Arrchange} into \eqref{eq:Ham}, we solve the Hamiltonian constraint for $\psi$, for strong field initial data.\par
The Poisson constraint is automatically satisfied with $\bar{E}^r_0 = 0$, because the charge density is initially 0. Only if we were to consider a perturbation both in the real and imaginary parts initially would we have to care with the Poisson constraint. An initial perturbation on the imaginary part would require an analogous condition to \eqref{eq:cpiinitial} for $\bar{d}_\phi$, with some extra terms appearing in \eqref{eq:hamchisimple}.

\subsection{RN BH Initial Data}
For a RN BH, $\bar{E}^r \neq 0 $. The other solution of \eqref{eq:poissonconst} besides the trivial one is,
\begin{equation}
    \bar{E}^r_0 = \frac{Q}{r^2}\chi^{3/2}_0\,,
    \label{eq:electricrn}
\end{equation}
where $Q$ is an integration constant that we physically interpret as the charge of the BH. When assuming a non-vanishing initial electric field in the Hamiltonian constraint, we get that the only solution that solves it is the same as \eqref{eq:electricrn}, confirming the consistency of the Einstein+Maxwell system of equations. Looking back at \eqref{eq:compacfactor}, we see that when $r \rightarrow 0$, $\tilde{r} \rightarrow R_0$, implying that $\bar{\Omega} \sim r/R_0$ there. From \eqref{eq:indatagen2}, we know that $\chi_0 = \bar{\Omega}^2/\Omega^2$. Therefore, we conclude that $\chi_0 \sim r^2$ as $r \rightarrow 0$. This means that near the origin $\bar{E}^r_0 \rightarrow 0$, avoiding the singularity that is present in the electric field as well\footnote{The covariant electric field diverges unavoidably at $r=0$. Hence, we stick with the contravariant formulation.}. Furthermore, we choose $\bar{A}_{3r} = 0$ initially, given the absence of magnetic field. $\bar{\Phi}_0$ will be given by setting the right-hand-side (RHS) of \eqref{eq:A3reveq} to 0. This is solved with the following equality,
\begin{equation*}
    \bar{E}_{r0} = - \frac{\partial_r (\bar{\Phi}_0 \alpha_0)}{\alpha_0}\,,
\end{equation*}
where note we are using the covariant version of the electric field here. Defining the quantity $V(r) = \bar{\Phi}_0\alpha_0$, the above equation can be rewritten as,
\begin{equation*}
    V^\prime = - \frac{Q}{r^2}\chi^{1/2}_0\alpha_0 = - \frac{Q}{r^2} \frac{\bar{\Omega}}{\Omega}(\bar{\Omega}-r\bar{\Omega}^\prime)\frac{\Omega}{\bar{\Omega}} = \partial_r \left(\bar{\Omega}\frac{Q}{r}\right)\,.
\end{equation*}
We conclude then, 
\begin{equation*}
    V(r) = \bar{\Omega}\frac{Q}{r} + C\,,
\end{equation*}
with $C$ being an integration constant. To write $\alpha_0$ as a function of $\bar{\Omega}$ and $\Omega$, we used the conformally flat initial data expressions \eqref{eq:initialdataoverall}. At the origin, $V \rightarrow Q/R_0 + C$. If we do not choose correctly the constant $C$, $\bar{\Phi}_0 = V/\alpha_0$ will blow up at the origin, because $\alpha_0 \rightarrow 0$ there as well. Therefore, the only possible choice is to choose $C = -Q/R_0$, such that $V \rightarrow 0$, as $r \rightarrow 0$. The initial data for $\bar{\Phi}$ becomes,
\begin{equation}
    \label{eq:potentialindata}
    \bar{\Phi}_0 = \frac{Q}{r}\left(\bar{\Omega}-\frac{r}{R_0}\right)\frac{\bar{\Omega}}{\Omega}\frac{1}{\bar{\Omega}-r\bar{\Omega}^\prime}\,.
\end{equation}
It is not clear at first that $\bar{\Phi}_0$ does not blow up at the origin so now we derive the limit to show that it does not. For the derivation, we need to take into account that $\bar{\Omega} \rightarrow r/R_0$, as $r \rightarrow 0$, and this implies that $\bar{\Omega}^\prime \rightarrow 1/R_0$. And we also use that $\bar{\Omega}^{\prime \prime} \rightarrow 0$, but nothing can be concluded about $\bar{\Omega}^{\prime \prime \prime}$. And the Taylor expansions,
\begin{equation*}
    \begin{aligned}
        \bar{\Omega} &= \bar{\Omega}(0) + r \bar{\Omega}^\prime (0) + \frac{r^2}{2}\bar{\Omega}^{\prime\prime}(0) + \frac{r^3}{6}\bar{\Omega}^{\prime\prime\prime}(0) + O(r^4)\\
        \bar{\Omega}^\prime &= \bar{\Omega}^\prime (0) + r \bar{\Omega}^{\prime\prime}(0) + \frac{r^2}{2}\bar{\Omega}^{\prime\prime\prime}(0) + O(r^3)\,.
    \end{aligned}
\end{equation*}
Substituting these expressions in \eqref{eq:potentialindata}, together with the fact that $\bar{\Omega}(0) = \bar{\Omega}^{\prime\prime}(0) = 0$ and $\bar{\Omega}^\prime(0) = 1/R_0$, 
\begin{equation}
    \bar{\Phi}_0(r \rightarrow 0) = \frac{1}{2}\frac{Q}{R_0}\frac{6}{K_{\text{CMC}}} = \frac{Q}{R_0}\frac{3}{K_{\text{CMC}}}\,.
    \label{eq:philimit}
\end{equation}
To simulate the Schwarzschild BH, it is just a matter of setting $Q = 0$ in the initial data.

\subsection{Charged Scalar Field + RN BH Initial Data}
By perturbing a RN BH spacetime with the charged scalar field, we have a non-trivial solution of the Poisson constraint, because $\bar{\Phi} \neq 0$ and thus $\bar{q}_{\text{dens.}} \neq 0$. To tackle this problem, one possibility is to extend the conformal transverse traceless method \cite{zilhao_nonlinear_2015} to the electric field, $\bar{E}^r \rightarrow \bar{E}^r \psi^{-6}$. However, to simplify further the equations, we will make the following substitution instead,
\begin{equation}
    \bar{E}^r = \chi^{3/2}\frac{Q}{r^2} + \delta \bar{E}^r \psi^{-6},\quad \bar{\Phi} \rightarrow \bar{\Phi} \psi^{-6}\,.
    \label{eq:addel}
\end{equation}
This way, the Poisson constraint gives us an elliptic equation independent of $\psi$,
\begin{equation*}
    (\delta \bar{E}^r)^\prime = -\frac{2 \delta \bar{E}^r}{r}+\frac{3 \delta \bar{E}^r \bar{\Omega}^\prime}{\bar{\Omega}}-\frac{3 \delta \bar{E}^r \Omega^\prime}{\Omega}+4\pi q^2 \bar{c}_\phi^2 \bar{\Phi}\,,
\end{equation*}
where the last term in the RHS comes from writing the charge density explicitly as a function of the terms that do not vanish initially, considering an initial perturbation only on the real part of the scalar field. A differential equation for $\psi$ is again given by solving the Hamiltonian constraint. This latter one is more lengthy, due to the different initial data we have to input for each variable \eqref{eq:indatagen2}.

\section{Numerical Setup}
\label{sec:numericalsetup}
To evolve the system, we developed a code adapted to spherically symmetric equations, based on the method of lines with 4th-order finite-differencing in the spatial derivatives. A 4th-order Runge-Kutta method is used to integrate in time. While this numerical setup is simple in comparison to more sophisticated ones like~\cite{verma2014,jiwari2020,JIWARI2012600}, it is perfectly appropriate for the formally-singular EMKG system under consideration. We use a staggered grid to avoid evaluating the equations exactly at the origin and $r = r_{\mathscr{I}^+}$, because the RHS of our equations are formally singular at these points. Regarding the boundary conditions at \scrip, we follow \cite{carcione_boundary_1994, calabrese_discrete_2006} and impose an outflow condition on the ghost points of the grid beyond $r_{\mathscr{I}^+}$. This condition allows information to be radiated away at \scrip, motivated by the fact that there is no physical ingoing radiation at future null infinity given its character as ingoing null hypersurface. The boundary condition at the origin, on the other hand, is dealt with differently according to the simulation. For regular initial data, we use parity conditions according to the nature of each variable, while for strong field initial data and for the collapse, simulations we use an inflow boundary condition equivalent to that used at \scrip. We use Kreiss-Oliger dissipation to attenuate high-frequency noise \cite{babiuc_implementation_2008, kindelan_optimized_2016}. For all simulations except the collapse, a Kreiss-Oliger amplitude of $0.08$ was more than enough to keep the simulations from blowing up early. Too much or too low dissipation in the strong field initial data simulations would make the simulation crash as soon as the scalar field entered the BH. For collapse simulations, the tuning had to be done more carefully. A $0.15$ amplitude worked for several grid resolutions for the purpose of convergence tests.

\section{Charged Scalar Field and RN BH in Hyperboloidal Slices}
\label{sec:results}
In this section, three scenarios will show the EMKG system working in hyperboloidal slices. We start with a charged scalar field perturbation (subsec.~\ref{subsec:chargedsc}) of Minkowksi, then we do a similar perturbation in a RN spacetime (subsec.~\ref{subsec:chargedscBH}) and we end by presenting the collapse of the charged scalar field into a charged BH (subsec.~\ref{subsec:chargedscCollapse}).

\subsection{Charged Scalar Field in Regular Initial Data}
\label{subsec:chargedsc}
\begin{figure*}[t]
    \centering
    \subfloat{\includegraphics[width=0.485\textwidth]{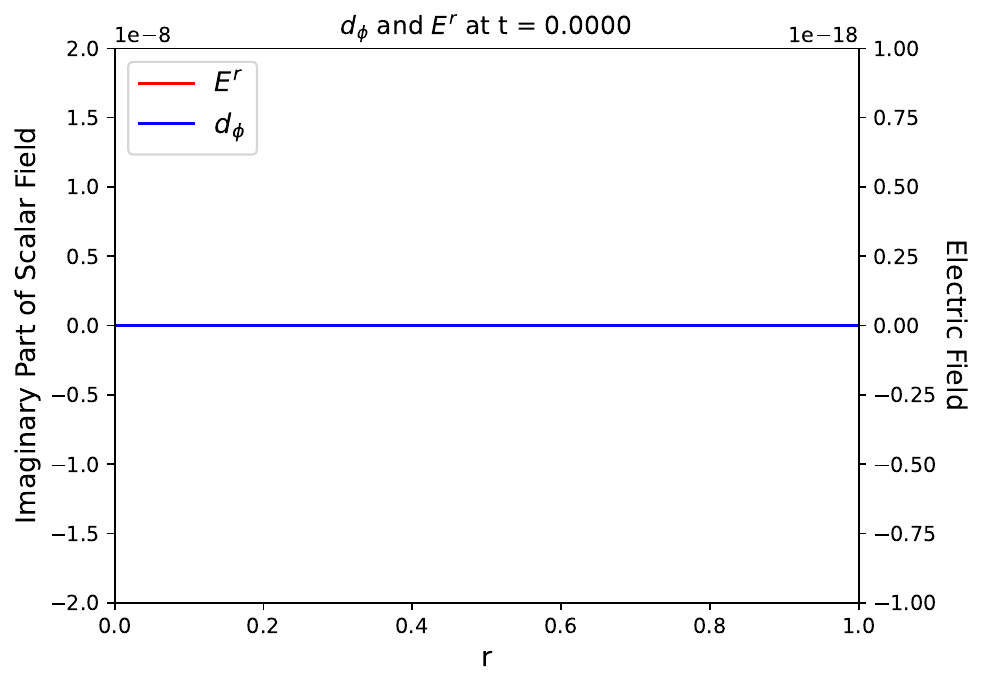}}
    \subfloat{\includegraphics[width=0.41\textwidth]{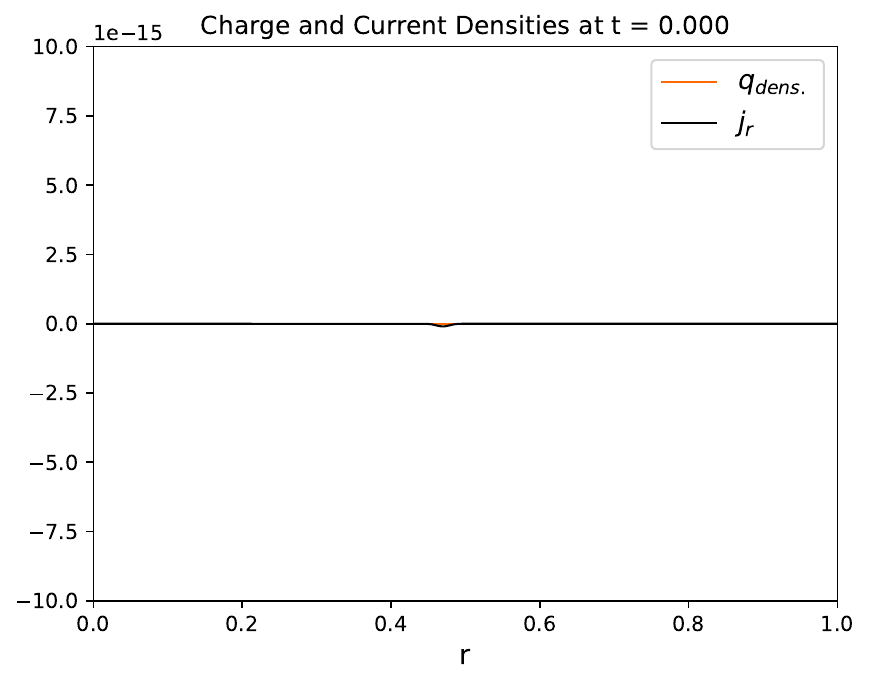}}\\[-6mm]
    \subfloat{\includegraphics[width=0.485\textwidth]{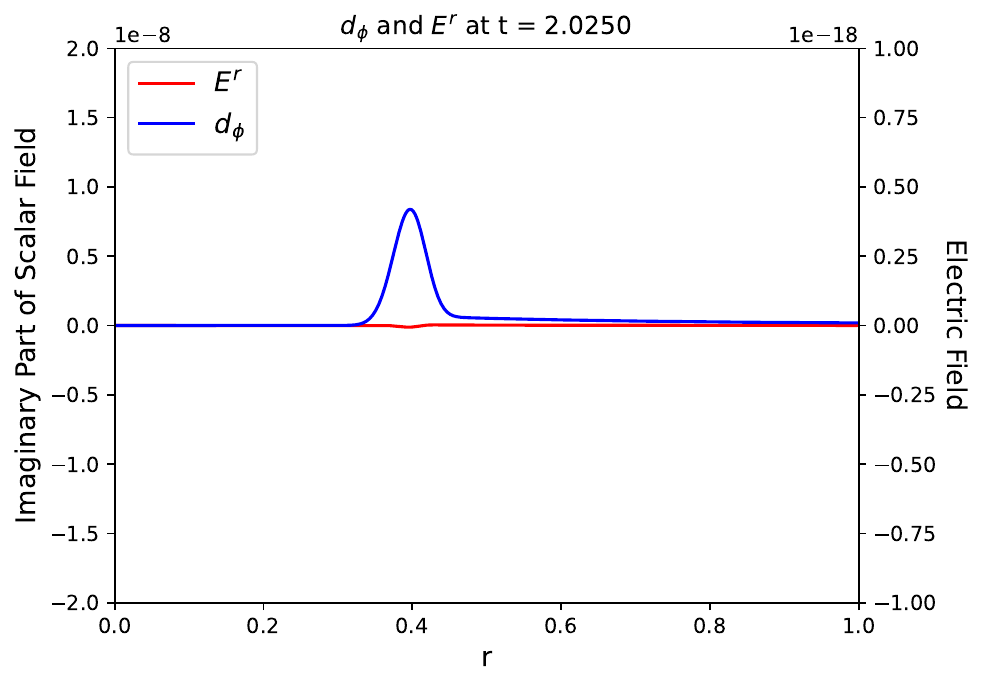}}
    \subfloat{\includegraphics[width=0.41\textwidth]{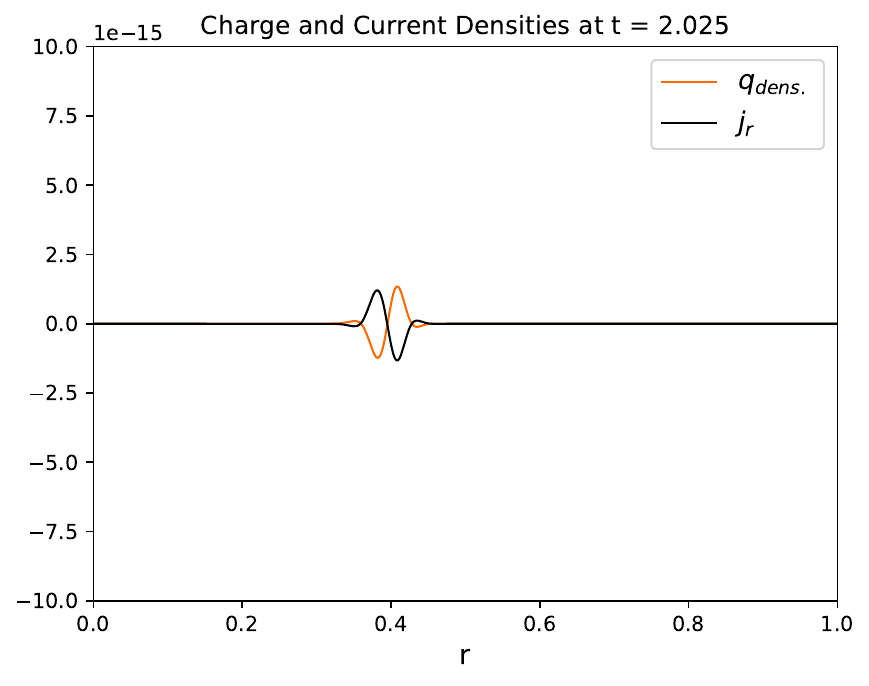}}\\[-6mm]
    \subfloat{\includegraphics[width=0.485\textwidth]{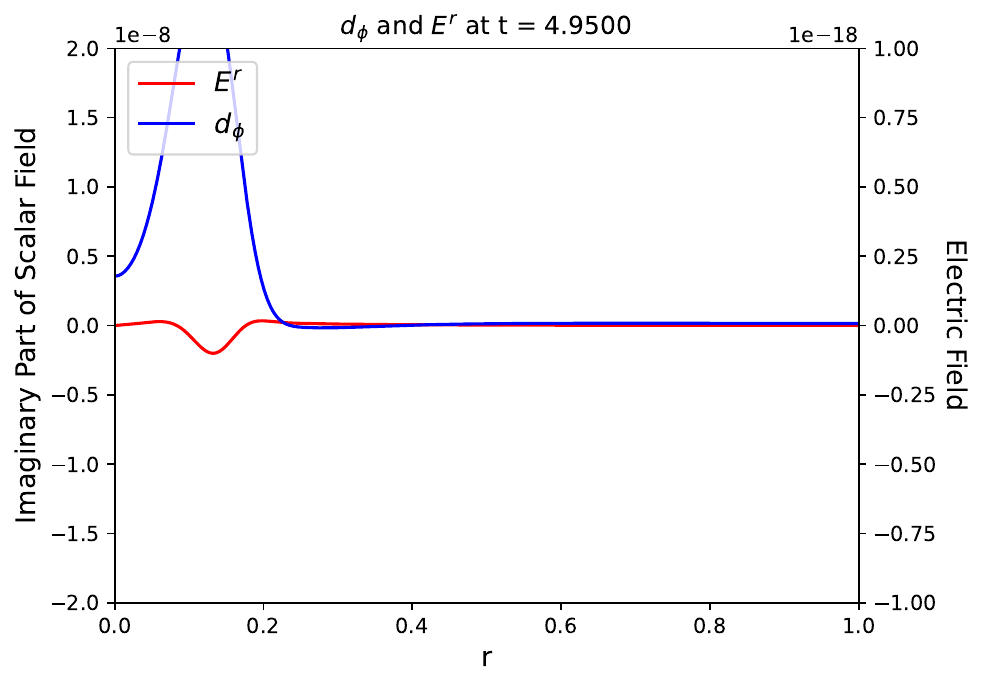}}
    \subfloat{\includegraphics[width=0.41\textwidth]{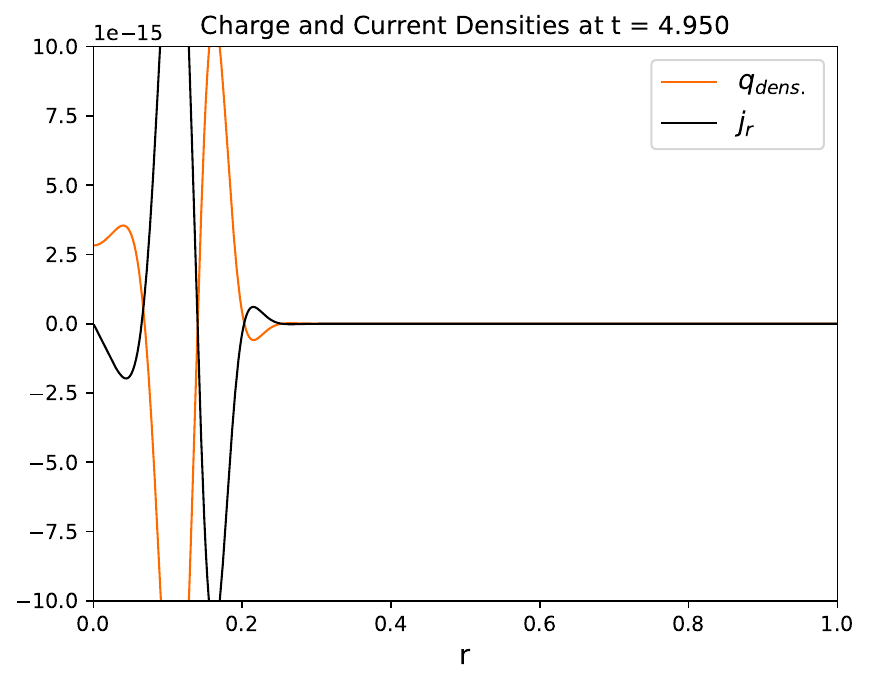}}\\[-6mm]
    \subfloat{\includegraphics[width=0.485\textwidth]{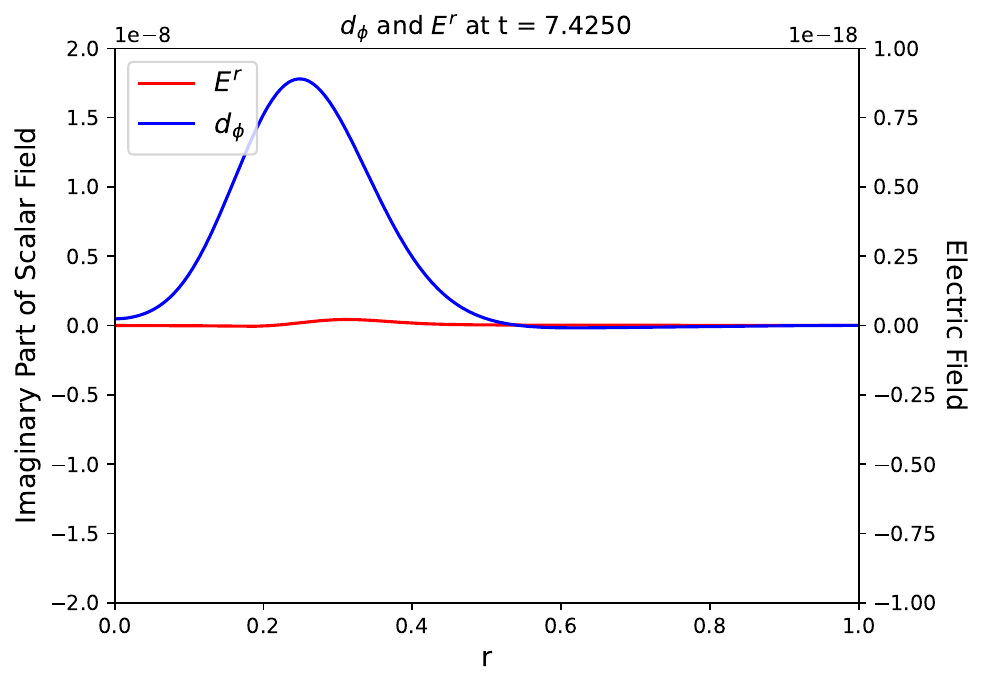}}
    \subfloat{\includegraphics[width=0.41\textwidth]{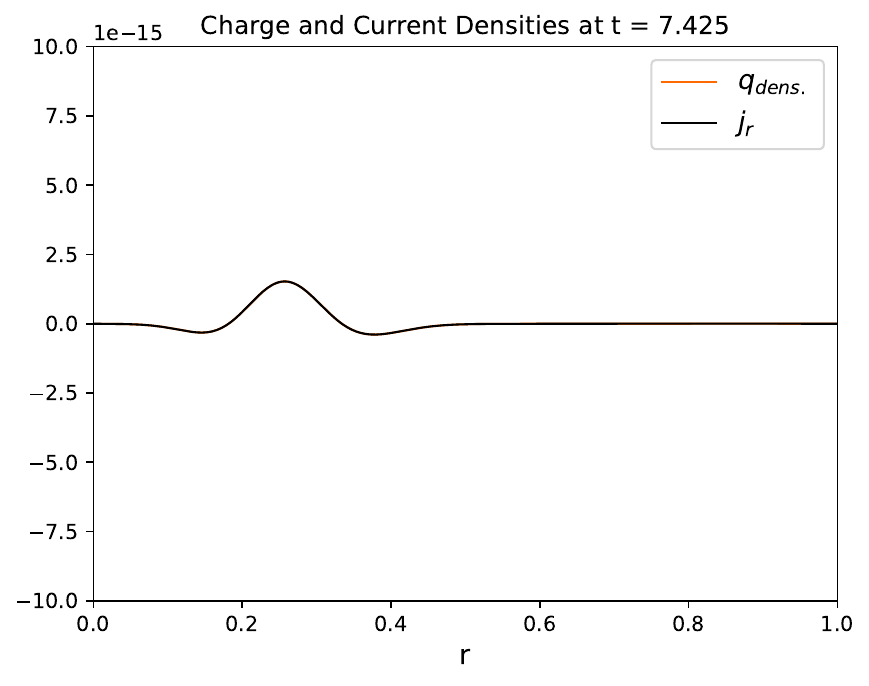}}
    \caption{Evolution of initial perturbations on $\bar{c}_\phi$ and $A_{3r}$, for regular initial data as given in sections~\ref{sec:regularinitialdata} and~\ref{subsec:chargedscalarinitialdata}. The plots show $\bar{d}_\phi$, $\bar{E}_r$, $\bar{j}_r$ and $\bar{q}_{\text{dens.}}$ evolving until the scalar field is mostly radiated out through future null infinity, at $r_{\mathscr{I}^+} = 1$.}
    \label{fig:scalarmax}
\end{figure*}

For a scalar field perturbation of Minkowski, we use the initial data described in subsec.~\ref{subsec:chargedscalarinitialdata}, setting the charge parameter $q=2$. The perturbations to the (rescaled) scalar field and the radial part of the vector potential $\bar{A}_{3r}$ have a Gaussian-like profile:
\begin{equation}
    A r^2 \exp\left(-\frac{r^2-\mu^2}{4\sigma^4}\right)\,,
    \label{eq:cphiprofile}
\end{equation}
where $A$, $\mu$ and $\sigma$ are constants. Both perturbations are centered at different $\mu$ ($0.5$ for $\bar{c}_\phi$ and $0.4$ for $\bar{A}_{3r}$). The other parameters used were $\sigma_\phi = \sigma_{A_3} = 0.1$ and $A_\phi = 0.01$ and $A_{A_3} = 0.001$. The evolution of $\bar{d}_\phi$, $\bar{E}_r$, $\bar{j}_r$ and $\bar{q}_{\text{dens.}}$ is shown in Fig.~\ref{fig:scalarmax}. We opted for only plotting these quantities, for they are enough to show that the coupling between the three different parts (Einstein, Maxwell, KG) of our system is working. What is happening beyond what we are showing in the plots is the following: $\bar{c}_\phi$ is being perturbed slightly, but it is also affected by $\bar{A}_{3r}$, which then couples to the imaginary part through its evolution equation \eqref{eq:dpitensorial}. Thus, the imaginary part ends up being excited. The behaviour of the variables from the EFEs is very similar to what was already shown in \cite{vano-vinuales_free_2015} (Figs. 8.3 and 8.4), hence we refrain from presenting it again.

Fig.~\ref{fig:scalarmax} shows that initially neither electric field nor any imaginary part of the scalar field are excited. However, as time goes on, we see two things: first, that $\bar{d}_\phi$ becomes non-zero (second row, on the left) due to the initial perturbations done; second, that when we have a non-vanishing current density initially (first row, on the right), the electric field will be excited even if it is initially zero.\par
\begin{figure}
    \centering
    \includegraphics[width=0.4\textwidth]{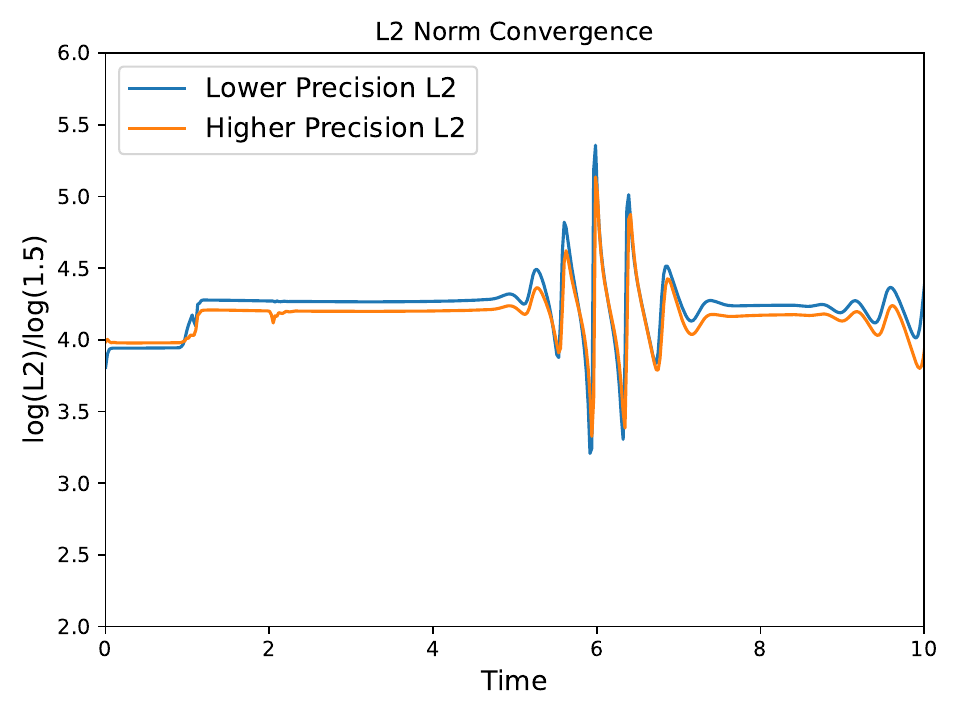}
    \caption{L2 norm convergence plot of the simulation shown in Fig.~\ref{fig:scalarmax}, here plotting the quantity~\eqref{convorder}. The ideal result is a horizontal line at 4, for the expected 4th order convergence of our setup -- the orange line, which uses a series of runs with higher resolution, is generally closer to 4 than the blue one, which is a good indication.}
    \label{fig:L2NormScalarMax}
\end{figure}

 As a way of measuring the reliability of our implementation, we show the L2 norm-convergence of the simulation in Fig.~\ref{fig:scalarmax} plotted in Fig.~\ref{fig:L2NormScalarMax}. The quantity shown corresponds to
\begin{equation}\label{convorder}
    \log_f\sqrt{(\sum (u_{f^2 h} - u_{fh})^2)/(\sum (u_{f h} - u_{h})^2)}\, ,
\end{equation}
where $u_{f^n h}$ corresponds to any variable of the simulation being evolved, in which a spatial step $f^n h$ is being used. The sum is done in the whole grid and over all the variables. The $f$ logarithm of the quantity should then give the order of convergence of the code, which in our case is 4. We have decided to do 4 runs, each with progressively higher precision (320, 480, 720 and 1080 points). The 3 higher precision runs (the orange line) are closer to 4, as expected, indicating the appropriate behaviour of the simulation.

\subsection{Charged Scalar Field in Strong Field Initial Data}
\label{subsec:chargedscBH}
\begin{figure}
    \centering
    \includegraphics[width=0.4\textwidth]{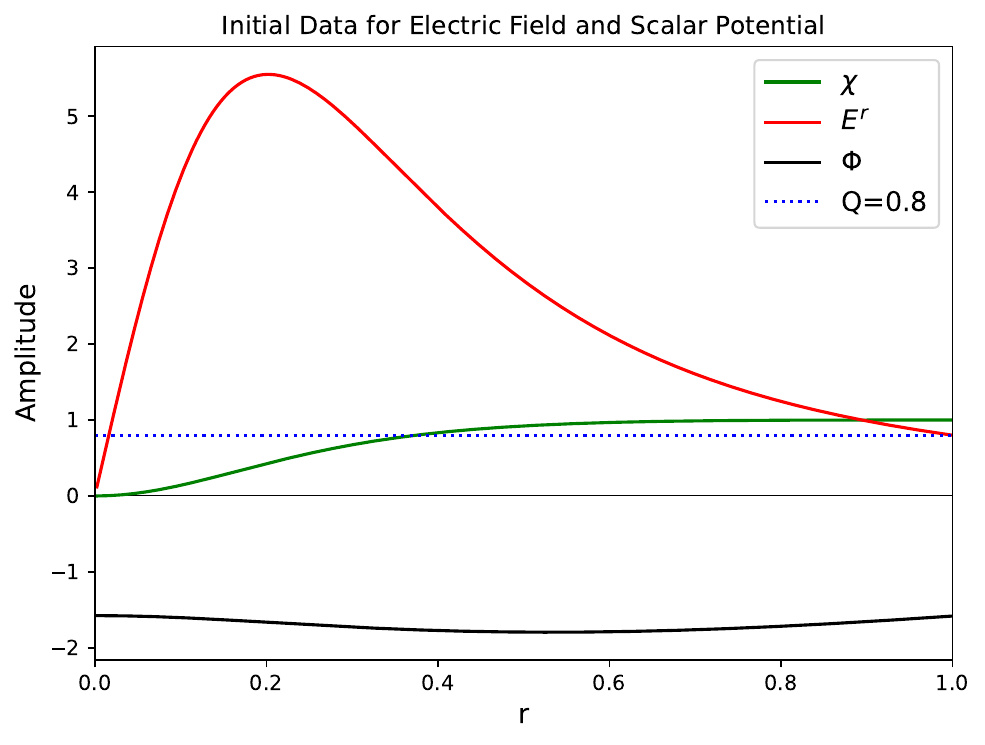}
    \caption{Initial data for the conformal factor $\chi$, the BH's electric field $\bar{E}^r$, \eqref{eq:electricrn}, and the scalar potential $\Phi$, \eqref{eq:potentialindata}. We set $Q=0.8$, $K_{\text{CMC}}=-1$ and $M=1$.}
    \label{fig:inmaxdata}
\end{figure}

With the electric field generated by the BH, a perturbation on $\bar{c}_\phi$ is enough to excite the whole system. The initial data corresponding to \eqref{eq:electricrn} and \eqref{eq:potentialindata} is shown in Fig.~\ref{fig:inmaxdata}, for a BH with $M = 1$ and $Q = 0.8$, and a scalar perturbation with $\mu = 0.5$, $\sigma = 0.1$, and $A = 0.0025$.\par
For $r > 0.3$, the (rescaled) electric field decreases with $r^2$. At \scrip, it attains a value different from 0 according to \eqref{eq:electricrn}. There, $r=r_{\scri^+}=1$ and $\chi_0 = 1$ and thus $E_0^r(r_{\scri^+}) = Q$. The horizonal line $y = Q = 0.8$ in Fig.~\ref{fig:inmaxdata} was included to facilitate this comparison. Regarding the initial data for $\bar{\Phi}$, at the origin it attains a value almost identical to \eqref{eq:philimit}, with any discrepancy due to numerical error.\par
\begin{figure}[ht]
    \centering
    \includegraphics[width=0.4\textwidth]{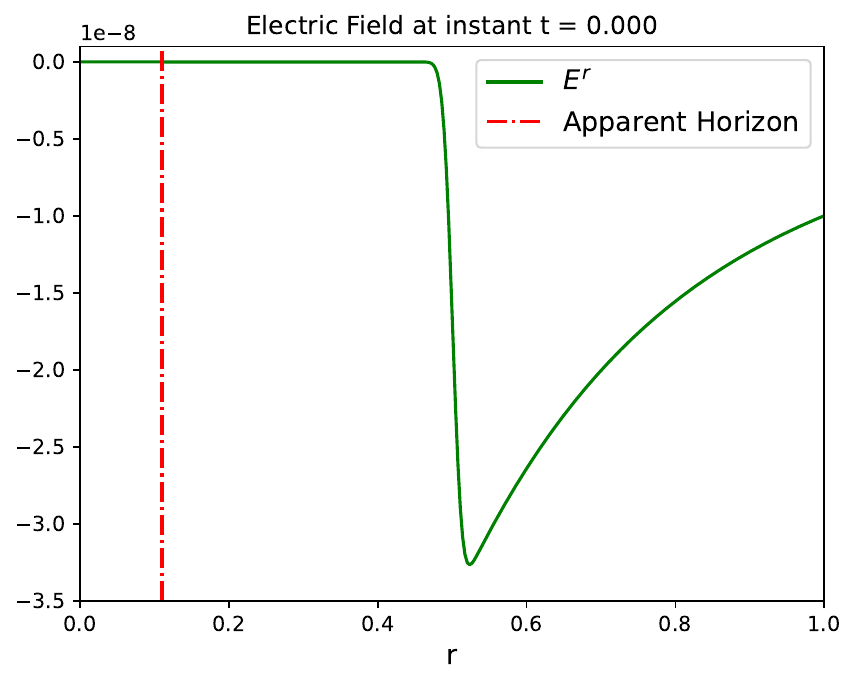}
        \caption{Additional electric field coming from the scalar field's initial interaction with the scalar potential from the BH.}
    \label{fig:ErScalar}
\end{figure}

The additional electric field coming from the presence of the scalar field \eqref{eq:addel} is shown in Fig.~\ref{fig:ErScalar}.  Note that a spherical perturbation can only produce an electric field outside of the sphere determined by the radius where the perturbation lies. That is why for $r < 0.5$ there is no electric field (besides the one generated by the BH). At \scrip, we see again that the (rescaled) electric field is different from 0.\par
\begin{figure*}[h]
    \centering
    \subfloat{\includegraphics[width=0.4\textwidth]{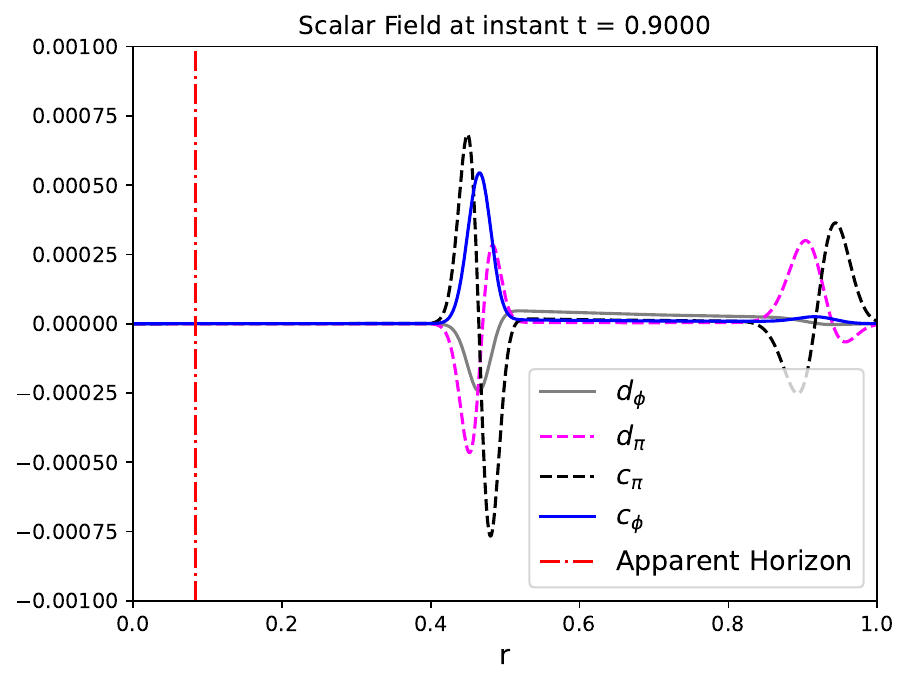}}
    \subfloat{\includegraphics[width=0.4\textwidth]{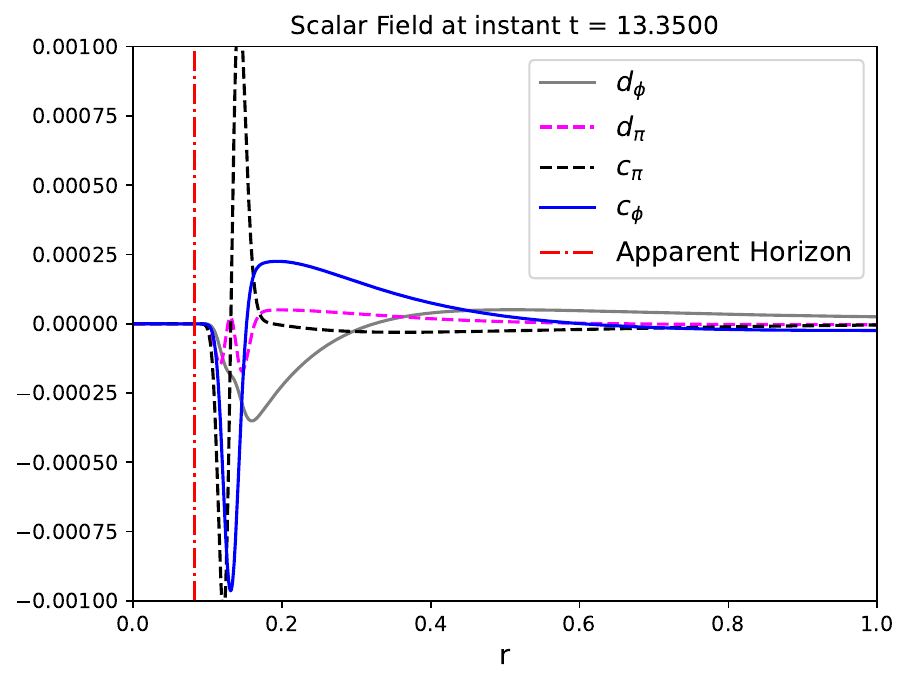}}\\[-2mm]
    \caption{Charged scalar field interacting with a charged BH. We have set the BH charge to 0.7. An initial perturbation in the real part of the scalar field was given.}
    \label{fig:rnscalar}
\end{figure*}

We further add two frames of the evolution in Fig.~\ref{fig:rnscalar}, where it is possible to see the real and imaginary parts interplaying with each other.

\begin{figure*}[h]
    \centering
    \subfloat{\includegraphics[width=0.4\textwidth]{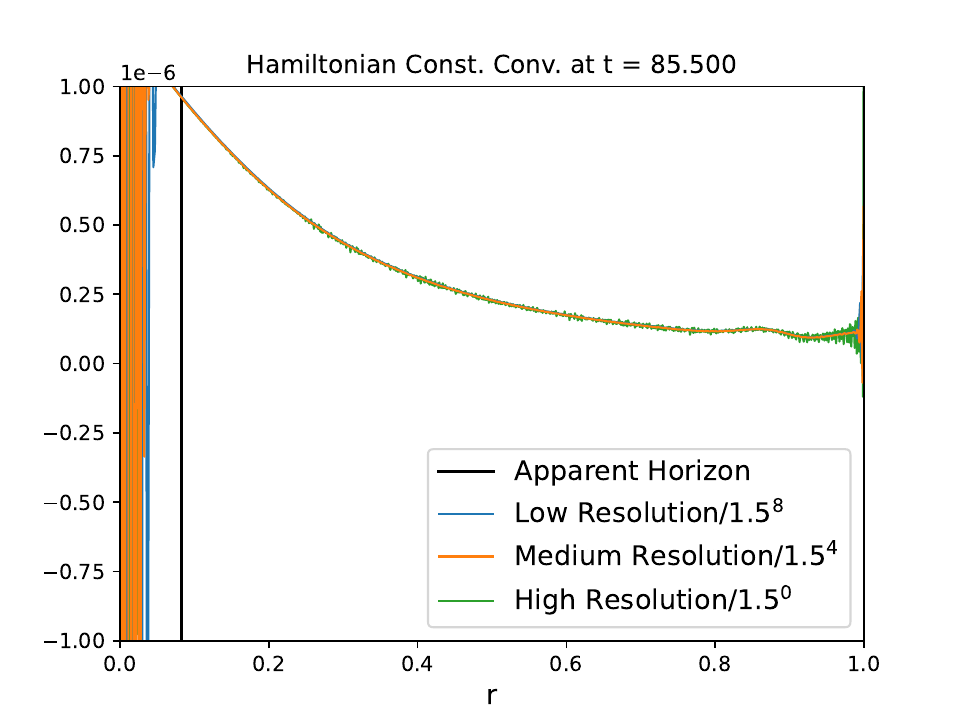}}
    \subfloat{\includegraphics[width=0.4\textwidth]{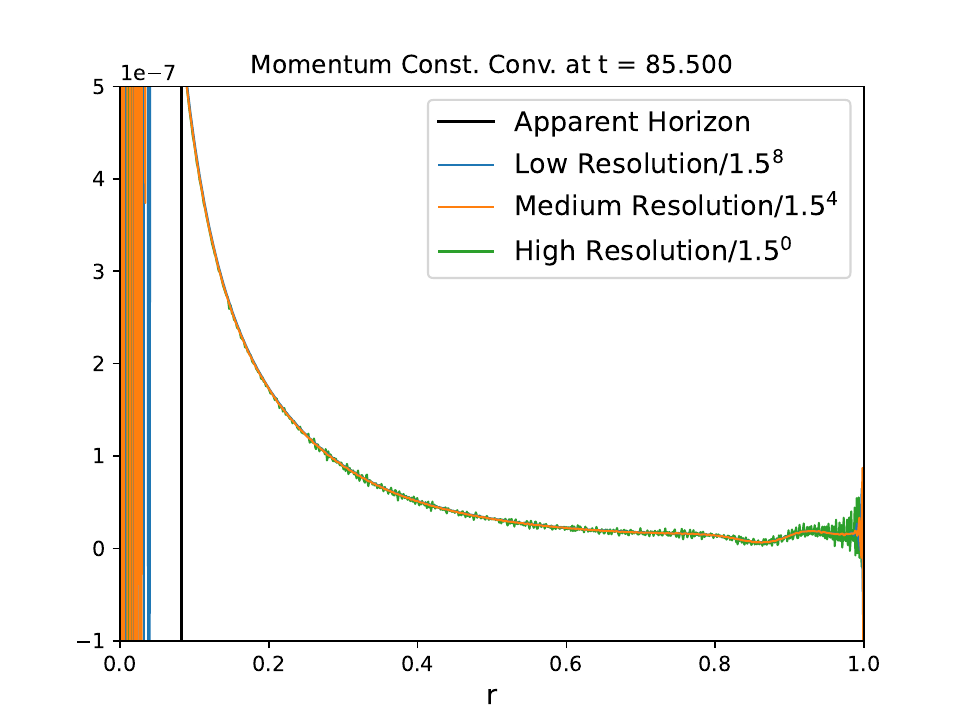}}\\[-2mm]
    \subfloat{\includegraphics[width=0.4\textwidth]{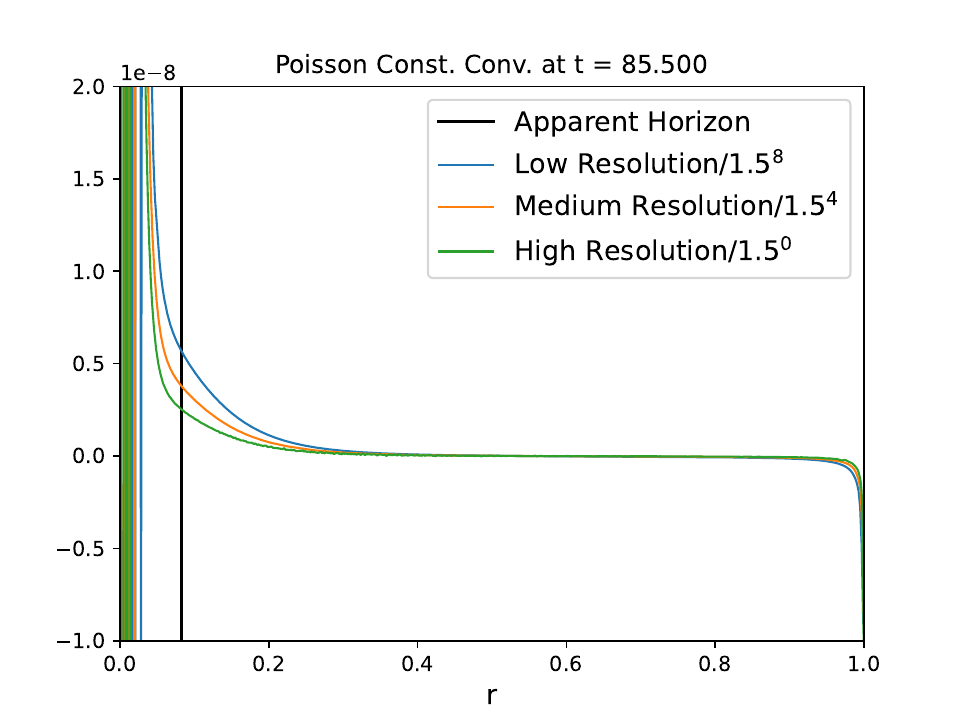}}
    \subfloat{\includegraphics[width=0.4\textwidth]{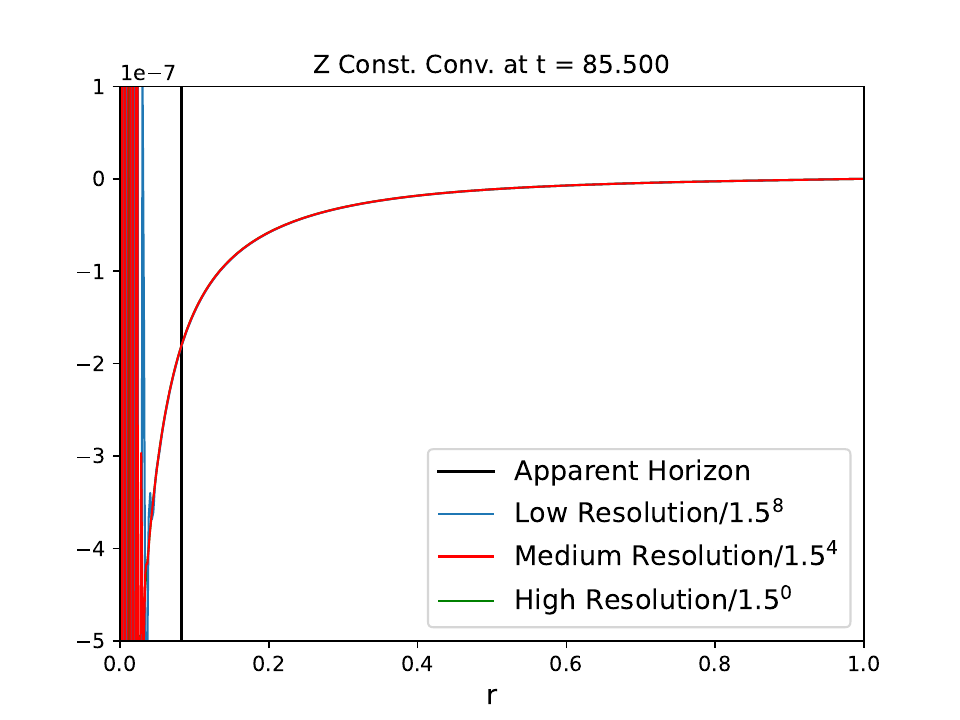}}
    \caption{Pointwise convergence of the constraints at a late-time for a charged scalar field perturbation of a RN BH spacetime. It corresponds to the simulation shown in Fig.~\ref{fig:rnscalar}. The low, medium, and high resolutions are on top of each other, when they cannot be distinguished in the graphs.}
    \label{fig:ConstConvRNSccalar}
\end{figure*}
The convergence of the constraints at a late-time when the perturbation has been absorbed by the BH is shown in Fig.~\ref{fig:ConstConvRNSccalar}.
Near the origin, the constraints do not converge as expected due to the presence of the puncture in trumpet form, where the spacetime is less smooth. This, however, happens inside the apparent horizon only and does not affect the rest of the points in the simulation.

\subsection{Collapse of Scalar Field into a RN BH}
\label{subsec:chargedscCollapse}

\begin{figure*}[h]
    \centering
    \subfloat{\includegraphics[width=0.4\textwidth]{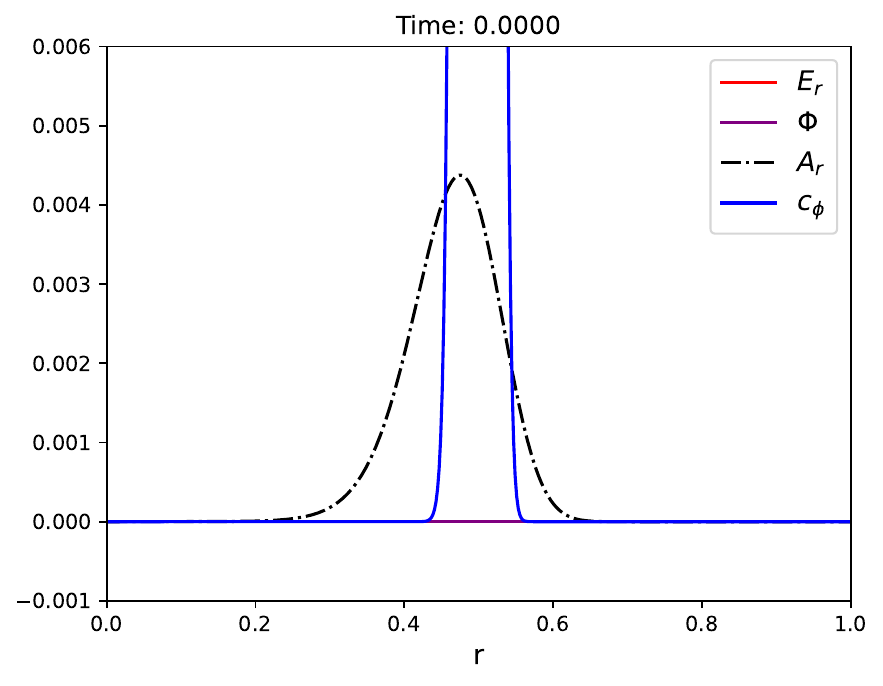}}
    \subfloat{\includegraphics[width=0.4\textwidth]{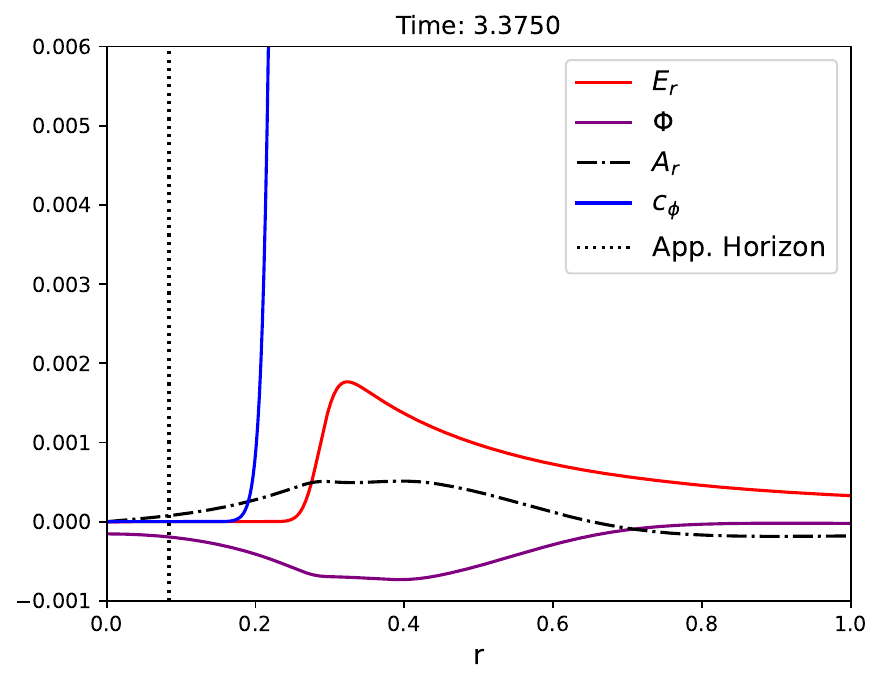}}\\[-2mm]
    \subfloat{\includegraphics[width=0.4\textwidth]{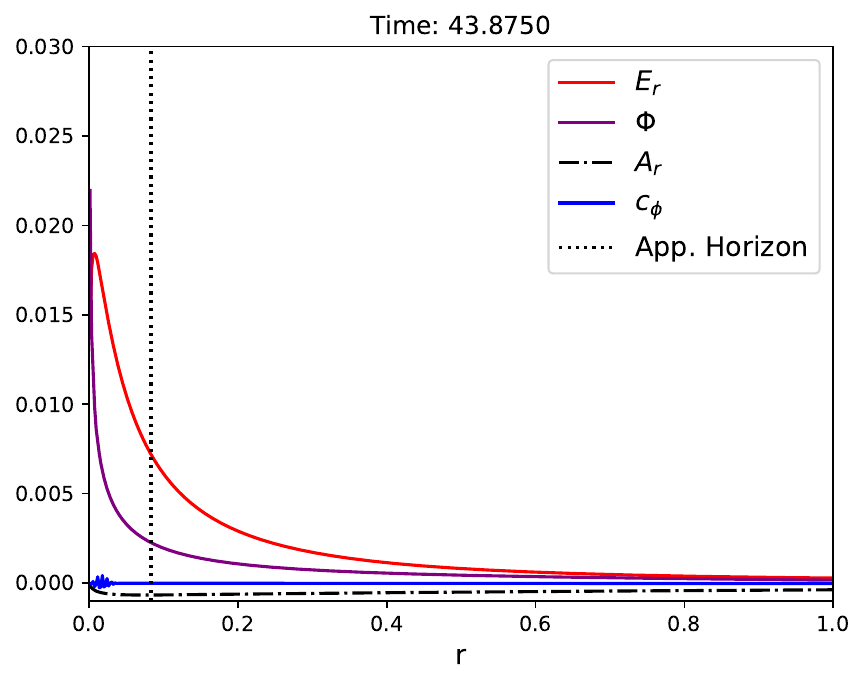}}
    \subfloat{\includegraphics[width=0.4\textwidth]{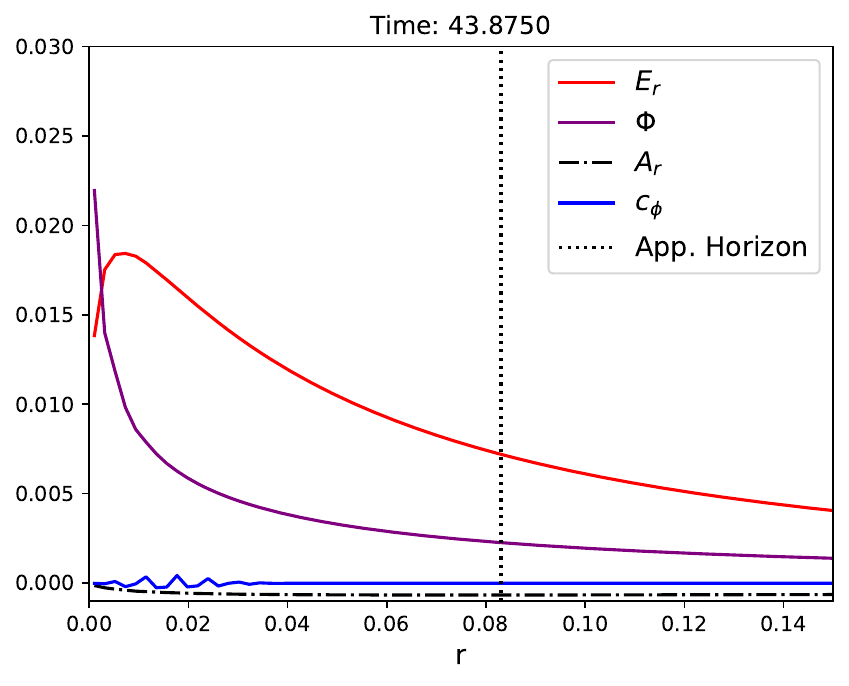}}
    \caption{$\bar{c}_\phi$, $E_r$, $\Phi$ and ${A_3}_r$ evolving until the charged scalar field collapses into a BH at the origin. The last plot is a zoom-in of the first-to-last plot, for a clearer picture of the electromagnetic fields near the origin. The ranges in the vertical axes on the top figures are different from the lower ones.}
    \label{fig:chargedcollapse}
\end{figure*}

\begin{figure*}[h]
    \centering
    \subfloat{\includegraphics[width=0.4\textwidth]{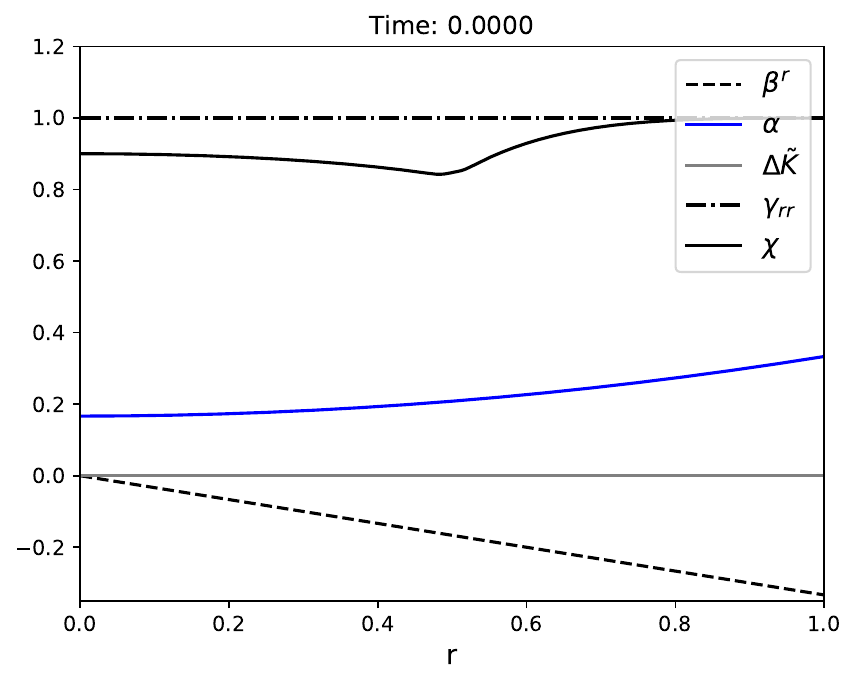}}
    \subfloat{\includegraphics[width=0.4\textwidth]{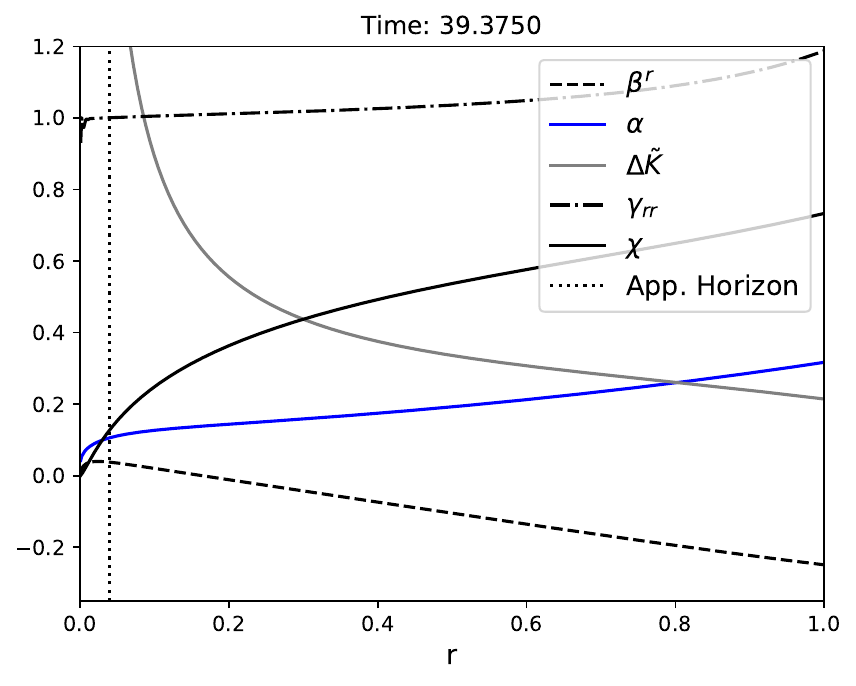}}
    \caption{Evolution of $\alpha$, $\beta^r$, $\Delta \tilde{K}$, $\gamma_{rr}$ and $\chi$, during the collapse of a charged scalar field into a RN BH (corresponding to the same evolution shown in Fig.~\ref{fig:chargedcollapse}). The fact that the lapse $\alpha$ goes to 0 at the origin means that a BH has formed.}
    \label{fig:chargedcollapse2}
\end{figure*}

The collapse of a scalar field into a RN BH has been of interest throughout the last decades \cite{misner_relativistic_1964, bicak_gravitational_1972, torres_gravitational_2014}. Therefore, we will apply a similar perturbation as was done in \ref{subsec:chargedsc}, but strong enough so that there is BH formation. For that, we use $A = 0.05$, $\mu=0.5$, and $\sigma=0.1$ for the real part of the physical scalar field. For the radial part of the vector potential, $A= 0.02$, $\mu = 0.46$, and $\sigma = 0.2$. We also opted for increasing the scalar field's charge to 5.0, for a higher coupling with the imaginary part. The result is shown in Fig.~\ref{fig:chargedcollapse}. We plot the electromagnetic fields, together with the real part of the scalar field. Note that, contrary to what happens in Fig.~\ref{fig:scalarmax}, here the scalar field eventually is dissipated away inside the BH's horizon, without reaching the origin or getting reflected there. This is the expected behaviour on the trumpet-like slice inside the horizon of the BH that has formed. Its creation can be observed in the lapse $\alpha$ becoming 0, which has already happened in the right plot in Fig.~\ref{fig:chargedcollapse2}.\par
In Fig.~\ref{fig:chargedcollapse}, we also see that a static electric field is formed after the collapse, matching the physical expectations. The apparent horizon is formed at around $r \approx 0.039$, and the Misner-Sharp mass of the system becomes $M_{MS} \approx 0.33$. The electric field generated by the ingoing charged scalar field is positive, due to the positive sign in the current density \eqref{eq:poissonconst}, but it continues to go to 0 near the origin, analogous to the electric field generated by the charged BH (Fig.~\ref{fig:inmaxdata}).\par
\begin{figure*}[h]
    \centering
    \subfloat{\includegraphics[width=0.4\textwidth]{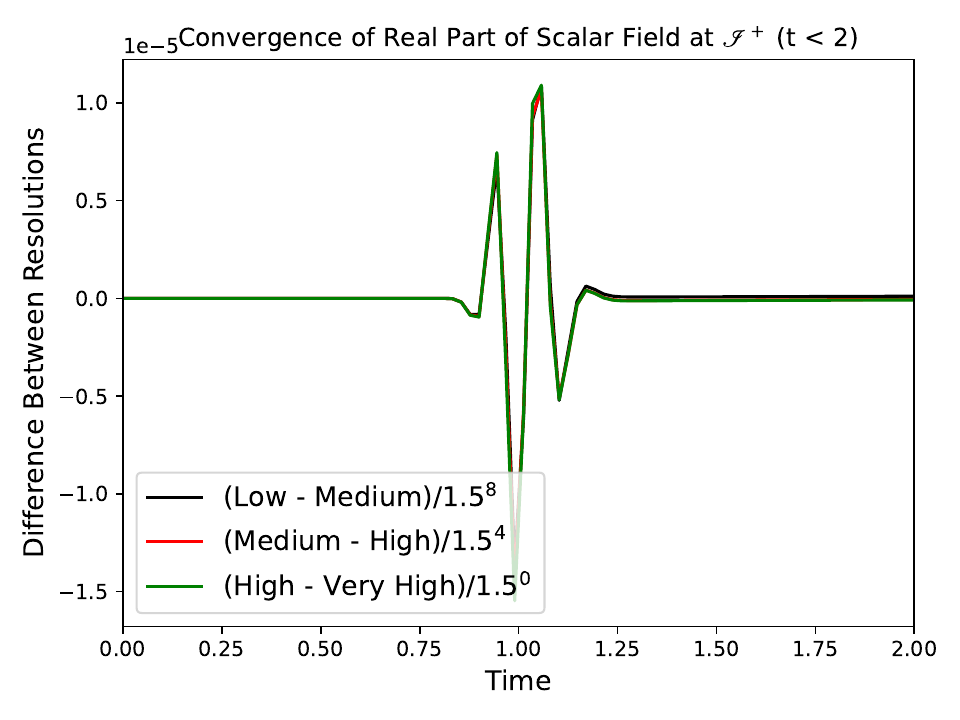}}
    \subfloat{\includegraphics[width=0.4\textwidth]{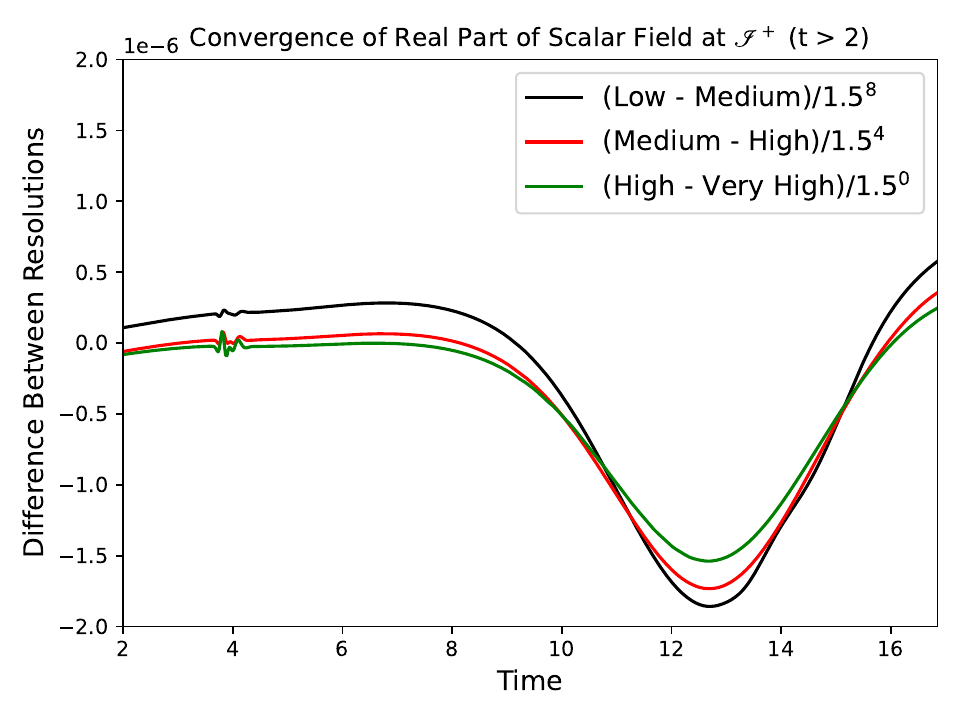}}\\[-2mm]
    \caption{Pointwise convergence of the real part of the scalar field at \scrip, as the scalar field collapses into a charged BH. On the left, a focus on the time window $t < 2$. On the right, for $t > 2$.}
    \label{fig:charged_scri}
\end{figure*}

To check whether the signal of the scalar field at \scrip is converging appropriately, we extrapolate the data half a spatial step, since we are using a staggered grid. Running several grid resolutions as done for the previous convergence plots, we obtain Fig.~\ref{fig:charged_scri}. The fact that the lines coincide means that the signal is converging at the appropriate order.
It is remarkable that even in a situation that has to be carefully numerically tuned in the code, the scalar field signal at \scrip converges nicely.

\section{Conclusions}
\label{sec:conclusions}
The takeaway message is that the Einstein-Maxwell-Klein-Gordon system has been implemented with success on hyperboloidal slices, in spherical symmetry. This was done using the BSSN/conformal Z4 formalism for the EFEs, together with the prescription in \cite{bona_elements_2009} for Maxwell's equations. An adapted Lorenz gauge was found to regularize the electromagnetic potentials at \scrip, and we made sure that all evolution quantities were finite and well-behaved everywhere in the domain.
To the best of our knowledge, these are the first hyperboloidal evolutions including electric and changed scalar fields with the BSSN/Z4 formulations. 

Here we showcase the capabilities of our implementation, by presenting evolutions of a charged complex scalar field perturbing a regular spacetime and one including a charged black hole. We have also shown a successful collapse of the scalar field into a RN BH in hyperboloidal slices. All of these results were accompanied by convergence tests to show reliability of the results. In a companion paper \cite{resultspaper} we will exploit our access to future null infinity to study the behaviour of the scalar field there in the fully nonlinear regime. 
The hyperboloidal approach can also be applied to other systems of equations, such as the Klein-Gordon-Zakharov, as treated mathematically in~\cite{Dong_2021}, or the Klein-Gordon-Dirac systems. The main suitability condition is that the evolved fields fall off fast enough towards null infinity. 
Our implementation of the Einstein-Maxwell-Klein-Gordon system opens the door to other physically interesting scenarios while allowing access to null infinity. This is relevant for the weak cosmic censorship conjecture \cite{1969NCimR...1..252P} if exploring collapsing (charged or uncharged) spacetimes close to the threshold of criticallity. Extensions to boson stars would only require the addition of a massive term in the KG equation. This however needs to be done carefully, because the massive term does not fall off with the appropriate rate at \scrip \cite{Gautam:2021ilg}. Also very relevant is to expand beyond spherical symmetry (currently ongoing work for the hyperboloidal approach), and include magnetic fields, which, in nature, seem to have more influence in the BH's dynamics. The Gertsenshtein effect \cite{gertsenshtein1961wave}, the conversion between electromagnetic and gravitational waves and viceversa, is also worth looking into. 

\section*{Acknowledgements}

The authors would like to thank Edgar Gasperín, Chris Stevens, David Hilditch and Miguel Zilhão for useful comments on the manuscript. We also thank Miguel Zilhão and Jorge Expósito for helping to understand the coupling between GR and electromagnetism.\par
AVV thanks the Fundac\~ao para a  Ci\^encia e Tecnologia (FCT), Portugal, for the financial support to the Center for Astrophysics and Gravitation (CENTRA/IST/ULisboa) through the Grant Project~No.~UIDB/00099/2020. 
This work was also supported by the Universitat de les Illes Balears (UIB); the Spanish Agencia Estatal de Investigación grants PID2022-138626NB-I00, RED2022-134204-E, RED2022-134411-T, funded by MICIU/AEI/10.13039/501100011033 and the ERDF/EU; and the Comunitat Autònoma de les Illes Balears through the Conselleria d'Educació i Universitats with funds from the European Union - NextGenerationEU/PRTR-C17.I1 (SINCO2022/6719) and from the European Union - European Regional Development Fund (ERDF) (SINCO2022/18146). JDA thanks Fundação para a Ciência e Tecnologia (FCT), Portugal, for the financial support through the Grant Project  2024.04456.CERN.

\appendix
\onecolumngrid

\section{Einstein's Field Equations in GBSSN+Z4 Formalism}
\label{appendix:gbssneinstein}
The conformally rescaled EFEs, when written in the generalized BSSN formalism, together with the Z4 quantities, give the following set of equations:
\begin{align}
    \partial_{\perp} \chi &= \frac{2}{3} \alpha \chi (K + 2 \Theta) + \frac{1}{3} \chi \partial_{\perp} \ln \gamma, \label{eq:chitensorial}\\
    \partial_{\perp} \gamma_{ab} &= -2 \alpha A_{ab} + \frac{1}{3} \gamma_{ab} \partial_{\perp} \ln \gamma\\
    \partial_{\perp} A_{ab} &= \Bigg[\alpha \chi \Bigg( R_{ab} + 2 D_{(a} Z_{b)} \Bigg) - \chi D_a D_b \alpha - D_{(a} \alpha D_{b)} \chi - \frac{\alpha D_{a} \chi D_{b} \chi}{4 \chi} + \frac{1}{2} \alpha D_a D_b \chi \nonumber \\
    & \quad + 2 Z_{(a} \alpha D_{b)} \chi + \frac{2 \alpha D_{(a} \chi D_{b)} \Omega}{\Omega} + \frac{2 \alpha \chi D_a D_b \Omega}{\Omega} + \frac{4 \alpha \chi Z_{(a} D_{b)} \Omega}{\Omega} - 8 \pi \alpha S_{ab} \Bigg]^{\text{TF}}, \\
    \partial_{\perp} K &= \alpha \Bigg( A^{ab} A_{ab} + \frac{1}{3} (K + 2 \Theta)^2 + \frac{\kappa_1 (1 - \kappa_2) \Theta}{\Omega} \Bigg) - \chi \Delta \alpha + \frac{1}{2} D^a \alpha D_a \chi + 2 C_{Z_4 c} Z^a D_a \alpha \nonumber \\
    & \quad + \frac{3[(\partial_\perp \Omega)^2 - \alpha^2 \chi D^a \Omega D_a \Omega]}{\Omega^2 \alpha} - \frac{2 C_{Z_4 c} \alpha Z_a D_a \Omega}{\Omega} + \frac{3 \chi D^a \alpha D_a \Omega}{\Omega} - \frac{\alpha D^a \chi D_a \Omega}{2 \Omega} \nonumber \\
    & \quad + \frac{\alpha \chi \Delta \Omega}{\Omega} + \frac{[K + 4 \Theta] \partial_{\perp} \Omega}{\Omega} + \frac{3 \partial_{\perp} \alpha \partial_{\perp} \Omega}{\Omega \alpha^2} - \frac{3 \partial_\perp \partial_{\perp} \Omega}{\Omega \alpha} + 4 \pi \alpha (\rho + S), \\
    \partial_{\perp} \Lambda_a &= \frac{2 Z^b \tilde{D}_{b}\beta^a}{\chi} + \alpha \Bigg( 2 A^{bc} \Delta \Gamma^a_{bc} - \frac{2}{3} D^a (K + 2 \Theta) - \frac{3 A^{ab} D_b \chi}{\chi} - \frac{4 Z^a (K + 2 \Theta)}{3 \chi} - \frac{2 \kappa_1 Z^a}{\Omega \chi} \Bigg) \nonumber \\
    & \quad + \gamma^{bc} \hat{D}_{b} \hat{D}_{c} \beta_a - \gamma^{bc} R[\hat{\gamma}]^a_{bcd}  \beta^{d} - 2 A^{ab} D_b \alpha - 2 C_{Z_4 c} \Theta D^a \alpha - \frac{4 \alpha A^{ab} D_b \Omega}{\Omega} \nonumber \\
    & \quad - \frac{2\alpha (2K + \Theta)D^a \Omega}{3\Omega} + \frac{2C_{Z4c} \alpha \Theta D^a \Omega}{\Omega} - \frac{4 D^a \partial_\perp \Omega}{\Omega} + \frac{4D^a \alpha \partial_\perp \Omega}{\Omega \alpha} - \frac{4Z^a \partial_\perp \Omega}{\Omega \chi} \nonumber \\
    & \quad -\frac{1}{6}D^a \partial_\perp \ln \gamma - \frac{1}{3}\Delta \Gamma^a \partial_\perp \ln \gamma - \frac{2Z^a \partial_\perp \ln \gamma}{3 \chi} - \frac{16\pi J^a \alpha}{\chi}\\
    \partial_{\perp} \Theta &= \frac{\alpha}{2} \Bigg[ \chi(R[\gamma] + 2 D^a Z_a) - A^{ab} A_{ab} + \frac{2}{3} (K + 2 \Theta)^2 - 2 C_{Z_4 c} \Theta (K+2 \Theta) - \frac{2 \kappa_1 (2 + \kappa_2) \Theta}{\Omega} \Bigg]\nonumber \\
    & \quad +\alpha \Delta \chi - \frac{5 \alpha D^a \chi D_a \chi}{4 \chi} - C_{Z4c} Z^a D_a \alpha - \frac{C_{Z4c} \alpha Z_a D_a \chi}{2\chi} + \frac{2 \alpha \chi \Delta \Omega}{\Omega} - \frac{\alpha D^a \chi D_a \Omega}{\Omega} \nonumber\\ 
    & \quad \frac{3[(\partial_\perp \Omega)^2 - \alpha^2 \chi D^a \Omega D_a \Omega]}{\Omega^2 \alpha} + \frac{2 K \partial_\perp \Omega}{\Omega} - 8 \pi \alpha \rho,
    \label{eq:z4GBSSN}
\end{align}
together with the constraints,
\vspace{-2mm}
\begin{align}
    \mathcal{H} &= \chi R[\gamma] - A_{ab} A^{ab} + \frac{2}{3}(K + 2\Theta)^2 + 2 \triangle \chi - \frac{5D^a \chi D_a \chi}{2\chi} + \frac{6[(\partial_\perp \Omega)^2 - \alpha^2 \chi D^a \Omega D_a \Omega]}{\Omega^2 \alpha^2} \nonumber \\
    &\quad - \frac{2 D^a \chi D_a \Omega}{\Omega} + \frac{4 \chi \Delta \Omega}{\Omega} + \frac{4(K + 2\Theta)\partial_\perp \Omega}{\Omega \alpha} - 16 \pi \rho,\label{eq:Ham}  \\
    \mathcal{M}_a &= D_b A^b_a - \frac{2}{3} D_a (K + 2\Theta) - \frac{3 A^b_a D_b \chi}{2\chi} - \frac{2 A^b_a D_b \Omega}{\Omega} - \frac{2(K + 2\Theta)D_a \Omega}{3\Omega} - \frac{2 D_a \partial_\perp \Omega}{\Omega \alpha} \nonumber \\
    &\quad + \frac{2 D_a \alpha \partial_\perp \Omega}{\Omega \alpha^2} - 8 \pi J_a,\label{eq:Mom} \\
    Z_a &= \frac{\gamma_{ab}}{2} \left(\Lambda^b - \Delta \Gamma^b \right),
\end{align}
where the third constraint comes from writing the Z4 quantity $Z_a$ in terms of $\Lambda^a$ and $\Delta \Gamma^a$. The equations we evolve use $\Delta \tilde{K} = \tilde{K}-K_{\text{CMC}}$ as evolution variable instead of $K$ and the ``physical'' $\tilde{\Theta}$ variable instead of $\Theta$, as is done in \cite{vano-vinuales_free_2015}. The transformations are given by \eqref{eq:ktransformations}. Regarding Maxwell's and KG equations, they become (written already as functions of the physical extrinsic curvature and the physical $\tilde \Theta$):
\vspace{-1mm}
\begin{align}
     \partial_t E_j =& -4 \pi \bar{j}_{j} \alpha + \frac{K_{CMC} E_{j} \alpha}{3 \Omega} + \frac{\Delta\tilde{K} E_{j} \alpha}{3 \Omega} + \frac{2 C_{Z4} E_{j} \tilde{\Theta} \alpha}{3 \Omega} + \beta^{i} D_{i}E_{j} - E_{i} D_{j} \beta^i + \frac{\alpha D_{j}\psi}{\Omega^2} - \frac{E_{j} \partial_\perp \Omega}{\Omega}\\
    \partial_t \psi =& -4 \pi \bar{q}_{\text{dens.}} \alpha + 4 \pi k \alpha \Omega^2 \psi +  \alpha \Omega^2 D_{i}E^{i} - \frac{3 E^{i} \alpha \Omega^2 D_{i}\chi}{2 \chi} + \beta^{i} D_{i}\psi \label{eq:electrictensorial}\\
    \partial_t {A_3}_j =&\, E_{j} \alpha + \beta^{i} D_{i}{A_3}_{j} -  \Phi D_{j}\alpha + {A_3}_{i} D_{j}\beta^{i} -  \alpha D_{j}\Phi\\
    \partial_t \Phi =& \frac{K_{CMC} \alpha \Phi}{\Omega} + \frac{\Delta \tilde K \alpha \Phi}{\Omega} + \frac{2 C_{Z4} \tilde{\Theta} \alpha \Phi}{\Omega} -  \alpha D_{k}{A_3}^{k} -  {A_3}^{k} D_{k}\alpha + \beta^{k} D_{k}\Phi + \frac{3 {A_3}^{k} \alpha D_{k}\chi}{2 \chi} -  \frac{3 \Phi \partial_\perp \Omega}{\Omega}\\
    \partial_t \bar{c}_\phi =& \, \bar{c}_\Pi \\
    \partial_t \bar{d}_\phi =& \,\bar{d}_\Pi \\
    \partial_t \bar{c}_\Pi =& \frac{\bar{c}_\Pi \partial_t \alpha}{\alpha} -  q \bar{d}_\phi \partial_t \Phi \alpha -  q^2 {A_3}_{m} {A_3}^{m} \bar{c}_\phi \alpha^2 + \frac{K_{CMC} \bar{c}_\Pi \alpha}{\Omega} + \frac{\bar{c}_\Pi \Delta\tilde{K} \alpha}{\Omega} + \frac{2 C_{Z4} \bar{c}_\Pi \tilde{\Theta} \alpha}{\Omega} - 2 q \bar{d}_\Pi \alpha \Phi \nonumber\\
    &+ \frac{K_{CMC} q \bar{d}_\phi \alpha^2 \Phi}{\Omega} + \frac{q \Delta\tilde{K} \bar{d}_\phi \alpha^2 \Phi}{\Omega} + \frac{2 q C_{Z4} \bar{d}_\phi \tilde{\Theta} \alpha^2 \Phi}{\Omega} + q^2 \bar{c}_\phi \alpha^2 \Phi^2 + 2 \beta^{i} D_{i}\bar{c}_\Pi + \partial_t \beta^{i} D_{i}\bar{c}_\phi \nonumber\\
    &-  \frac{\partial_t \alpha \beta^{i} D_{i}\bar{c}_\phi}{\alpha} -  \frac{K_{CMC} \alpha \beta^{i} D_{i}\bar{c}_\phi}{\Omega} -  \frac{\Delta\tilde{K} \alpha \beta^{i} D_{i}\bar{c}_\phi}{\Omega} -  \frac{2 C_{Z4} \tilde{\Theta} \alpha \beta^{i} D_{i}\bar{c}_\phi}{\Omega} + 2 q \alpha \beta^{i} \Phi D_{i}\bar{d}_\phi \nonumber\\
    &-  \frac{\bar{c}_\Pi \beta^{i} D_{i}\alpha}{\alpha} -  \frac{K_{CMC} \bar{c}_\phi \alpha \beta^{i} D_{i}\Omega}{\Omega^2} -  \frac{\bar{c}_\phi \Delta\tilde{K} \alpha \beta^{i} D_{i}\Omega}{\Omega^2} -  \frac{2 C_{Z4} \bar{c}_\phi \tilde{\Theta} \alpha \beta^{i} D_{i}\Omega}{\Omega^2} + \frac{\bar{c}_\phi \partial_t \beta^{i} D_{i}\Omega}{\Omega}\nonumber\\
    & -  \frac{\bar{c}_\phi \partial_t \alpha \beta^{i} D_{i}\Omega}{\alpha \Omega} + q \bar{d}_\phi \alpha \beta^{i} D_{i}\Phi + \frac{3 q {A_3}^{i} \bar{d}_\phi \alpha^2 D_{i}\chi}{2 \chi} + \frac{\beta^{i} \beta^{j} D_{i}\alpha D_{j}\bar{c}_\phi}{\alpha} + \frac{\beta^{i} \beta^{j} D_{i}\Omega D_{j}\bar{c}_\phi}{\Omega}\nonumber\\
    & -  \beta^{j} D_{i}\bar{c}_\phi D_{j}\beta^{i} -  \frac{\bar{c}_\phi \beta^{j} D_{i}\Omega D_{j}\beta^{i}}{\Omega} -  \frac{\beta^{i} \beta^{j} D_{i}\bar{c}_\phi D_{j}\Omega}{\Omega} + \frac{\bar{c}_\phi \beta^{i} \beta^{j} D_{i}\alpha D_{j}\Omega}{\alpha \Omega} + \frac{2 \bar{c}_\phi \beta^{i} \beta^{j} D_{i}\Omega D_{j}\Omega}{\Omega^2}\nonumber\\
    & -  \tfrac{3}{2} \gamma^{ij} \alpha^2 D_{i}\bar{c}_\phi D_{j}\chi -  \frac{3 \bar{c}_\phi \gamma^{ij} \alpha^2 D_{i}\Omega D_{j}\chi}{2 \Omega} -  \beta^{i} \beta^{j} D_{j}D_{i}\bar{c}_\phi -  \frac{\bar{c}_\phi \beta^{i} \beta^{j} D_{j}D_{i}\Omega}{\Omega} -  q {A_3}^{k} \bar{d}_\phi \alpha D_{k}\alpha\nonumber\\
    & + \frac{4 q {A_3}^{k} \bar{d}_\phi \alpha^2 D_{k}\Omega}{\Omega} + \gamma^{lk} \alpha \chi D_{k}\alpha D_{l}\bar{c}_\phi -  \frac{4 \gamma^{kl} \alpha^2 \chi D_{k}\bar{c}_\phi D_{l}\Omega}{\Omega} + \frac{\bar{c}_\phi \gamma^{lk} \alpha \chi D_{k}\alpha D_{l}\Omega}{\Omega}\nonumber\\
    & -  \frac{4 \bar{c}_\phi \gamma^{kl} \alpha^2 \chi D_{k}\Omega D_{l}\Omega}{\Omega^2} -  q \bar{d}_\phi \gamma^{mn} \alpha^2 \chi D_{m}{A_3}_{n} - 2 q {A_3}^{m} \alpha^2 D_{m}\bar{d}_\phi -  \frac{4 q {A_3}^{m} \bar{d}_\phi \alpha^2 D_{m}\Omega}{\Omega}\nonumber\\
    & -  \frac{q {A_3}^{m} \bar{d}_\phi \alpha^2 D_{m}\chi}{2 \chi} + \frac{2 \gamma^{mn} \alpha^2 \chi D_{m}\Omega D_{n}\bar{c}_\phi}{\Omega} + \tfrac{1}{2} \gamma^{mn} \alpha^2 D_{m}\chi D_{n}\bar{c}_\phi + \frac{2 \gamma^{mn} \alpha^2 \chi D_{m}\bar{c}_\phi D_{n}\Omega}{\Omega}\nonumber\\
    & + \frac{2 \bar{c}_\phi \gamma^{mn} \alpha^2 \chi D_{m}\Omega D_{n}\Omega}{\Omega^2} + \frac{\bar{c}_\phi \gamma^{mn} \alpha^2 D_{m}\chi D_{n}\Omega}{2 \Omega} -  \frac{q {A_3}^{n} \bar{d}_\phi \alpha^2 D_{n}\chi}{2 \chi} + \tfrac{1}{2} \gamma^{mn} \alpha^2 D_{m}\bar{c}_\phi D_{n}\chi\nonumber\\
    & + \frac{\bar{c}_\phi \gamma^{mn} \alpha^2 D_{m}\Omega D_{n}\chi}{2 \Omega} + \gamma^{mn} \alpha^2 \chi D_{n}D_{m}\bar{c}_\phi + \frac{\bar{c}_\phi \gamma^{mn} \alpha^2 \chi D_{n}D_{m}\Omega}{\Omega} -  \frac{3 \bar{c}_\Pi \partial_\perp \Omega}{\Omega} -  \frac{3 q \bar{d}_\phi \alpha \Phi \partial_\perp \Omega}{\Omega}\nonumber\\
    & + \frac{3 \beta^{i} D_{i}\bar{c}_\phi \partial_\perp \Omega}{\Omega} + \frac{3 \bar{c}_\phi \beta^{i} D_{i}\Omega \partial_\perp \Omega}{\Omega^2}\\
    \partial_t \bar{d}_\Pi &= \frac{\bar{d}_\Pi \partial_t \alpha}{\alpha} + q \bar{c}_\phi \partial_t \Phi \alpha -  q^2 {A_3}_{m} {A_3}^{m} \bar{d}_\phi \alpha^2 + \frac{K_{CMC} \bar{d}_\Pi \alpha}{\Omega} + \frac{\Delta\tilde{K} \bar{d}_\Pi \alpha}{\Omega} + \frac{2 C_{Z4} \bar{d}_\Pi \tilde{\Theta} \alpha}{\Omega} + 2 q \bar{c}_\Pi \alpha \Phi\nonumber\\
    & -  \frac{K_{CMC} q \bar{c}_\phi \alpha^2 \Phi}{\Omega} -  \frac{q \bar{c}_\phi \Delta\tilde{K} \alpha^2 \Phi}{\Omega} -  \frac{2 q C_{Z4} \bar{c}_\phi \tilde{\Theta} \alpha^2 \Phi}{\Omega} + q^2 \bar{d}_\phi \alpha^2 \Phi^2 - 2 q \alpha \beta^{i} \Phi D_{i}\bar{c}_\phi + 2 \beta^{i} D_{i}\bar{d}_\Pi\nonumber\\
    & + \partial_t\beta^{i} D_{i}\bar{d}_\phi -  \frac{\partial_t \alpha \beta^{i} D_{i}\bar{d}_\phi}{\alpha} -  \frac{K_{CMC} \alpha \beta^{i} D_{i}\bar{d}_\phi}{\Omega} -  \frac{\Delta\tilde{K} \alpha \beta^{i} D_{i}\bar{d}_\phi}{\Omega} -  \frac{2 C_{Z4} \tilde{\Theta} \alpha \beta^{i} D_{i}\bar{d}_\phi}{\Omega} -  \frac{\bar{d}_\Pi \beta^{i} D_{i}\alpha}{\alpha}\nonumber\\
    & -  \frac{K_{CMC} \bar{d}_\phi \alpha \beta^{i} D_{i}\Omega}{\Omega^2} -  \frac{\Delta\tilde{K} \bar{d}_\phi \alpha \beta^{i} D_{i}\Omega}{\Omega^2} -  \frac{2 C_{Z4} \bar{d}_\phi \tilde{\Theta} \alpha \beta^{i} D_{i}\Omega}{\Omega^2} + \frac{\bar{d}_\phi RHS\beta^{i} D_{i}\Omega}{\Omega} -  \frac{\bar{d}_\phi \partial_t \alpha \beta^{i} D_{i}\Omega}{\alpha \Omega}\nonumber\\
    & -  q \bar{c}_\phi \alpha \beta^{i} D_{i}\Phi -  \frac{3 q {A_3}^{i} \bar{c}_\phi \alpha^2 D_{i}\chi}{2 \chi} + \frac{\beta^{i} \beta^{j} D_{i}\alpha D_{j}\bar{d}_\phi}{\alpha} + \frac{\beta^{i} \beta^{j} D_{i}\Omega D_{j}\bar{d}_\phi}{\Omega} -  \beta^{j} D_{i}\bar{d}_\phi D_{j}\beta^{i}\nonumber\\
    & -  \frac{\bar{d}_\phi \beta^{j} D_{i}\Omega D_{j}\beta^{i}}{\Omega} -  \frac{\beta^{i} \beta^{j} D_{i}\bar{d}_\phi D_{j}\Omega}{\Omega} + \frac{\bar{d}_\phi \beta^{i} \beta^{j} D_{i}\alpha D_{j}\Omega}{\alpha \Omega} + \frac{2 \bar{d}_\phi \beta^{i} \beta^{j} D_{i}\Omega D_{j}\Omega}{\Omega^2} -  \tfrac{3}{2} \gamma^{ij} \alpha^2 D_{i}\bar{d}_\phi D_{j}\chi \nonumber\\
    &-  \frac{3 \bar{d}_\phi \gamma^{ij} \alpha^2 D_{i}\Omega D_{j}\chi}{2 \Omega} -  \beta^{i} \beta^{j} D_{j}D_{i}\bar{d}_\phi -  \frac{\bar{d}_\phi \beta^{i} \beta^{j} D_{j}D_{i}\Omega}{\Omega} + q {A_3}^{k} \bar{c}_\phi \alpha D_{k}\alpha -  \frac{4 q {A_3}^{k} \bar{c}_\phi \alpha^2 D_{k}\Omega}{\Omega}\nonumber\\
    &+ \gamma^{lk} \alpha \chi D_{k}\alpha D_{l}\bar{d}_\phi -  \frac{4 \gamma^{kl} \alpha^2 \chi D_{k}\bar{d}_\phi D_{l}\Omega}{\Omega} + \frac{\bar{d}_\phi \gamma^{lk} \alpha \chi D_{k}\alpha D_{l}\Omega}{\Omega} -  \frac{4 \bar{d}_\phi \gamma^{kl} \alpha^2 \chi D_{k}\Omega D_{l}\Omega}{\Omega^2}\nonumber\\
    & + q \bar{c}_\phi \gamma^{mn} \alpha^2 \chi D_{m}{A_3}_{n} + 2 q {A_3}^{m} \alpha^2 D_{m}\bar{c}_\phi + \frac{4 q {A_3}^{m} \bar{c}_\phi \alpha^2 D_{m}\Omega}{\Omega} + \frac{q {A_3}^{m} \bar{c}_\phi \alpha^2 D_{m}\chi}{2 \chi}+ \frac{2 \gamma^{mn} \alpha^2 \chi D_{m}\Omega D_{n}\bar{d}_\phi}{\Omega}\nonumber\\
    & + \tfrac{1}{2} \gamma^{mn} \alpha^2 D_{m}\chi D_{n}\bar{d}_\phi + \frac{2 \gamma^{mn} \alpha^2 \chi D_{m}\bar{d}_\phi D_{n}\Omega}{\Omega} + \frac{2 \bar{d}_\phi \gamma^{mn} \alpha^2 \chi D_{m}\Omega D_{n}\Omega}{\Omega^2} + \frac{\bar{d}_\phi \gamma^{mn} \alpha^2 D_{m}\chi D_{n}\Omega}{2 \Omega}\nonumber\\
    & + \frac{q {A_3}^{n} \bar{c}_\phi \alpha^2 D_{n}\chi}{2 \chi} + \tfrac{1}{2} \gamma^{mn} \alpha^2 D_{m}\bar{d}_\phi D_{n}\chi + \frac{\bar{d}_\phi \gamma^{mn} \alpha^2 D_{m}\Omega D_{n}\chi}{2 \Omega} + \gamma^{mn} \alpha^2 \chi D_{n}D_{m}\bar{d}_\phi + \frac{\bar{d}_\phi \gamma^{mn} \alpha^2 \chi D_{n}D_{m}\Omega}{\Omega}\nonumber\\
    & -  \frac{3 \bar{d}_\Pi \partial_\perp \Omega}{\Omega} + \frac{3 q \bar{c}_\phi \alpha \Phi \partial_\perp \Omega}{\Omega} + \frac{3 \beta^{i} D_{i}\bar{d}_\phi \partial_\perp \Omega}{\Omega} + \frac{3 \bar{d}_\phi \beta^{i} D_{i}\Omega \partial_\perp \Omega}{\Omega^2}\,.\label{eq:dpitensorial}
 \end{align}
We took out the overbars from $A_3$, $E^i$ and $\Phi$ for better readibility.\par
We are lacking the definitions of the projections of the stress-energy tensor, present in some of the EFEs. We have,
\begin{align*}
    \rho &= \bar{n}^a \bar{n}^b T_{ab}\,,\\
    J^a &= -\bar{\gamma}^{ab} \bar{n}^c T_{bc}\,,\\
    S_{ab} &= \bar{\gamma}^c_a \bar{\gamma}^d_b T_{cd}\,,\\
    S &= \bar{\gamma}^{ab} S_{ab}\,.
\end{align*}
We now display these projections separately for the Maxwell and KG parts, for better readability. Some cross terms will appear in the KG equations, due to the natural coupling arising from the charge. We will directly show them already in the GBSSN+Z4 formulation.

\subsection*{Scalar Field Part}
\vspace{-3mm}
\begin{align}
    \rho =& \,\frac{\bar{c}_\Pi^2 \Omega^2}{2 \alpha^2} + \frac{q^2 \bar{d}_\Pi^2 \Omega^2}{2 \alpha^2} + \tfrac{1}{2} q^2 {A_3}_{i} {A_3}^{i} \bar{c}_\phi^2 \Omega^6 + \tfrac{1}{2} q^4 {A_3}_{i} {A_3}^{i} \bar{d}_\phi^2 \Omega^6 -  \frac{q^2 \bar{c}_\phi \bar{d}_\Pi \Omega^4 \Phi}{\alpha} + \frac{q^2 \bar{c}_\Pi \bar{d}_\phi \Omega^4 \Phi}{\alpha}\nonumber\\
    & + \tfrac{1}{2} q^2 \bar{c}_\phi^2 \Omega^6 \Phi^2 + \tfrac{1}{2} q^4 \bar{d}_\phi^2 \Omega^6 \Phi^2 -  \frac{\bar{c}_\Pi \beta^{i} \Omega^2 D_{i}\bar{c}_\phi}{\alpha^2} -  q^2 {A_3}^{i} \bar{d}_\phi \Omega^4 D_{i}\bar{c}_\phi -  \frac{q^2 \bar{d}_\phi \beta^{i} \Omega^4 \Phi D_{i}\bar{c}_\phi}{\alpha}\nonumber \\
    & -  \frac{q^2 \bar{d}_\Pi \beta^{i} \Omega^2 D_{i}\bar{d}_\phi}{\alpha^2} + q^2 {A_3}^{i} \bar{c}_\phi \Omega^4 D_{i}\bar{d}_\phi + \frac{q^2 \bar{c}_\phi \beta^{i} \Omega^4 \Phi D_{i}\bar{d}_\phi}{\alpha} -  \frac{\bar{c}_\Pi \bar{c}_\phi \beta^{i} \Omega D_{i}\Omega}{\alpha^2} -  \frac{q^2 \bar{d}_\Pi \bar{d}_\phi \beta^{i} \Omega D_{i}\Omega}{\alpha^2}\nonumber\\
    & + \frac{\beta^{i} \beta^{j} \Omega^2 D_{i}\bar{c}_\phi D_{j}\bar{c}_\phi}{2 \alpha^2} + \tfrac{1}{2} \gamma^{ji} \Omega^2 \chi D_{i}\bar{c}_\phi D_{j}\bar{c}_\phi + \frac{q^2 \beta^{i} \beta^{j} \Omega^2 D_{i}\bar{d}_\phi D_{j}\bar{d}_\phi}{2 \alpha^2} + \tfrac{1}{2} q^2 \gamma^{ji} \Omega^2 \chi D_{i}\bar{d}_\phi D_{j}\bar{d}_\phi \nonumber\\
    &+ \frac{\bar{c}_\phi \beta^{i} \beta^{j} \Omega D_{i}\bar{c}_\phi D_{j}\Omega}{\alpha^2} + \bar{c}_\phi \gamma^{ji} \Omega \chi D_{i}\bar{c}_\phi D_{j}\Omega + \frac{q^2 \bar{d}_\phi \beta^{i} \beta^{j} \Omega D_{i}\bar{d}_\phi D_{j}\Omega}{\alpha^2} + q^2 \bar{d}_\phi \gamma^{ji} \Omega \chi D_{i}\bar{d}_\phi D_{j}\Omega\nonumber\\
    & + \frac{\bar{c}_\phi^2 \beta^{i} \beta^{j} D_{i}\Omega D_{j}\Omega}{2 \alpha^2} + \frac{q^2 \bar{d}_\phi^2 \beta^{i} \beta^{j} D_{i}\Omega D_{j}\Omega}{2 \alpha^2} + \tfrac{1}{2} \bar{c}_\phi^2 \gamma^{ji} \chi D_{i}\Omega D_{j}\Omega + \tfrac{1}{2} q^2 \bar{d}_\phi^2 \gamma^{ji} \chi D_{i}\Omega D_{j}\Omega \\
    J^i =& - \frac{q^2 {A_3}^{i} \bar{c}_\phi \bar{d}_\Pi \Omega^4}{\alpha} + \frac{q^2 {A_3}^{i} \bar{c}_\Pi \bar{d}_\phi \Omega^4}{\alpha} + q^2 {A_3}^{i} \bar{c}_\phi^2 \Omega^6 \Phi + q^4 {A_3}^{i} \bar{d}_\phi^2 \Omega^6 \Phi -  \frac{q^2 {A_3}^{i} \bar{d}_\phi \beta^{j} \Omega^4 D_{j}\bar{c}_\phi}{\alpha}\nonumber\\
    & -  \frac{\bar{c}_\Pi \gamma^{ij} \Omega^2 \chi D_{j}\bar{c}_\phi}{\alpha} -  q^2 \bar{d}_\phi \gamma^{ij} \Omega^4 \Phi \chi D_{j}\bar{c}_\phi + \frac{q^2 {A_3}^{i} \bar{c}_\phi \beta^{j} \Omega^4 D_{j}\bar{d}_\phi}{\alpha} -  \frac{q^2 \bar{d}_\Pi \gamma^{ij} \Omega^2 \chi D_{j}\bar{d}_\phi}{\alpha}\nonumber \\
    &+ q^2 \bar{c}_\phi \gamma^{ij} \Omega^4 \Phi \chi D_{j}\bar{d}_\phi -  \frac{\bar{c}_\Pi \bar{c}_\phi \gamma^{ij} \Omega \chi D_{j}\Omega}{\alpha} -  \frac{q^2 \bar{d}_\Pi \bar{d}_\phi \gamma^{ij} \Omega \chi D_{j}\Omega}{\alpha} + \frac{\gamma^{ij} \beta^{k} \Omega^2 \chi D_{j}\bar{c}_\phi D_{k}\bar{c}_\phi}{\alpha}\nonumber \\
    &+ \frac{\bar{c}_\phi \gamma^{ij} \beta^{k} \Omega \chi D_{j}\Omega D_{k}\bar{c}_\phi}{\alpha} + \frac{q^2 \gamma^{ij} \beta^{k} \Omega^2 \chi D_{j}\bar{d}_\phi D_{k}\bar{d}_\phi}{\alpha} + \frac{q^2 \bar{d}_\phi \gamma^{ij} \beta^{k} \Omega \chi D_{j}\Omega D_{k}\bar{d}_\phi}{\alpha} + \frac{\bar{c}_\phi \gamma^{ij} \beta^{k} \Omega \chi D_{j}\bar{c}_\phi D_{k}\Omega}{\alpha}\nonumber\\
    & + \frac{q^2 \bar{d}_\phi \gamma^{ij} \beta^{k} \Omega \chi D_{j}\bar{d}_\phi D_{k}\Omega}{\alpha} + \frac{\bar{c}_\phi^2 \gamma^{ij} \beta^{k} \chi D_{j}\Omega D_{k}\Omega}{\alpha} + \frac{q^2 \bar{d}_\phi^2 \gamma^{ij} \beta^{k} \chi D_{j}\Omega D_{k}\Omega}{\alpha}\\
    S_{ij} =& \, q^2 {A_3}_{i} {A_3}_{j} \bar{c}_\phi^2 \Omega^6 + q^4 {A_3}_{i} {A_3}_{j} \bar{d}_\phi^2 \Omega^6 + \frac{\bar{c}_\Pi^2 \gamma_{ij} \Omega^2}{2 \alpha^2 \chi} + \frac{q^2 \bar{d}_\Pi^2 \gamma_{ij} \Omega^2}{2 \alpha^2 \chi} -  \frac{q^2 {A_3}_{k} {A_3}^{k} \bar{c}_\phi^2 \gamma_{ij} \Omega^6}{2 \chi} -  \frac{q^4 {A_3}_{k} {A_3}^{k} \bar{d}_\phi^2 \gamma_{ij} \Omega^6}{2 \chi} \nonumber\\
    &-  \frac{q^2 \bar{c}_\phi \bar{d}_\Pi \gamma_{ij} \Omega^4 \Phi}{\alpha \chi} + \frac{q^2 \bar{c}_\Pi \bar{d}_\phi \gamma_{ij} \Omega^4 \Phi}{\alpha \chi} + \frac{q^2 \bar{c}_\phi^2 \gamma_{ij} \Omega^6 \Phi^2}{2 \chi} + \frac{q^4 \bar{d}_\phi^2 \gamma_{ij} \Omega^6 \Phi^2}{2 \chi} -  q^2 {A_3}_{j} \bar{d}_\phi \Omega^4 D_{i}\bar{c}_\phi \nonumber\\
    & + q^2 {A_3}_{j} \bar{c}_\phi \Omega^4 D_{i}\bar{d}_\phi -  q^2 {A_3}_{i} \bar{d}_\phi \Omega^4 D_{j}\bar{c}_\phi + \Omega^2 D_{i}\bar{c}_\phi D_{j}\bar{c}_\phi + \bar{c}_\phi \Omega D_{i}\Omega D_{j}\bar{c}_\phi + q^2 {A_3}_{i} \bar{c}_\phi \Omega^4 D_{j}\bar{d}_\phi\nonumber\\
    & + q^2 \Omega^2 D_{i}\bar{d}_\phi D_{j}\bar{d}_\phi + q^2 \bar{d}_\phi \Omega D_{i}\Omega D_{j}\bar{d}_\phi + \bar{c}_\phi \Omega D_{i}\bar{c}_\phi D_{j}\Omega + q^2 \bar{d}_\phi \Omega D_{i}\bar{d}_\phi D_{j}\Omega + \bar{c}_\phi^2 D_{i}\Omega D_{j}\Omega \nonumber\\
    &+ q^2 \bar{d}_\phi^2 D_{i}\Omega D_{j}\Omega -  \frac{\bar{c}_\Pi \gamma_{ij} \beta^{k} \Omega^2 D_{k}\bar{c}_\phi}{\alpha^2 \chi} + \frac{q^2 {A_3}^{k} \bar{d}_\phi \gamma_{ij} \Omega^4 D_{k}\bar{c}_\phi}{\chi} -  \frac{q^2 \bar{d}_\phi \gamma_{ij} \beta^{k} \Omega^4 \Phi D_{k}\bar{c}_\phi}{\alpha \chi}\nonumber\\
    & -  \frac{q^2 \bar{d}_\Pi \gamma_{ij} \beta^{k} \Omega^2 D_{k}\bar{d}_\phi}{\alpha^2 \chi} -  \frac{q^2 {A_3}^{k} \bar{c}_\phi \gamma_{ij} \Omega^4 D_{k}\bar{d}_\phi}{\chi} + \frac{q^2 \bar{c}_\phi \gamma_{ij} \beta^{k} \Omega^4 \Phi D_{k}\bar{d}_\phi}{\alpha \chi} -  \frac{\bar{c}_\Pi \bar{c}_\phi \gamma_{ij} \beta^{k} \Omega D_{k}\Omega}{\alpha^2 \chi} \nonumber\\
    &-  \frac{q^2 \bar{d}_\Pi \bar{d}_\phi \gamma_{ij} \beta^{k} \Omega D_{k}\Omega}{\alpha^2 \chi} -  \tfrac{1}{2} \gamma_{ij} \gamma^{kl} \Omega^2 D_{k}\bar{c}_\phi D_{l}\bar{c}_\phi + \frac{\gamma_{ij} \beta^{k} \beta^{l} \Omega^2 D_{k}\bar{c}_\phi D_{l}\bar{c}_\phi}{2 \alpha^2 \chi} -  \tfrac{1}{2} q^2 \gamma_{ij} \gamma^{kl} \Omega^2 D_{k}\bar{d}_\phi D_{l}\bar{d}_\phi\nonumber\\
    & + \frac{q^2 \gamma_{ij} \beta^{k} \beta^{l} \Omega^2 D_{k}\bar{d}_\phi D_{l}\bar{d}_\phi}{2 \alpha^2 \chi} -  \bar{c}_\phi \gamma_{ij} \gamma^{kl} \Omega D_{k}\bar{c}_\phi D_{l}\Omega + \frac{\bar{c}_\phi \gamma_{ij} \beta^{k} \beta^{l} \Omega D_{k}\bar{c}_\phi D_{l}\Omega}{\alpha^2 \chi} -  q^2 \bar{d}_\phi \gamma_{ij} \gamma^{kl} \Omega D_{k}\bar{d}_\phi D_{l}\Omega\nonumber\\
    & + \frac{q^2 \bar{d}_\phi \gamma_{ij} \beta^{k} \beta^{l} \Omega D_{k}\bar{d}_\phi D_{l}\Omega}{\alpha^2 \chi} -  \tfrac{1}{2} \bar{c}_\phi^2 \gamma_{ij} \gamma^{kl} D_{k}\Omega D_{l}\Omega -  \tfrac{1}{2} q^2 \bar{d}_\phi^2 \gamma_{ij} \gamma^{kl} D_{k}\Omega D_{l}\Omega + \frac{\bar{c}_\phi^2 \gamma_{ij} \beta^{k} \beta^{l} D_{k}\Omega D_{l}\Omega}{2 \alpha^2 \chi}\nonumber\\
    & + \frac{q^2 \bar{d}_\phi^2 \gamma_{ij} \beta^{k} \beta^{l} D_{k}\Omega D_{l}\Omega}{2 \alpha^2 \chi}\\
    S =& \, \frac{3 \bar{c}_\Pi^2 \Omega^2}{2 \alpha^2} + \frac{3 q^2 \bar{d}_\Pi^2 \Omega^2}{2 \alpha^2} -  \tfrac{1}{2} q^2 {A_3}_{i} {A_3}^{i} \bar{c}_\phi^2 \Omega^6 -  \tfrac{1}{2} q^4 {A_3}_{i} {A_3}^{i} \bar{d}_\phi^2 \Omega^6 -  \frac{3 q^2 \bar{c}_\phi \bar{d}_\Pi \Omega^4 \Phi}{\alpha} + \frac{3 q^2 \bar{c}_\Pi \bar{d}_\phi \Omega^4 \Phi}{\alpha}\nonumber\\
    & + \tfrac{3}{2} q^2 \bar{c}_\phi^2 \Omega^6 \Phi^2 + \tfrac{3}{2} q^4 \bar{d}_\phi^2 \Omega^6 \Phi^2 -  \frac{3 \bar{c}_\Pi \beta^{i} \Omega^2 D_{i}\bar{c}_\phi}{\alpha^2} + q^2 {A_3}^{i} \bar{d}_\phi \Omega^4 D_{i}\bar{c}_\phi -  \frac{3 q^2 \bar{d}_\phi \beta^{i} \Omega^4 \Phi D_{i}\bar{c}_\phi}{\alpha}\nonumber\\
    & -  \frac{3 q^2 \bar{d}_\Pi \beta^{i} \Omega^2 D_{i}\bar{d}_\phi}{\alpha^2} -  q^2 {A_3}^{i} \bar{c}_\phi \Omega^4 D_{i}\bar{d}_\phi + \frac{3 q^2 \bar{c}_\phi \beta^{i} \Omega^4 \Phi D_{i}\bar{d}_\phi}{\alpha} -  \frac{3 \bar{c}_\Pi \bar{c}_\phi \beta^{i} \Omega D_{i}\Omega}{\alpha^2} -  \frac{3 q^2 \bar{d}_\Pi \bar{d}_\phi \beta^{i} \Omega D_{i}\Omega}{\alpha^2}\nonumber\\
    & -  \tfrac{3}{2} \Omega^2 \chi D_{i}\bar{c}_\phi D^{i}\bar{c}_\phi - 3 \bar{c}_\phi \Omega \chi D_{i}\Omega D^{i}\bar{c}_\phi -  \tfrac{3}{2} q^2 \Omega^2 \chi D_{i}\bar{d}_\phi D^{i}\bar{d}_\phi - 3 q^2 \bar{d}_\phi \Omega \chi D_{i}\Omega D^{i}\bar{d}_\phi -  \tfrac{3}{2} \bar{c}_\phi^2 \chi D_{i}\Omega D^{i}\Omega\nonumber\\
    & -  \tfrac{3}{2} q^2 \bar{d}_\phi^2 \chi D_{i}\Omega D^{i}\Omega + \frac{3 \beta^{i} \beta^{j} \Omega^2 D_{i}\bar{c}_\phi D_{j}\bar{c}_\phi}{2 \alpha^2} + \gamma^{ij} \Omega^2 \chi D_{i}\bar{c}_\phi D_{j}\bar{c}_\phi + \bar{c}_\phi \gamma^{ij} \Omega \chi D_{i}\Omega D_{j}\bar{c}_\phi + \frac{3 q^2 \beta^{i} \beta^{j} \Omega^2 D_{i}\bar{d}_\phi D_{j}\bar{d}_\phi}{2 \alpha^2}\nonumber\\
    & + q^2 \gamma^{ij} \Omega^2 \chi D_{i}\bar{d}_\phi D_{j}\bar{d}_\phi + q^2 \bar{d}_\phi \gamma^{ij} \Omega \chi D_{i}\Omega D_{j}\bar{d}_\phi + \frac{3 \bar{c}_\phi \beta^{i} \beta^{j} \Omega D_{i}\bar{c}_\phi D_{j}\Omega}{\alpha^2} + \bar{c}_\phi \gamma^{ij} \Omega \chi D_{i}\bar{c}_\phi D_{j}\Omega\nonumber\\
    & + \frac{3 q^2 \bar{d}_\phi \beta^{i} \beta^{j} \Omega D_{i}\bar{d}_\phi D_{j}\Omega}{\alpha^2} + q^2 \bar{d}_\phi \gamma^{ij} \Omega \chi D_{i}\bar{d}_\phi D_{j}\Omega + \frac{3 \bar{c}_\phi^2 \beta^{i} \beta^{j} D_{i}\Omega D_{j}\Omega}{2 \alpha^2} + \frac{3 q^2 \bar{d}_\phi^2 \beta^{i} \beta^{j} D_{i}\Omega D_{j}\Omega}{2 \alpha^2}\nonumber\\
    & + \bar{c}_\phi^2 \gamma^{ij} \chi D_{i}\Omega D_{j}\Omega + q^2 \bar{d}_\phi^2 \gamma^{ij} \chi D_{i}\Omega D_{j}\Omega 
\end{align}

\subsection*{Maxwell Part}
There are two ways to write Faraday's tensor, $F_{\mu\nu}$, one, which is the more standard way, is saying, $F_{\mu\nu} = \nabla_\mu A_\nu - \nabla_\nu A_\mu$. The other way is to write it as a function of the electric field \eqref{eq:maxwellelectric}. When writing out the stress-energy tensor part related to electromagnetism, this latter definition simplifies to a greater extent the expressions. We will work with the definition in terms of the electric field. This has been thoroughly discussed in \cite{alcubierre_einstein-maxwell_2009} and the authors of the paper even advise against using the electric and magnetic potentials. We, however, have to evolve these potentials because they enter directly into the scalar field's evolution equations. The stress-energy tensor, when using the definition in terms of the electric field becomes,
\begin{equation}
    (T_{\mu\nu})_{\text{Max}} = \Omega^2 \left(-\frac{E_\mu E_\nu}{4\pi}  + \frac{E_\alpha E^\alpha \bar{g}_{\mu\nu}}{8\pi}+ \frac{E_\alpha E^\alpha n_\mu n_\nu}{4\pi}\right)\,.
\end{equation}
The projections then become,
\begin{align}
    \rho =&\, \frac{E_{i} E^{i} \Omega^2}{8 \pi}\\
    J^i =& \, 0\\
    S_{ij} =&\, -\frac{E_i E_j \Omega^2}{4\pi} + \frac{E_k E^k \gamma_{ij} \Omega^2}{8\pi \chi}\\
    S = &\, \frac{E_i E^i \Omega^2}{8\pi}\, .
\end{align}
Note that $S = \rho$, which is according to the fact that the electromagnetic part of the stress-energy tensor is trace-free.

\bibliographystyle{apsrev}
\bibliography{references}

@article{zenginoglu_hyperboloidal_2008,
	title = {Hyperboloidal foliations and scri-fixing},
	volume = {25},
	issn = {0264-9381, 1361-6382},
	url = {http://arxiv.org/abs/0712.4333},
	doi = {10.1088/0264-9381/25/14/145002},
	number = {14},
	urldate = {2023-12-16},
	journal = {Class. Quantum Grav.},
	author = {Zenginoğlu, Anıl},
	month = jul,
	year = {2008},
	note = {arXiv:0712.4333 [gr-qc]},
	pages = {145002},
}

@misc{vano-vinuales_free_2015,
	title = {Free evolution of the hyperboloidal initial value problem in spherical symmetry},
	url = {http://arxiv.org/abs/1512.00776},
	doi = {10.48550/arXiv.1512.00776},
	urldate = {2023-12-16},
	publisher = {arXiv},
	author = {Vañó-Viñuales, Alex},
	month = nov,
	year = {2015},
	note = {arXiv:1512.00776 [gr-qc]},
}

@article{purrer_news_2005,
	title = {News from {Critical} {Collapse}: {Bondi} {Mass}, {Tails} and {Quasinormal} {Modes}},
	volume = {71},
	issn = {1550-7998, 1550-2368},
	shorttitle = {News from {Critical} {Collapse}},
	url = {http://arxiv.org/abs/gr-qc/0411078},
	doi = {10.1103/PhysRevD.71.104005},
	number = {10},
	urldate = {2023-12-16},
	journal = {Phys. Rev. D},
	author = {Pürrer, Michael and Husa, Sascha and Aichelburg, Peter C.},
	month = may,
	year = {2005},
	note = {arXiv:gr-qc/0411078},
	pages = {104005},
}

@article{frauendiener_numerical_2000,
	title = {Numerical treatment of the hyperboloidal initial value problem for the vacuum {Einstein} equations. {III}. {On} the determination of radiation},
	volume = {17},
	issn = {0264-9381, 1361-6382},
	url = {http://arxiv.org/abs/gr-qc/9808072},
	doi = {10.1088/0264-9381/17/2/308},
	number = {2},
	urldate = {2023-12-16},
	journal = {Class. Quantum Grav.},
	author = {Frauendiener, J.},
	month = jan,
	year = {2000},
	note = {arXiv:gr-qc/9808072},
	pages = {373--387},
}

@article{alcubierre_einstein-maxwell_2009,
	title = {The {Einstein}-{Maxwell} system in 3+1 form and initial data for multiple charged black holes},
	volume = {80},
	issn = {1550-7998, 1550-2368},
	url = {http://arxiv.org/abs/0907.1151},
	doi = {10.1103/PhysRevD.80.104022},
	number = {10},
	urldate = {2023-12-16},
	journal = {Phys. Rev. D},
	author = {Alcubierre, Miguel and Degollado, Juan Carlos and Salgado, Marcelo},
	month = nov,
	year = {2009},
	note = {arXiv:0907.1151 [gr-qc]},
	pages = {104022},
}

@misc{lopez_charged_2023,
	title = {Charged boson stars revisited},
	url = {http://arxiv.org/abs/2303.04066},
	doi = {10.48550/arXiv.2303.04066},
	urldate = {2023-12-16},
	publisher = {arXiv},
	author = {López, José Damián and Alcubierre, Miguel},
	month = mar,
	year = {2023},
	note = {arXiv:2303.04066 [gr-qc]},
}

@article{torres_gravitational_2014,
	title = {Gravitational collapse of charged scalar fields},
	volume = {46},
	issn = {0001-7701, 1572-9532},
	url = {http://arxiv.org/abs/1407.7885},
	doi = {10.1007/s10714-014-1773-4},
	number = {9},
	urldate = {2023-12-16},
	journal = {Gen Relativ Gravit},
	author = {Torres, Jose M. and Alcubierre, Miguel},
	month = sep,
	year = {2014},
	note = {arXiv:1407.7885 [gr-qc]},
	pages = {1773},
}

@article{cote_revisiting_2019,
	title = {Revisiting the conformal invariance of {Maxwell}'s equations in curved spacetime},
	volume = {51},
	issn = {0001-7701, 1572-9532},
	url = {http://arxiv.org/abs/1905.09968},
	doi = {10.1007/s10714-019-2599-x},
	number = {9},
	urldate = {2023-12-16},
	journal = {Gen Relativ Gravit},
	author = {Côté, Jeremy and Faraoni, Valerio and Giusti, Andrea},
	month = sep,
	year = {2019},
	note = {arXiv:1905.09968 [gr-qc, physics:hep-th]},
	pages = {117},
}

@article{leaver_spectral_1986,
	title = {Spectral decomposition of the perturbation response of the {Schwarzschild} geometry},
	volume = {34},
	url = {https://link.aps.org/doi/10.1103/PhysRevD.34.384},
	doi = {10.1103/PhysRevD.34.384},
	number = {2},
	urldate = {2023-12-17},
	journal = {Phys. Rev. D},
	author = {Leaver, Edward W.},
	month = jul,
	year = {1986},
	note = {Publisher: American Physical Society},
	pages = {384--408},
}

@book{bona_elements_2009,
	address = {Berlin, Heidelberg},
	series = {Lecture {Notes} in {Physics}},
	title = {Elements of {Numerical} {Relativity} and {Relativistic} {Hydrodynamics}: {From} {Einstein}' s {Equations} to {Astrophysical} {Simulations}},
	volume = {783},
	isbn = {978-3-642-01163-4 978-3-642-01164-1},
	shorttitle = {Elements of {Numerical} {Relativity} and {Relativistic} {Hydrodynamics}},
	url = {https://link.springer.com/10.1007/978-3-642-01164-1},
	language = {en},
	urldate = {2023-12-17},
	publisher = {Springer},
	author = {Bona, Carles and Palenzuela-Luque, Carlos and Bona-Casas, Carles},
	year = {2009},
	doi = {10.1007/978-3-642-01164-1},
}

@article{jetzer_charged_1989,
	title = {Charged boson stars},
	volume = {227},
	issn = {0370-2693},
	url = {https://www.sciencedirect.com/science/article/pii/0370269389909416},
	doi = {10.1016/0370-2693(89)90941-6},
	number = {3},
	urldate = {2023-12-17},
	journal = {Physics Letters B},
	author = {Jetzer, Ph. and Van Der Bij, J. J.},
	month = aug,
	year = {1989},
	pages = {341--346},
}

@misc{moxon_spectre_2021,
	title = {The {SpECTRE} {Cauchy}-characteristic evolution system for rapid, precise waveform extraction},
	url = {http://arxiv.org/abs/2110.08635},
	doi = {10.48550/arXiv.2110.08635},
	urldate = {2023-12-18},
	publisher = {arXiv},
	author = {Moxon, Jordan and Scheel, Mark A. and Teukolsky, Saul A. and Deppe, Nils and Fischer, Nils and Hébert, Francois and Kidder, Lawrence E. and Throwe, William},
	month = oct,
	year = {2021},
	note = {arXiv:2110.08635 [gr-qc]},
}

@misc{ma_fully_2023,
	title = {Fully relativistic three-dimensional {Cauchy}-characteristic matching},
	url = {http://arxiv.org/abs/2308.10361},
	doi = {10.48550/arXiv.2308.10361},
	urldate = {2023-12-18},
	publisher = {arXiv},
	author = {Ma, Sizheng and Moxon, Jordan and Scheel, Mark A. and Nelli, Kyle C. and Deppe, Nils and Bonilla, Marceline S. and Kidder, Lawrence E. and Kumar, Prayush and Lovelace, Geoffrey and Throwe, William and Vu, Nils L.},
	month = aug,
	year = {2023},
	note = {arXiv:2308.10361 [gr-qc]},
}

@book{wald_general_1984,
	address = {Chicago},
	edition = {UK ed. edition},
	title = {General {Relativity}},
	isbn = {978-0-226-87033-5},
	language = {English},
	publisher = {University of Chicago Press},
	author = {Wald, Robert M.},
	month = jun,
	year = {1984},
}

@misc{bishop_cauchy-characteristic_1997,
	title = {Cauchy-characteristic extraction in numerical relativity},
	url = {http://arxiv.org/abs/gr-qc/9705033},
	doi = {10.48550/arXiv.gr-qc/9705033},
	urldate = {2024-07-29},
	publisher = {arXiv},
	author = {Bishop, N. and Gomez, R. and Lehner, L. and Winicour, J.},
	month = may,
	year = {1997},
	note = {arXiv:gr-qc/9705033},
}

@article{arbab_conformal_2021,
	title = {Conformal electrodynamics in curved space},
	volume = {241},
	issn = {00304026},
	url = {https://linkinghub.elsevier.com/retrieve/pii/S0030402621006975},
	doi = {10.1016/j.ijleo.2021.167009},
	language = {en},
	urldate = {2024-08-06},
	journal = {Optik},
	author = {Arbab, A.I.},
	month = sep,
	year = {2021},
	pages = {167009},
}

@article{babiuc_implementation_2008,
	title = {Implementation of standard testbeds for numerical relativity},
	volume = {25},
	issn = {0264-9381, 1361-6382},
	url = {http://arxiv.org/abs/0709.3559},
	doi = {10.1088/0264-9381/25/12/125012},
	number = {12},
	urldate = {2024-08-17},
	journal = {Classical and Quantum Gravity},
	author = {Babiuc, M. C. and Husa, S. and Alic, D. and Hinder, I. and Lechner, C. and Schnetter, E. and Szilagyi, B. and Zlochower, Y. and Dorband, N. and Pollney, D. and Winicour, J.},
	month = jun,
	year = {2008},
	note = {arXiv:0709.3559 [gr-qc]},
	pages = {125012},
}

@article{carcione_boundary_1994,
	title = {Boundary conditions for wave propagation problems},
	volume = {16},
	copyright = {https://www.elsevier.com/tdm/userlicense/1.0/},
	issn = {0168874X},
	url = {https://linkinghub.elsevier.com/retrieve/pii/0168874X94900744},
	doi = {10.1016/0168-874X(94)90074-4},
	language = {en},
	number = {3-4},
	urldate = {2024-08-17},
	journal = {Finite Elements in Analysis and Design},
	author = {Carcione, J.M.},
	month = jun,
	year = {1994},
	pages = {317--327},
}

@misc{calabrese_discrete_2006,
	title = {Discrete boundary treatment for the shifted wave equation},
	url = {http://arxiv.org/abs/gr-qc/0509119},
	doi = {10.48550/arXiv.gr-qc/0509119},
	urldate = {2024-06-28},
	publisher = {arXiv},
	author = {Calabrese, Gioel and Gundlach, Carsten},
	month = jul,
	year = {2006},
	note = {arXiv:gr-qc/0509119
version: 2},
}

@article{kindelan_optimized_2016,
	title = {Optimized {Finite} {Difference} {Formulas} for {Accurate} {High} {Frequency} {Components}},
	volume = {2016},
	doi = {10.1155/2016/7860618},
	journal = {Mathematical Problems in Engineering},
	author = {Kindelan, Manuel and Moscoso, Miguel and Gonzalez-Rodriguez, Pedro},
	month = jan,
	year = {2016},
	pages = {1--15},
}

@misc{vano-vinuales_wave_2023,
	title = {Wave equation: analysis and discretization},
	language = {en},
	author = {Vañó-Viñuales, Alex},
	year = {2023},
}

@article{bona_symmetry-breaking_2004,
	title = {A symmetry-breaking mechanism for the {Z4} general-covariant evolution system},
	volume = {69},
	issn = {1550-7998, 1550-2368},
	url = {http://arxiv.org/abs/gr-qc/0307067},
	doi = {10.1103/PhysRevD.69.064036},
	number = {6},
	urldate = {2024-08-27},
	journal = {Physical Review D},
	author = {Bona, C. and Ledvinka, T. and Palenzuela, C. and Zacek, M.},
	month = mar,
	year = {2004},
	note = {arXiv:gr-qc/0307067},
	pages = {064036},
}

@article{bona_general-covariant_2003,
	title = {General-covariant evolution formalism for numerical relativity},
	volume = {67},
	copyright = {http://link.aps.org/licenses/aps-default-license},
	issn = {0556-2821, 1089-4918},
	url = {https://link.aps.org/doi/10.1103/PhysRevD.67.104005},
	doi = {10.1103/PhysRevD.67.104005},
	language = {en},
	number = {10},
	urldate = {2024-08-27},
	journal = {Physical Review D},
	author = {Bona, C. and Ledvinka, T. and Palenzuela, C. and Žáček, M.},
	month = may,
	year = {2003},
	pages = {104005},
}

@book{john_david_jackson_classical_1975,
	title = {Classical {Electrodynamics}, 2nd {Edition}},
	url = {http://archive.org/details/ClassicalElectrodynamics2nd},
	language = {eng},
	urldate = {2024-08-27},
	author = {{John David Jackson}},
	month = oct,
	year = {1975},
}

@article{brown_bssn_2008,
	title = {{BSSN} in {Spherical} {Symmetry}},
	volume = {25},
	issn = {0264-9381, 1361-6382},
	url = {http://arxiv.org/abs/0705.3845},
	doi = {10.1088/0264-9381/25/20/205004},
	number = {20},
	urldate = {2024-08-31},
	journal = {Classical and Quantum Gravity},
	author = {Brown, David},
	month = oct,
	year = {2008},
	note = {arXiv:0705.3845 [gr-qc]},
	pages = {205004},
}

@article{gentle_constant_2001,
	title = {Constant {Crunch} {Coordinates} for {Black} {Hole} {Simulations}},
	volume = {63},
	issn = {0556-2821, 1089-4918},
	url = {http://arxiv.org/abs/gr-qc/0005113},
	doi = {10.1103/PhysRevD.63.064024},
	number = {6},
	urldate = {2024-09-22},
	journal = {Physical Review D},
	author = {Gentle, Adrian P. and Holz, Daniel E. and Kheyfets, Arkady and Laguna, Pablo and Miller, Warner A. and Shoemaker, Deirdre M.},
	month = feb,
	year = {2001},
	note = {arXiv:gr-qc/0005113},
	pages = {064024},
}

@article{baumgarte_analytical_2007,
	title = {Analytical representation of a black hole puncture solution},
	volume = {75},
	copyright = {http://link.aps.org/licenses/aps-default-license},
	issn = {1550-7998, 1550-2368},
	url = {https://link.aps.org/doi/10.1103/PhysRevD.75.067502},
	doi = {10.1103/PhysRevD.75.067502},
	language = {en},
	number = {6},
	urldate = {2024-12-09},
	journal = {Physical Review D},
	author = {Baumgarte, Thomas W. and Naculich, Stephen G.},
	month = mar,
	year = {2007},
	pages = {067502},
}

@article{majumdar_residual_2023,
	title = {Residual gauge symmetry in light-cone electromagnetism},
	volume = {2023},
	issn = {1029-8479},
	url = {http://arxiv.org/abs/2212.10637},
	doi = {10.1007/JHEP02(2023)215},
	number = {2},
	urldate = {2025-01-14},
	journal = {Journal of High Energy Physics},
	author = {Majumdar, Sucheta},
	month = feb,
	year = {2023},
	note = {arXiv:2212.10637 [hep-th]},
}

@article{mccartor_light-cone_1994,
	title = {Light-{Cone} {Quantization} of {Gauge} {Fields}},
	volume = {62},
	issn = {0170-9739, 1434-6052},
	url = {http://arxiv.org/abs/hep-th/9311065},
	doi = {10.1007/BF01560250},
	number = {2},
	urldate = {2025-01-14},
	journal = {Zeitschrift für Physik C Particles and Fields},
	author = {McCartor, Gary and Robertson, David G.},
	month = jun,
	year = {1994},
	note = {arXiv:hep-th/9311065},
	pages = {349--355},
}

@misc{jaramillo_full_2024,
	title = {Full {3D} nonlinear dynamics of charged and magnetized boson stars},
	url = {http://arxiv.org/abs/2411.07284},
	doi = {10.48550/arXiv.2411.07284},
	urldate = {2025-01-20},
	publisher = {arXiv},
	author = {Jaramillo, Víctor and Núñez, Darío and Ruiz, Milton and Zilhão, Miguel},
	month = nov,
	year = {2024},
	note = {arXiv:2411.07284 [gr-qc]},
}

@article{zilhao_nonlinear_2015,
	title = {Nonlinear interactions between black holes and {Proca} fields},
	volume = {32},
	issn = {0264-9381, 1361-6382},
	url = {http://arxiv.org/abs/1505.00797},
	doi = {10.1088/0264-9381/32/23/234003},
	number = {23},
	urldate = {2025-01-25},
	journal = {Classical and Quantum Gravity},
	author = {Zilhão, Miguel and Witek, Helvi and Cardoso, Vitor},
	month = dec,
	year = {2015},
	note = {arXiv:1505.00797 [gr-qc]},
	pages = {234003},
}

@article{hilditch_introduction_2013,
	title = {An {Introduction} to {Well}-posedness and {Free}-evolution},
	volume = {28},
	issn = {0217-751X, 1793-656X},
	url = {http://arxiv.org/abs/1309.2012},
	doi = {10.1142/S0217751X13400150},
	number = {22n23},
	urldate = {2025-01-20},
	journal = {International Journal of Modern Physics A},
	author = {Hilditch, David},
	month = sep,
	year = {2013},
	note = {arXiv:1309.2012 [gr-qc]},
	pages = {1340015},
}

@article{vano-vinuales_spherically_2023,
	title = {Spherically symmetric black hole spacetimes on hyperboloidal slices},
	volume = {9},
	issn = {2297-4687},
	url = {http://arxiv.org/abs/2304.05384},
	doi = {10.3389/fams.2023.1206017},
	urldate = {2025-01-10},
	journal = {Frontiers in Applied Mathematics and Statistics},
	author = {Vañó-Viñuales, Alex},
	month = aug,
	year = {2023},
	note = {arXiv:2304.05384 [gr-qc]},
	pages = {1206017},
}

@article{misner_relativistic_1964,
	title = {Relativistic {Equations} for {Adiabatic}, {Spherically} {Symmetric} {Gravitational} {Collapse}},
	volume = {136},
	url = {https://link.aps.org/doi/10.1103/PhysRev.136.B571},
	doi = {10.1103/PhysRev.136.B571},
	number = {2B},
	urldate = {2025-02-09},
	journal = {Physical Review},
	author = {Misner, Charles W. and Sharp, David H.},
	month = oct,
	year = {1964},
	note = {Publisher: American Physical Society},
	pages = {B571--B576},
}

@article{zenginoglu_gravitational_2008,
	title = {Gravitational perturbations of {Schwarzschild} spacetime at null infinity and the hyperboloidal initial value problem},
	volume = {26},
	doi = {10.1088/0264-9381/26/3/035009},
	journal = {Classical and Quantum Gravity},
	author = {Zenginoğlu, Anıl and Nunez, Dario and Husa, Sascha},
	month = nov,
	year = {2008},
}

@article{bicak_gravitational_1972,
	title = {Gravitational collapse with charge and small asymmetries {I}. {Scalar} perturbations},
	volume = {3},
	copyright = {http://www.springer.com/tdm},
	issn = {0001-7701, 1572-9532},
	url = {http://link.springer.com/10.1007/BF00759172},
	doi = {10.1007/BF00759172},
	language = {en},
	number = {4},
	urldate = {2025-03-01},
	journal = {General Relativity and Gravitation},
	author = {Bičák, Jiří},
	month = dec,
	year = {1972},
	pages = {331--349},
}

@article{Ma:2024hzq,
    author = "Ma, Sizheng and Scheel, Mark A. and Moxon, Jordan and Nelli, Kyle C. and Deppe, Nils and Kidder, Lawrence E. and Throwe, William and Vu, Nils L.",
    title = "{Merging black holes with Cauchy-characteristic matching: Computation of late-time tails}",
    eprint = "2412.06906",
    archivePrefix = "arXiv",
    primaryClass = "gr-qc",
    month = "12",
    year = "2024"
}

@misc{madler_characteristic_2025,
	title = {Characteristic initial value problems for the {Einstein}-{Maxwell}-scalar field equations in spherical symmetry},
	url = {http://arxiv.org/abs/2503.24162},
	doi = {10.48550/arXiv.2503.24162},
	urldate = {2025-04-01},
	publisher = {arXiv},
	author = {Mädler, Thomas and Gannouj, Radouane and Gallo, Emanuel},
	month = mar,
	year = {2025},
	note = {arXiv:2503.24162 [gr-qc]},
}

@inproceedings{friedrich_cauchy_2000,
	address = {Berlin, Heidelberg},
	title = {The {Cauchy} {Problem} for the {Einstein} {Equations}},
	isbn = {978-3-540-46580-5},
	doi = {10.1007/3-540-46580-4_2},
	language = {en},
	booktitle = {Einstein’s {Field} {Equations} and {Their} {Physical} {Implications}},
	publisher = {Springer},
	author = {Friedrich, Helmut and Rendall, Alan},
	editor = {Schmidt, Bernd G.},
	year = {2000},
	pages = {127--223},
}

@article{Patino:2025epy,
    author = {Pati\~no, Jorge Exp\'osito and R\"uter, Hannes R. and Hilditch, David},
    title = "{Pseudospectral implementation of the Einstein-Maxwell system}",
    eprint = "2504.11069",
    archivePrefix = "arXiv",
    primaryClass = "gr-qc",
    month = "4",
    year = "2025"
}

@misc{brown_covariant_2009,
	title = {Covariant formulations of {BSSN} and the standard gauge},
	url = {https://arxiv.org/abs/0902.3652v2},
	language = {en},
	urldate = {2025-04-23},
	journal = {arXiv.org},
	author = {Brown, J. David},
	month = feb,
	year = {2009},
	doi = {10.1103/PhysRevD.79.104029},
}

@article{malec_constant_2003,
	title = {Constant mean curvature slices in the extended {Schwarzschild} solution and collapse of the lapse: {Part} {I}},
	volume = {68},
	issn = {0556-2821, 1089-4918},
	shorttitle = {Constant mean curvature slices in the extended {Schwarzschild} solution and collapse of the lapse},
	url = {http://arxiv.org/abs/gr-qc/0307046},
	doi = {10.1103/PhysRevD.68.124019},
	number = {12},
	urldate = {2025-04-23},
	journal = {Physical Review D},
	author = {Malec, Edward and Murchadha, Niall {\'O}},
	month = dec,
	year = {2003},
	note = {arXiv:gr-qc/0307046},
	pages = {124019},
}

@article{Gautam:2021ilg,
    author = "Gautam, Shalabh and Va\~n\'o-Vi\~nuales, Alex and Hilditch, David and Bose, Sukanta",
    title = "{Summation by Parts and Truncation Error Matching on Hyperboloidal Slices}",
    eprint = "2101.05038",
    archivePrefix = "arXiv",
    primaryClass = "gr-qc",
    reportNumber = "LIGO preprint number: LIGO-DCC-P2000514",
    doi = "10.1103/PhysRevD.103.084045",
    journal = "Phys. Rev. D",
    volume = "103",
    number = "8",
    pages = "084045",
    year = "2021"
}

@book{Misner1973,
  adsurl = {http://adsabs.harvard.edu/abs/1973grav.book.....M},
  author = {{Misner}, C. W. and {Thorne}, K. S. and {Wheeler}, J. A.},
  title = {Gravitation},
  publisher = {San Francisco: W.H.~Freeman and Co., 1973}, 
  editor = {{Misner, C.~W., Thorne, K.~S., \& Wheeler, J.~A.}},
  username = {mbird},
  year = "1973"
}

@article{resultspaper,
    title = {Charged {Scalar} {Field} at {Future} {Null} {Infinity} via {Nonlinear} {Hyperboloidal} {Evolution}},
	url = {http://arxiv.org/abs/2506.15311},
	doi = {10.48550/arXiv.2506.15311},
	urldate = {2025-06-24},
	publisher = {arXiv},
	author = {Álvares, João D. and Vaño-Vinũales, Alex},
	month = jun,
	year = {2025},
	note = {arXiv:2506.15311 [gr-qc]},
}

@ARTICLE{1969NCimR...1..252P,
       author = {{Penrose}, Roger},
        title = "{Gravitational Collapse: the Role of General Relativity}",
      journal = {Nuovo Cimento Rivista Serie},
         year = 1969,
        month = jan,
       volume = {1},
        pages = {252},
       adsurl = {https://ui.adsabs.harvard.edu/abs/1969NCimR...1..252P}
}

@article{gertsenshtein1961wave,
  title={Wave resonance of light and gravitational waves},
  author={Gertsenshtein, M. E.},
  journal={Sov. Phys. JETP},
  volume={41},
  pages={113--114},
  year={1961}
}

@mastersthesis{masterlachlancampion,
	author 	= "Campion, Lachlan",
	title	= "{The generalised conformal field equations in the presence of trace-free matter fields}",
	year	= "2025",
	school  = "{University of Canterbury}"
}

@article{Bishop:1996gt,
      author         = "Bishop, Nigel T. and Gomez, Roberto and Lehner, Luis and
                        Winicour, Jeffrey",
      title          = "{Cauchy characteristic extraction in numerical
                        relativity}",
      journal        = "Phys.Rev.",
      volume         = "D54",
      pages          = "6153-6165",
      doi            = "10.1103/PhysRevD.54.6153",
      year           = "1996",
      reportNumber   = "PRINT-96-225 (SOUTH-AFRICA)",
      SLACcitation   = "%%CITATION = PHRVA,D54,6153;%%",
}

@article{Taylor:2013zia,
      author         = "Taylor, Nicholas W. and Boyle, Michael and Reisswig,
                        Christian and Scheel, Mark A. and Chu, Tony and others",
      title          = "{Comparing Gravitational Waveform Extrapolation to
                        Cauchy-Characteristic Extraction in Binary Black Hole
                        Simulations}",
      journal        = "Phys.Rev.",
      number         = "12",
      volume         = "D88",
      pages          = "124010",
      doi            = "10.1103/PhysRevD.88.124010",
      year           = "2013",
      eprint         = "1309.3605",
      archivePrefix  = "arXiv",
      primaryClass   = "gr-qc",
      SLACcitation   = "%%CITATION = ARXIV:1309.3605;%%",
}

@article{PhysRevLett.10.66,
  title = {Asymptotic Properties of Fields and Space-Times},
  author = {Penrose, Roger},
  journal = {Phys. Rev. Lett.},
  volume = {10},
  issue = {2},
  pages = {66--68},
  numpages = {0},
  year = {1963},
  month = {Jan},
  publisher = {American Physical Society},
  doi = {10.1103/PhysRevLett.10.66},
  url = {http://link.aps.org/doi/10.1103/PhysRevLett.10.66}
}

@article{friedrich1983,
	journal = "Comm. Math. Phys.",
	author = "Friedrich, Helmut",
	ajournal = "Communications in Mathematical Physics",
	number = "4",
	pages = "445--472",
	publisher = "Springer",
	title = "Cauchy problems for the conformal vacuum field equations in general relativity",
	url = "http://projecteuclid.org/euclid.cmp/1103940664",
	volume = "91",
	year = "1983"
}

@article{Hubner:1998hn,
      author         = {H\"ubner, Peter},
      title          = "{How to avoid artificial boundaries in the numerical calculation of black hole spacetimes}",
      journal        = "Class.Quant.Grav.",
      volume         = "16",
      pages          = "2145",
      doi            = "10.1088/0264-9381/16/7/301",
      year           = "1999",
      eprint         = "gr-qc/9804065",
      archivePrefix  = "arXiv",
      primaryClass   = "gr-qc",
      reportNumber   = "AEI-062",
      SLACcitation   = "%%CITATION = GR-QC/9804065;%%",
}

@article{Frauendiener:1997zc,
      author         = {Frauendiener, J\"org},
      title          = "{Numerical treatment of the hyperboloidal initial value
                        problem for the vacuum Einstein equations. 1. The
                        Conformal field equations}",
      journal        = "Phys.Rev.",
      volume         = "D58",
      pages          = "064002",
      doi            = "10.1103/PhysRevD.58.064002",
      year           = "1998",
      eprint         = "gr-qc/9712050",
      archivePrefix  = "arXiv",
      primaryClass   = "gr-qc",
      SLACcitation   = "%%CITATION = GR-QC/9712050;%%",
}

@article{Rinne:2009qx,
      author         = "Rinne, Oliver",
      title          = "{An Axisymmetric evolution code for the Einstein
                        equations on hyperboloidal slices}",
      journal        = "Class.Quant.Grav.",
      volume         = "27",
      pages          = "035014",
      doi            = "10.1088/0264-9381/27/3/035014",
      year           = "2010",
      eprint         = "0910.0139",
      archivePrefix  = "arXiv",
      primaryClass   = "gr-qc",
      reportNumber   = "DAMTP-2009-63",
      SLACcitation   = "%%CITATION = ARXIV:0910.0139;%%",
}

@article{Rinne:2013qc,
      author         = "Rinne, Oliver and Moncrief, Vincent",
      title          = "{Hyperboloidal Einstein-matter evolution and tails for
                        scalar and Yang-Mills fields}",
      journal        = "Class.Quant.Grav.",
      volume         = "30",
      pages          = "095009",
      doi            = "10.1088/0264-9381/30/9/095009",
      year           = "2013",
      eprint         = "1301.6174",
      archivePrefix  = "arXiv",
      primaryClass   = "gr-qc",
      reportNumber   = "AEI-2013-043",
      SLACcitation   = "%%CITATION = ARXIV:1301.6174;%%",
}

@article{Frauendiener:2025xcj,
    author = {Frauendiener, J\"org and Stevens, Chris and Thwala, Sebenele},
    title = "{Fully Nonlinear Gravitational Wave Simulations from Past to Future Null Infinity}",
    eprint = "2504.02188",
    archivePrefix = "arXiv",
    primaryClass = "gr-qc",
    doi = "10.1103/PhysRevLett.134.161401",
    journal = "Phys. Rev. Lett.",
    volume = "134",
    number = "16",
    pages = "161401",
    year = "2025"
}

@article{Vano-Vinuales:2014koa,
      author         = "Va{\~n}{\'o}-Vi{\~n}uales, Alex and Husa, Sascha and Hilditch,
                        David",
      title          = "{Spherical symmetry as a test case for unconstrained
                        hyperboloidal evolution}",
      journal        = "Class. Quant. Grav.",
      volume         = "32",
      year           = "2015",
      number         = "17",
      pages          = "175010",
      doi            = "10.1088/0264-9381/32/17/175010",
      eprint         = "1412.3827",
      archivePrefix  = "arXiv",
      primaryClass   = "gr-qc",
      SLACcitation   = "%%CITATION = ARXIV:1412.3827;%%"
}

@article{Vano-Vinuales:2017qij,
      author         = "Va{\~n}{\'o}-Vi{\~n}uales, Alex and Husa, Sascha",
      title          = "{Spherical symmetry as a test case for unconstrained
                        hyperboloidal evolution II: gauge conditions}",
      journal        = "Class. Quant. Grav.",
      volume         = "35",
      year           = "2018",
      number         = "4",
      pages          = "045014",
      doi            = "10.1088/1361-6382/aaa4e2",
      eprint         = "1705.06298",
      archivePrefix  = "arXiv",
      primaryClass   = "gr-qc",
      SLACcitation   = "%%CITATION = ARXIV:1705.06298;%%"
}

@article{Vano-Vinuales:2023yzs,
    author = "Va\~n\'o-Vi\~nuales, Alex",
    title = "{Spherically symmetric black hole spacetimes on hyperboloidal slices}",
    eprint = "2304.05384",
    archivePrefix = "arXiv",
    primaryClass = "gr-qc",
    journal = "Front. Appl. Math. Stat., Sec. Statistical and Computational Physics",
    volume = "9",
    doi = "10.3389/fams.2023.1206017",
    year = "2023"
}

@article{Vano-Vinuales:2023pum,
    author = "Va\~n\'o-Vi\~nuales, Alex",
    title = "{Conformal diagrams for stationary and dynamical strong-field hyperboloidal slices}",
    eprint = "2311.04972",
    archivePrefix = "arXiv",
    primaryClass = "gr-qc",
    doi = {10.1088/1361-6382/ad3aca},
    url = {https://dx.doi.org/10.1088/1361-6382/ad3aca},
    journal = "Class. Quant. Grav.",
    year = {2024},
    month = {apr},
    publisher = {IOP Publishing},
    volume = {41},
    number = {10},
    pages = {105003}
}

@article{Vano-Vinuales:2024tat,
    author = "Va\~n\'o-Vi\~nuales, Alex and Valente, Tiago",
    title = "{Height-function-based 4D reference metrics for hyperboloidal evolution}",
    eprint = "2408.08952",
    archivePrefix = "arXiv",
    primaryClass = "gr-qc",
    month = "11",
    year = "2024",
    doi = "10.1007/s10714-024-03323-8",
    journal = "General Relativity and Gravitation",
    volume = "56",
    number = "135",
    publisher = "Springer"
}

@article{Morales:2016rgt,
      author         = "Morales, Manuel D. and Sarbach, Olivier",
      title          = "{Evolution of scalar fields surrounding black holes on
                        compactified constant mean curvature hypersurfaces}",
      journal        = "Phys. Rev.",
      volume         = "D95",
      year           = "2017",
      number         = "4",
      pages          = "044001",
      doi            = "10.1103/PhysRevD.95.044001",
      eprint         = "1609.05756",
      archivePrefix  = "arXiv",
      primaryClass   = "gr-qc",
      SLACcitation   = "%%CITATION = ARXIV:1609.05756;%%"
}

@article{Frauendiener:2021eyv,
    author = {Frauendiener, J\"org and Stevens, Chris},
    title = "{The non-linear perturbation of a black hole by gravitational waves. I. The Bondi\textendash{}Sachs mass loss}",
    eprint = "2105.09515",
    archivePrefix = "arXiv",
    primaryClass = "gr-qc",
    doi = "10.1088/1361-6382/ac1be3",
    journal = "Class. Quant. Grav.",
    volume = "38",
    number = "19",
    pages = "194002",
    year = "2021"
}

@article{Frauendiener:2022bkj,
    author = {Frauendiener, J\"org and Stevens, Chris},
    title = "{The non-linear perturbation of a black hole by gravitational waves. II. Quasinormal modes and the compactification problem}",
    eprint = "2211.13276",
    archivePrefix = "arXiv",
    primaryClass = "gr-qc",
    doi = "10.1088/1361-6382/acd4b1",
    journal = "Class. Quant. Grav.",
    volume = "40",
    number = "12",
    pages = "125006",
    year = "2023"
}

@article{Frauendiener:2023ltp,
    author = {Frauendiener, J\"org and Goodenbour, Alex and Stevens, Chris},
    title = "{The non-linear perturbation of a black hole by gravitational waves. III. Newman\textendash{}Penrose constants}",
    eprint = "2301.05268",
    archivePrefix = "arXiv",
    primaryClass = "gr-qc",
    doi = "10.1088/1361-6382/ad2288",
    journal = "Class. Quant. Grav.",
    volume = "41",
    number = "6",
    pages = "065005",
    year = "2024"
}

@article{Peterson:2024bxk,
    author = "Peterson, Christian and Gautam, Shalabh and Va\~n\'o-Vi\~nuales, Alex and Hilditch, David",
    title = "{Spherical Evolution of the Generalized Harmonic Gauge Formulation of General Relativity on Compactified Hyperboloidal Slices}",
    eprint = "2409.02994",
    archivePrefix = "arXiv",
    primaryClass = "gr-qc",
  journal = {Phys. Rev. D},
  volume = {110},
  issue = {12},
  pages = {124033},
  numpages = {22},
  year = {2024},
  month = {Dec},
  publisher = {American Physical Society},
  doi = {10.1103/PhysRevD.110.124033},
  url = {https://link.aps.org/doi/10.1103/PhysRevD.110.124033}
}

@article{Peterson:2023bha,
    author = "Peterson, Christian and Gautam, Shalabh and Rainho, In\^es and Va\~n\'o-Vi\~nuales, Alex and Hilditch, David",
    title = "{3D evolution of a semilinear wave model for the Einstein field equations on compactified hyperboloidal slices}",
    eprint = "2303.16190",
    archivePrefix = "arXiv",
    primaryClass = "gr-qc",
    doi = "10.1103/PhysRevD.108.024067",
    journal = "Phys. Rev. D",
    volume = "108",
    number = "2",
    pages = "024067",
    year = "2023"
}

@article{Hilditch:2015qea,
      author         = "Hilditch, David",
      title          = "{Dual Foliation Formulations of General Relativity}",
      year           = "2015",
      eprint         = "1509.02071",
      archivePrefix  = "arXiv",
      primaryClass   = "gr-qc",
      SLACcitation   = "%%CITATION = ARXIV:1509.02071;%%"
}

@article{Gasperin:2019rjg,
    author = "Gasperin, Edgar and Gautam, Shalabh and Hilditch, David and Va\~n\'o-Vi\~nuales, Alex",
    title = "{The Hyperboloidal Numerical Evolution of a Good-Bad-Ugly Wave Equation}",
    eprint = "1909.11749",
    archivePrefix = "arXiv",
    primaryClass = "gr-qc",
    doi = "10.1088/1361-6382/ab5f21",
    journal = "Class. Quant. Grav.",
    volume = "37",
    number = "3",
    pages = "035006",
    year = "2020"
}

@article{NOK,
      author         = "Nakamura, T. and Oohara, K. and Kojima, Y.",
      title          = "{General relativistic collapse to black holes and gravitational waves from black holes}",
      journal        = "Prog. Theor. Phys. Suppl.",
      volume         = "90",
      pages          = "1-218",
      year           = "1987"
}

@article{PhysRevD.52.5428,
  title = {Evolution of three-dimensional gravitational waves: Harmonic slicing case},
  author = {Shibata, Masaru and Nakamura, Takashi},
  journal = {Phys. Rev. D},
  volume = {52},
  issue = {10},
  pages = {5428--5444},
  numpages = {0},
  year = {1995},
  month = {Nov},
  publisher = {American Physical Society},
  doi = {10.1103/PhysRevD.52.5428},
  url = {http://link.aps.org/doi/10.1103/PhysRevD.52.5428}
}

@article{Baumgarte:1998te,
      author         = "Baumgarte, Thomas W. and Shapiro, Stuart L.",
      title          = "{On the numerical integration of Einstein's field
                        equations}",
      journal        = "Phys.Rev.",
      volume         = "D59",
      pages          = "024007",
      doi            = "10.1103/PhysRevD.59.024007",
      year           = "1999",
      eprint         = "gr-qc/9810065",
      archivePrefix  = "arXiv",
      primaryClass   = "gr-qc",
      SLACcitation   = "%%CITATION = GR-QC/9810065;%%",
}

@article{verma2014,
author = {Verma, Anjali and Jiwari, Ram and Kumar, Satish},
year = {2014},
month = {08},
pages = {},
title = {A numerical scheme based on differential quadrature method for numerical simulation of nonlinear Klein-Gordon equation},
volume = {24},
journal = {International Journal of Numerical Methods for Heat \& Fluid Flow},
doi = {10.1108/HFF-01-2013-0014}
}

@article{jiwari2020,
author = {Jiwari, Ram},
year = {2020},
month = {11},
pages = {},
title = {Barycentric rational interpolation and local radial basis functions based numerical algorithms for multidimensional sine‐Gordon equation},
volume = {37},
journal = {Numerical Methods for Partial Differential Equations},
doi = {10.1002/num.22636}
}

@article{JIWARI2012600,
title = {Numerical simulation of two-dimensional sine-Gordon solitons by differential quadrature method},
journal = {Computer Physics Communications},
volume = {183},
number = {3},
pages = {600-616},
year = {2012},
issn = {0010-4655},
doi = {https://doi.org/10.1016/j.cpc.2011.12.004},
url = {https://www.sciencedirect.com/science/article/pii/S0010465511003894},
author = {Ram Jiwari and Sapna Pandit and R.C. Mittal}
}

@misc{VandeMoortel:2017ztd,
  author        = {Van de Moortel, Maxime},
  title         = {Stability and Instability of the Sub-extremal Reissner–Nordström Black Hole Interior for the Einstein–Maxwell–Klein–Gordon Equations in Spherical Symmetry},
  note          = {in Commun. Math. Phys. 360, 103--168 (2018)},
  year          = {2018},
  eprint        = {1704.05790},
  archivePrefix = {arXiv},
  primaryClass  = {gr-qc}
}

@article{Kauffman:2021hgs,
    author = "Kauffman, Christopher and Lindblad, Hans",
    title = "{Global Stability of Minkowski Space for the Einstein{\textendash}Maxwell{\textendash}Klein{\textendash}Gordon System in Generalized Wave Coordinates}",
    eprint = "2109.03270",
    archivePrefix = "arXiv",
    primaryClass = "gr-qc",
    doi = "10.1007/s00023-023-01331-z",
    journal = "Annales Henri Poincare",
    volume = "24",
    number = "11",
    pages = "3837--3919",
    year = "2023"
}

@article{Dong_2021,
   title={Asymptotic Behavior of the Solution to the Klein–Gordon–Zakharov Model in Dimension Two},
   volume={384},
   ISSN={1432-0916},
   url={http://dx.doi.org/10.1007/s00220-021-04003-3},
   DOI={10.1007/s00220-021-04003-3},
   number={1},
   journal={Communications in Mathematical Physics},
   publisher={Springer Science and Business Media LLC},
   author={Dong, Shijie},
   year={2021},
   month=feb, pages={587–607} }

\newpage

\end{document}